\documentclass[12pt,a4paper]{article}   
\usepackage[dvips]{graphicx}
\usepackage{amsmath}
\usepackage{amssymb}
\usepackage{amsthm}
\usepackage{amscd}

\def\m@th{\mathsurround=0pt}

\newcommand{\np}{\nonumber\\}
\newtheorem{theorem}{Theorem}
\newtheorem{lemma}{Lemma}
\newtheorem{proposition}{Proposition}
\newtheorem{corollary}{Corollary}

\newtheorem{conjecture}{Conjecture}
\def\pn{\par\noindent}
\def\CL{ {\cal L}}
\def\ch{{\rm ch}}

\def\nt{\negthickspace}

\title{
The dilute $A_L$ models and the $\Phi_{1,2}$ perturbation of unitary minimal CFTs. 
}
\author{ J. Suzuki\thanks{e-mail: sjsuzuk@ipc.shizuoka.ac.jp}\\
        \parbox{0.9\textwidth}{
        {\em
        \begin{center}
      Department of Physics,\\
      Faculty of Science, Shizuoka University \\
      Ohya 836, Shizuoka\\
       Japan
        \end{center}
        }}
       }
\date{}
\begin{document}
\maketitle
\begin{abstract}
Motivated by recent studies by Dorey, Pocklington and Tateo\cite{DPT1, DPT2}for 
 unitary minimal models perturbed by $\phi_{1,2}$, 
we examine the  thermodynamics of one dimensional quantum systems,
whose counterparts  in the 2D classical model are the dilute $A_L$ models in regime  2.
The functional relations for arbitrary values of $L$ are established.
Guided by numerical evidences, 
 we obtain  a set of coupled integral equations from the established relations, which 
 yields the  evaluation of the free energy at arbitrary temperature.
In the scaling limit, the integral equations coincide with the thermodynamic
Bethe ansatz equations (TBA) proposed in \cite{DPT2}, thereby support
their results.
The new Fermionic representations of the Virasoro
characters are shortly remarked.
\end{abstract}

\clearpage

%%%%%%%%%%%%%%%%%%%%%%%%%%%%%%%%%%%%%%%%%%%%%%%
%
\section{Introduction}
%
%%%%%%%%%%%%%%%%%%%%%%%%%%%%%%%%%%%%%%%%%%%%%%%%
%

The Andrews-Baxter Forrester model \cite{ABF} has been a prototype 
in the studies of   integrable lattice models as well as the integrable
$\Phi_{1,3}$  perturbation of the minimal unitary CFT.

The dilute $A_L$ model proposed in \cite{WNS1, WNS2}  deserves the same attention,
as it represents a lattice analogue of the 
$\phi_{1,2}$ and $\phi_{2,1}$   perturbations of the minimal unitary CFT.

Especially, there are abundant research results  on  the dilute $A_3$ model in regime 2,
which corresponds to the model of high interest,   $M_{3,4}+\Phi_{1,2}$ theory.
The latter belongs to the same universality class as the Ising model in a magnetic field at $T_c$,
and is characterized by the famous $E_8$ structure\cite{ZamolodchikovE8}-\cite{DelfMuss}.
There are several evidences for this coincidence, the central charge\cite{WNS1, WNS2}, surface exponents
\cite{BFZ}
universal amplitude ratio \cite{BS1, BS2, KS02}  scaling dimensions\cite{WPNS} and excitation spectra \cite{MO97}.
The most impressive demonstration may be presented in \cite{BWN}, which confirms
the hidden $E_8$ structure  behind the 
 thermodynamic Bethe ansatz equation (TBA, for short)
 of the  dilute $A_3$ model.  See also  \cite{GN1, GN2}.
 
Similarly,   $M_{4,5}+\Phi_{1,2}$($ M_{6,7}+\Phi_{1,2}$ ) theory
has  the hidden $E_7 (E_6)$ structure behind \cite{ZamolodchikovE8}. 
In spite of  lack of  relevant string hypothesis, 
  the technique of the quantum transfer matrix (QTM)
and the fusion relations make it  possible to manifest the desired Lie algebraic
structure in the TBA of  the dilute $A_{3,4,6}$ in regime 2\cite{SuzE8, SuzE7}.

The aim of the present report is to extend the result further; we will 
establish a closed set of fusion relations of the
dilute $A_L$ model  for {\itshape arbitrary} 
$L$.  From these relations, we derive the nonlinear integral equations (TBA) which yields
the evaluation of the free energy of the corresponding 1D quantum model
at arbitrary temperatures.
This is obviously motivated by the recent progress in the study of the models as perturbed
conformal field theories 
\cite{DPT1, DPT2}.
The perturbation theory by an operator  $\Phi_{1,2}$ or $ \Phi_{2,1}$  encountered serious 
technical difficulties.
The systematic studies on the bootstrap procedure on $S$ matrix have been
initiated in \cite{Smir} and \cite{ChimZam}.  See also \cite{ KMM92, Koubek94}.
The latter approach bases on the scaling $q-$ state Potts field theory and will be relevant 
in our view point.
It is only recently  that  the closed  bootstrap procedure, associated to the Potts field theory, 
has been obtained for a set of $S-$ matrices for a wide range of parameters
  by Dorey, Pocklington and Tateo\cite{DPT1}.
The check of the results against a finite size system however, again poses a challenging problem
because of  the non-diagonal nature of the scattering theory. 
In \cite{DPT2},  a set of  involved TBA is conjectured  by ingenious 
insights from special cases which have similarity to the TBA for the sine-Gordon model.
They prepare  pieces of TBA equations, one of which comes from $Z_n$ models\cite{DTT}, then glue them together by respecting symmetries and by physical requirements.
The resultant equations pass several nontrivial checks,  however, lack firm grounds.
It is thus interesting to see if it is possible to recover the $Y-$ system
\cite{AlBZamolodchiklovY, KunibaNakanishi,  TateoSG}, the algebraic version of the TBA,
from the fusion relations of the dilute $A_L$ model.

The functional relations among transfer matrices  
 (the $T-$ system)  are determined by the fusion relations\cite{KRS, KNS1}.
The decomposition of the product of two modules 
at the singular points of the Boltzmann weight  rules the latter.
Thus the problem seems to be straightforward, at the first sight.
This is , unfortunately, not true.
 There are often infinitely many  functional relations for a given model.
 One thus has to  choose a subset which offers the closed  $T-$ system
 with finitely many  entries.
Moreover,  among the sets, the only those of which both sides contain fusion transfer matrices
 with good analytic properties 
(to be specified)  will be relevant for the TBA.
Therefore the problem of finding the relevant  functional relations  so as to reach the TBA is far 
from trivial.

It is well known that the dilute $A_L$ model has $U_q(A^{(2)}_2)$ symmetry at the critical point. 
The off-critical extension,  deformed Virasoro algebra ${\cal V}_{x,r}(A^{(2)}_2)$ plays a role 
in the vertex operator approach \cite{HJKOS}.
This elegant result, however, is not of our direct relevance. 
The crucial fact  for the present study is, rather, that the  fusion structure which is valid
even away from criticality \cite{GPZ}; 
 it possesses the fusion structures 
similar to $U_q(A^{(1)}_1)$ and 
 $U_q(A^{(1)}_2)$ . 
 We will relate the former  to  breathers while the 
 latter to the pseudo-particles constituting ladders in their TBA diagrams.
 In addition, we will introduce a set of objects which we call "magnon-like", 
 characterized by  similar diagrams to those associated with simply-laced Lie algebra ($A,D$ and $E$).
It turns out that they constitute closed functional relations with the good
analyticity.
They are  three almost independent relations but for one common object
 $T_{B_3}$, which plays the role of glue.
 %
%Thus our natural interpretation of the gluing procedure is as follows; three  pieces ,
%each related to $U_q(A^{(1)}_1)$, 
%$U_q(A^{(1)}_2)$  and  possibly to ADE series,  are coupled together
%by the glue  $T_{B_3}$ .
%
Thus the construction employed here makes  the meaning of gluing in \cite{DPT2} very explicit.

We will introduce auxiliary functions,  fusion transfer matrices,
\footnote{We expect them to be  the eigenvalues of transfer matrices
of which fusion types are described by the diagrams.
This is however yet to be proved.
As our argument does not rely on whether this coincidence is true or not,
we simply call the  auxiliary functions  as (eigenvalues of)  fusion transfer matrices.}
labelled  by  skew Young diagrams\cite{BR, KS, SuzG2, Tsuboi97}.
In view of the latter diagrams,  proofs of the functional relations seem to be
similar to those of the decomposition rules in the symmetric group.
There are, however, extra complication due to the fact that 
our coupling constant in a sense lies at a "root of unity".
Sometimes two  (or even more)  diagrams, with completely different
disguise, 
correspond to an identical  fusion transfer matrix.
One must choose  relevant expressions (diagrams) case by case
in order to obtain suitable decompositions.
This makes the proof of the  functional relations simple but lengthy, unfortunately.
There are four families of functional relations depending on the values of $L$
and they  need separate treatment at the present stage.

The validity of the obtained algebraic equations does not necessary legitimate the
integral equations (TBA).  
We need another step to clarify  the analytic structure of $Y-$ functions.
This remains an important issue to be understood.
In this report, we employ the  direct numerical calculation by 
various range of parameters and adopt a simple conjecture  on the analyticity.
With the established  functional relations  and  the assumption on  the analyticity, 
it is then straightforward to transform the
$T-$ system into $Y-$ system and then $Y-$ system
 into coupled integral equations\cite{KluemperPearce92}.
In the scaling limit,  the TBA proposed in \cite{DPT2} is recovered from them.
A simple ansatz, suggested by numerics, on the zeros of   fusion transfer matrices
remarkably leads to the correct mass ratio of the particles determined by $S-$ 
matrix poles.

This paper is organized as follows.
In the next section, we give a brief review on the dilute
$A_L$ models and the QTM method. 
Some  obvious consistency of the lattice models and the CFT perturbed by
$\Phi_{1,2}$ will be remarked.
Section 3 is devoted to the discussion on  fusion transfer matrices
which are  parameterized by skew Young diagrams. 
We introduce there important kinds of objects in section 4, 5 and 6.
The first object introduced in section 4 seems to correspond to the breathers
in \cite{DPT1} while the second one in section 5  is responsible for the ladder structure in
the $Y-$ system diagram in  \cite{DPT2}.  The section 6 is divided into 4 subsections,
each is devoted to the discussion on "magnon-like"objects.
 In case of  the dilute $A_L$ model, $L$ even,  a fundamental role seems 
 to be played by a  "kink " transfer matrix, which is introduced in section 7.
 We will give a remark on the "reality property" of  the magnon-like objects in section 8.
 Once fusion relations, the $T-$ system, are established, it is readily transformed into
 $Y-$ system. The explicit procedure will be discussed in section 9.
 The proofs of the $T-$ system (or  the "magnon-like" $t-$ system) are simple but lengthy.
 Those for $L$ even (odd) are somewhat similar.
 For brevity,  we thus present the proof for $L=4k+2$ in the section 10 and $L=4k+1$
 in the appendix F.
In section 11,  it will be shown that the conjectured TBA 
 is naturally restored in a scaling limit.
We conclude the paper with brief summary and discussion in section 12.
A new fermionic formula for the Virasoro character found for $M_{5,6}$
is briefly commented.

%
%%%%%%%%%%%%%%%%%%%%%%%%%%%%%%%%%%%%%%%%%%%%%%%%%
%
\section{The dilute $A_L$ models and the quantum transfer matrix}
\begin{figure}[hbtp]
\centering
\includegraphics[width=6cm]{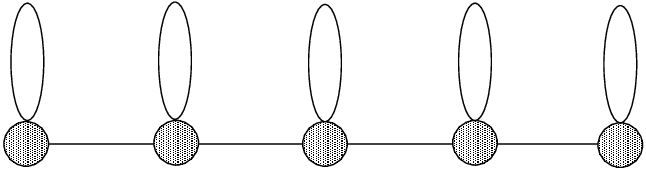}
\caption{ 
An incidence diagram for the dilute $A_5$ model. The local states corresponding to
connected nodes can be located to nearest neighbor sites on a square lattice.
  }
\label {adj}
\end{figure}

%
%%%%%%%%%%%%%%%%%%%%%%%%%%%%%%%%%%%%%%%%%%%%%%%%%
The dilute $A_L$ model is proposed in \cite{WNS1} as
an elliptic extension of the Izergin-Korepin model \cite{IK}.
The model is of the restricted SOS type with local 
variables $\in \{1,2,\cdots,L \}$.
The variables $\{a, b\} $ on neighboring sites
should satisfy the adjacency condition, $|a-b|\le 1$,
which is often described by a graph in fig.\ref{adj}.
%
%------------------------------------------------------------------
%
%
%-------------------------------------------------------------------
%
The following RSOS weights are found in \cite{WNS1} which  satisfy the Yang-Baxter relation.
They are parameterized by the spectral parameter $u$, the elliptic nome $q={\rm e}^{-\tau}$.

\begin{eqnarray}
\raise 2mm \vtop{\hbox{$a$}\hbox{$a$}}
\,\framebox[0.4cm][c]{$u$} \,
\raise 2mm \vtop{\hbox{$a$}\hbox{$a$}}
&=& \frac{\theta_1(6-u)\theta_1(3+u)}{\theta_1(6)\theta_1(3)}- 
    \frac{\theta_1(u)\theta_1(3-u)}{\theta_1(6)\theta_1(3)} \times \np
& & \Bigl(
           \frac{S_{a+1}}{S_a} \frac{\theta_4(2a-5)}{\theta_4(2a+1)}
          +\frac{S_{a-1}}{S_a} \frac{\theta_4(2a+5)}{\theta_4(2a-1)}
    \Bigr ), \np
%--------------------------------------------------1
\raise 2mm \vtop{\hbox{$a\pm 1$}\hbox{$\quad a$}}
\,\framebox[0.4cm][c]{$u$} \>
\raise 2mm \vtop{\hbox{$a$}\hbox{$a$}}&=&
\raise 2mm \vtop{\hbox{$a$}\hbox{$a$}}
\> \framebox[0.4cm][c]{$u$} \>
\raise 2mm \vtop{\hbox{$a$}\hbox{$a\pm 1$}}
=\frac{\theta_1(3-u)\theta_4(\pm 2a+1-u)}{\theta_1(3)\theta_4(\pm 2a +1)},
\np
%---------------------------------------------------2
\raise 2mm \vtop{\hbox{$\quad a$}\hbox{$a\pm 1$}}
\,\framebox[0.4cm][c]{$u$} \>
\raise 2mm \vtop{\hbox{$a$}\hbox{$a$}}&=&
\raise 2mm \vtop{\hbox{$a$}\hbox{$a$}}
\> \framebox[0.4cm][c]{$u$} \>
\raise 2mm \vtop{\hbox{$a\pm 1$}\hbox{$a$}}
=\Bigl ( \frac{S_{a\pm 1}}{S_a} \Bigr )^{1/2}
\frac{\theta_1(u)\theta_4(\pm 2a-2+u)}{\theta_1(3)\theta_4(\pm 2a +1)},
\np
%------------------------------------------------------3
\raise 2mm \vtop{\hbox{$a$}\hbox{$a$}}
\> \framebox[0.4cm][c]{$u$} \>
\raise 2mm \vtop{\hbox{$a\pm 1$}\hbox{$a\pm 1$}}&=&
\raise 2mm \vtop{\hbox{$a\pm 1$}\hbox{$\quad a$}}
\> \framebox[0.4cm][c]{$u$} \>
\raise 2mm \vtop{\hbox{$a\pm 1$}\hbox{$\quad a$}}
=\Bigl ( \frac{\theta_4(\pm 2a +3)\theta_4(\pm 2a-1)}
              {\theta_4^2(\pm 2a+1)} \Bigr )^{1/2}
\frac{\theta_1(u)\theta_1(3-u)}{\theta_1(2)\theta_1(3)},
\np
%---------------------------------------------------4
\raise 2mm \vtop{\hbox{$a\pm 1$}\hbox{$\quad a$}}
\> \framebox[0.4cm][c]{$u$} \>
\raise 2mm \vtop{\hbox{$\quad a$}\hbox{$a\mp 1$}}&=&
\frac{\theta_1(2-u)\theta_1(3-u)}{\theta_1(2)\theta_1(3)},
\np
%----------------------------------------------------5
\raise 2mm \vtop{\hbox{$\quad a$}\hbox{$a\pm 1$}}
\> \framebox[0.4cm][c]{$u$} \>
\raise 2mm \vtop{\hbox{$a\mp 1$}\hbox{$\quad a$}}&=&
-\Bigl(  \frac{S_{a-1} S_{a+1}}{S_a^2} \Bigr )^{1/2}
\frac{\theta_1(u)\theta_1(1-u)}{\theta_1(2)\theta_1(3)},
\np
%----------------------------------------------------6
\raise 2mm \vtop{\hbox{$\quad a$}\hbox{$a\pm 1$}}
\> \framebox[0.4cm][c]{$u$} \>
\raise 2mm \vtop{\hbox{$a\pm 1$}\hbox{$\quad a$}}&=&
\frac{\theta_1(3-u)\theta_1(\pm 4a+2+u)}{\theta_1(3)\theta_1(\pm 4a+2)} \np
& &+
\frac{S_{a\pm 1}}{S_a}
\frac{\theta_1(u)\theta_1(\pm 4a-1+u)}{\theta_1(3)\theta_1(\pm 4a+2)},  \quad
\hbox{ for } \pm 4a+2 \ne 0, \np
 &=&\frac{\theta_1(3+u)\theta_1(\pm4 a-4+u)}{\theta_1(3)\theta_1(\pm4 a-4)} \np
 & &+
 \Bigl (
        \frac{S_{a\mp1}\theta_1(4)}{S_a \theta_1(2)}-
	    	  \frac{\theta_4(\pm 2 a-5)}{ \theta_4(\pm 2a+1)}
                      \Bigr )
		\frac{\theta_1(u) \theta_1(\pm 4a-1+u)}
		     {\theta_1(3) \theta_1(\pm 4a-4)},  \hbox{  otherwise }. 
%			 
%-------------------------------------------------------7
\end{eqnarray}
Here $\theta_{1,4}(x) =\vartheta_{1,4}(\lambda x, \tau)$, 
\begin{eqnarray*}
\vartheta_1(x,\tau) &=&2 q^{1/4}\sin x
\prod_{n=1}^{\infty}(1-2q^{2n} \cos 2x+q^{4n})(1-q^{2n}), \\
\vartheta_4(x,\tau) &=&
\prod_{n=1}^{\infty} (1-2q^{2n-1} \cos 2x+q^{4n-2})(1-q^{2n}),\\
S_a&=&(-1)^a \frac{\theta_1(4a)}{\theta_4(2a)}.
\end{eqnarray*}
Note an important property, the standard initial condition,  that
only those weights with identical variables on the NE and the SW diagonal 
remain nonzero when $u \rightarrow 0$.

The crossing parameter $\lambda$ needs to be a function of $L$ for the restriction. 
The model exhibits four different physical regimes depending on parameters\cite{WNS1, WNS2},
\begin{itemize}
 \item regime 1. $0<u<3 , \,\lambda=\frac{\pi L}{4(L+1)},\, L\ge 2 $
 \item regime 2. $0<u<3, \, \lambda=\frac{\pi (L+2)}{4(L+1)},\,  L \ge 3$
 \item regime 3. $3-\frac{\pi}{\lambda}<u<0, \,
  \lambda=\frac{\pi (L+2)}{4(L+1)}, \, L \ge 3 $
 \item regime 4.  $3-\frac{\pi}{\lambda}<u<0,\, 
  \lambda=\frac{\pi L}{4(L+1)}, \, L \ge 2$.
\end{itemize}
In the low temperature limit, the ground states correspond to ferromagnetic ordered states
in regime 1 and 2 in which we are interested.

As usual, we define the row to row transfer matrix 
$T_{\hbox{\footnotesize{RTR}}}(u)$
by
$$
(T_{\hbox{\footnotesize{RTR}}}(u))^{\{b\}}_{\{a\}} =\prod_{j=1}^M 
\raise 2mm \vtop{\hbox{$b_j$}\hbox{$a_j$}}
\> \framebox[0.4cm][c]{$u$} \>
\raise 2mm \vtop{\hbox{$b_{j+1}$}\hbox{$a_{j+1}$}}.
$$
Denoting its largest eigenvalue  by  the same symbol, $T_{\hbox{\footnotesize{RTR}}}(u)$,
the free energy per site of the 2D classical model in thermodynamic limit reads,
$
\beta f= -\lim_{M \rightarrow \infty} \frac{1}{M} \log  T_{\hbox{\footnotesize{RTR}}}(u).
$
Several other quantities other than the free energy have been
evaluated for the models.  
For example, we mention the explicit  one point functions elaborated in \cite{WPNS}.

We are not dealing with the dilute $A_L$ model directly but its
 one dimensional quantum counterpart.

The Hamiltonian for the latter is defined by
\begin{equation}
{\cal H} = - 2 \frac{\partial}{\partial u} 
\ln T_{\hbox{\footnotesize{RTR}}}(u) |_{u=0}
\label{hamiltransfer}
\end{equation}
as in \cite{BWN}.  
The explicit form of the Hamiltonian is not needed in the following argument.

We are interested in the free energy of the 1D model at finite temperatures.
The success in the previous studies on the thermodynamics of 1D quantum systems
depends on the efficient conjectures on the dominant solutions
 of the Bethe ansatz equation
in the thermodynamic limit ,  the string hypothesis.
It seems too difficult to perform the same procedure 
for arbitrary $L$.
Instead, we apply another machinery, the method of the quantum transfer matrix (QTM)\cite{Suz85}-\cite{Tak91}.

The quantity of our interest is the partition function of the 1D model,
$$
Z_{\rm 1D}= {\rm tr} {\rm e}^{-\beta {\cal H}}.
$$
The first step is to rewrite this as the partition function of the 2D model,
defined on $M \times N$ square lattice by a simple trick,
\begin{equation}
Z_{\rm 1D}= \lim_{N \rightarrow \infty} {\rm tr}  ({\rm e}^{-\frac{\beta}{N} {\cal H}} )^N
               = \lim_{N \rightarrow \infty} {\rm Tr} ({\cal T}^{\vee})^N
                =\lim_{N \rightarrow \infty} {\rm Tr}  Z_{\rm 2D}.
	\label{partition1D}
\end{equation}
The fictitious dimension,
$N$,  is
referred to as the Trotter number.
${\cal T}^{\vee}$ is a kind of transfer matrix , 
composed of matrix elements of  ${\rm e}^{-\frac{\beta}{N} {\cal H}} $.

In order to relate ${\cal T}^{\vee}$  to the row to row transfer matrix above,
we introduce a rotated transfer matrix,
$\tilde{T}_{\hbox{\footnotesize{RTR}}}(u)$ ,

$$
(\tilde{T}_{\hbox{\footnotesize{RTR}}}(u))^{\{b\}}_{\{a\}} =\prod_{j=1}^M 
\raise 2mm \vtop{\hbox{$b_j$}\hbox{$a_j$}}
\> \framebox[0.4cm][c]{\rotatebox{90}{$u$}} \>
\raise 2mm \vtop{\hbox{$b_{j+1}$}\hbox{$a_{j+1}$}}.
$$

Thank to  (\ref{hamiltransfer})  and the standard initial condition, 
we immediately see,
$$
T_{\hbox{\footnotesize{RTR}}}(u) \sim  P_R {\rm e}^{-\frac{u }{2} {\cal H} +O(u^2)}, \qquad
\tilde{T}_{\hbox{\footnotesize{RTR}}}(u) \sim P_L {\rm e}^{-\frac{u}{2} {\cal H} +O(u^2)},
$$
where $P_R (P_L)$ is the right (left) shift operator, commuting with the Hamiltonian
 and $P_L P_R=P_R P_L=$
the identity operator.
Then the desired expression for  ${\cal T}^{\vee}$ is obtained
$$
{\cal T}^{\vee}=   T_{\hbox{\footnotesize{RTR}}}(\frac{\beta}{N})  \tilde{T}_{\hbox{\footnotesize{RTR}}}(\frac{\beta}{N}) 
$$
where the limit $N \rightarrow \infty$ is assumed in the both sides.
The eigenvalues of ${\cal T}^{\vee}$ are highly degenerated  in $N \rightarrow \infty$ , which makes the summation
in  the partition function difficult to perform.  
The intriguing observation \cite{Suz85} is that if one defines yet another transfer matrix  ${\cal T}$,
the quantum transfer matrix (QTM), 
propagating in the horizontal direction,  then there is a gap in between the largest and the second largest
eigenvalues which remains non vanishing in the limit $N \rightarrow \infty$.
Now that $Z_{\rm 1D} \sim  {\rm Tr}  {\cal T}^M $ and $M \rightarrow \infty$, 
 the {\itshape  only} largest eigenvalue of ${\cal T}$ is necessary in the evaluation of  (\ref{partition1D}).

In the  most sophisticated formulation\cite{Klu92},  the QTM is defined with one further variable $x$,

$$
(T_{\hbox{\footnotesize{QTM}}}(u,x))^{\{b\}}_{\{a\}} =\prod_{j=1}^{N/2}
\mbox{\parbox[c][0.9cm]{0cm}{}}^{b_{2j-1}}_{a_{2j-1}}
\fbox{\parbox[c][0.7cm]{0.7cm}{ {\scriptsize $u\!+\! ix$}  }} \>
\mbox{\parbox[c][0.9cm]{0cm}{}}^{b_{2j}}_{a_{2j}}
\> \>
\mbox{\parbox[c][0.9cm]{0cm}{}}^{\,\,b_{2j}}_{a_{2j}}
\fbox{\parbox[c][0.7cm]{0.7cm}{ \; \rotatebox{90}{{\scriptsize $u\!-\! ix$}}  }} \>
\mbox{\parbox[c][0.9cm]{0cm}{}}^{b_{2j+1}}_{a_{2j+1}}.
$$
By taking $x=0$ one recovers  ${\cal T}$ .

The commutative property of QTMs with fixed $u$ is crucial in the QTM formulation,
$$
[T_{\hbox{\footnotesize{QTM}}}(u,x), T_{\hbox{\footnotesize{QTM}}}(u,x') ]=0.
$$
It represents the existence of infinitely many conserved quantities commuting with
QTM, which may be a key in the exact enumeration of the free energy at arbitrary temperatures.
Since we will adopt identical value for $u=\frac{\beta}{N}$ for commuting QTM,  it will be dropped
afterwards.

Let us explicitly write again that the free energy per site is represented {\it only} by the
largest eigenvalue of $T_{\hbox{\footnotesize{QTM}}}$ with
$x=0$ and
$u= \frac{\beta}{N}$
\begin{eqnarray*}
\beta f &=&
-\lim_{M\rightarrow \infty} \frac{1}{M} 
   \ln {\hbox{Tr }} \exp(-\beta {\cal H})  \\
 &=& -\lim_{N\rightarrow \infty}\ln  \Bigl( \hbox{the largest eigenvalue of }
 T_{\hbox{\footnotesize{QTM}}}(x=0) \Bigr ).
\end{eqnarray*}

Our aim is thus to evaluate the largest eigenvalue of QTM. 
This is again not so simple problem as it seems at the first sight.
The coupling constant $u$ depends on
the fictitious system size, Trotter number, which makes it difficult to take the
limit $N \rightarrow \infty$.

Our strategy is to introduce auxiliary transfer matrices (fusion transfer matrices) 
commuting with QTM and  to explore functional relations among them \cite{ SuzE8, 
SuzE7, Klu93, JKS98, KSS98,ST00}.
We find the latter relations, if suitable subset is chosen, are efficient enough to yields
the evaluation of  the largest eigenvalue of $T_{\hbox{\footnotesize{QTM}}}$ for
arbitrary $N$:  The functional relations can be transformed into coupled integral equations
whose solution yields the desired quantity.
The resultant equations admit to take the limit $N \rightarrow \infty$
without any difficulty.

In the next section, we will introduce the fusion transfer matrices.
Before doing so, we mention a simple  evidence that the  dilute $A_L$ model 
(or its 1D analogue) is a lattice analogue of the minimal unitary model perturbed by
$\Phi_{1,2}$ or $\Phi_{2,1}$ operator.
In TBA formulation, the dressed energy function $\epsilon$ and $Y=\exp(\epsilon)$ are
main objects.  
It is pointed out in \cite{KlassenMeltzer} 
that, within framework of TBA, it seems too difficult to show
analytically that the small $r$ expansion coincides with the result from
the conformal perturbation theory.
The hidden periodicity of the $Y-$ system, found by a brute-force  substitution,
partially resolves the problem \cite{AlBZamolodchiklovY}.

We will show, within our formalism, $Y$ is expressible by ratios of  fusion transfer matrices.
Therefore in our view point,  the periodicity of  $Y$ (and $\epsilon$ )
is  a consequence of the periodicity of the original Boltzmann weights.
 In view of the rapidity variable $\theta \equiv \frac{x \pi}{3}$,  the periods read

\begin{equation}
P_{\theta} =
\begin{cases}
\frac{4(L+1) \pi} {3 L}   ,& \hbox{ for regime 1} \\
 \frac{4(L+1) \pi}{ 3(L+2) }  ,&  \hbox{ for regime 2}.
\end{cases}
%\label{period}
\end{equation}

From this one can read off the scaling dimension of the perturbed field by
$ \triangle = 1-\frac{\pi}{P_{\theta}}$, explicitly, 

\begin{equation}
\triangle =
\begin{cases}
\frac{p+3} {4 p}   ,& \hbox{ for regime 1} \\
 \frac{p-2}{ 4(p+1) }  ,&  \hbox{ for regime 2}.
\end{cases}
\end{equation}
Note that we adopted a different identification,

\begin{equation}
p =
\begin{cases}
L+1  & \hbox{ for regime 1} \\
 L  &  \hbox{ for regime 2}.
 \label{p-Lindentify}
\end{cases}
\end{equation}

 They coincide with the scaling dimensions
 $\triangle_{2,1}$ and $\triangle_{1,2}$ in the unitary model 
 $M_{p, p+1}$. 
 This thus supports the hypothesis that the dilute $A_L$ models
 in regime 1 (regime 2) are lattice regularizations of 
 perturbed minimal models   $M_{p, p+1}$ under the identification
 (\ref{p-Lindentify}). 
 
 From now on, we restrict our argument to the regime 2.
 %This is not due to additional technical difficulties but simply
% because of lack of spaces
As is announced,   the fusion transfer matrices will be examined in the next
section.
%
%=========================================================
%
\section{The fusion transfer matrices and quantum Jacobi-Trudi formula}
%
%=========================================================
%
Any model integrable in the sense of the Yang-Baxter relation offers
a systematic way of generalizations, the fusion  procedure.
It utilizes the singular values of spectral parameter at which the Boltzmann weights reduce
to projectors to a subspace.
In \cite{GPZ},  a detailed description is given for 
the explicit procedures to obtain "generalized" Boltzmann weights.

The explicit construction of these weights for various fusion types is
not of issue in this report.  Rather, the main objects of our interest are
functional relations which involves QTM. 
In the following we introduce several auxiliary functions which generalizes
QTM.  We expect that they corresponds to eigenvalues of transfer matrices
made by the fusion procedure.
We, however, do not attempt to prove this, as it is not necessary for the evaluation
of free energy.

The most important building block in our argument is the following expression for
the (any) eigenvalues of QTM, which we will denote by $T_1(x)$,

\begin{eqnarray}
T_1(x)&=& 
w \phi(x+\frac{3}{2}i) \phi(x+\frac{1}{2}i)\frac{Q(x-\frac{5}{2} i)}{Q(x-\frac{1}{2} i)} +
\phi(x+\frac{3}{2}i) \phi(x-\frac{3}{2}i) 
\frac{Q(x-\frac{3}{2}i)\, Q(x+\frac{3}{2} i)}{Q(x-\frac{1}{2} i)\, Q(x+\frac{1}{2} i)} \nonumber \\
&+ &w^{-1} \phi(x-\frac{3}{2}i) \phi(x-\frac{1}{2}i)  
 \frac{Q(x+\frac{5}{2} i)}{Q(x+\frac{1}{2} i)},  \label{dvf1}\\
Q(x)&:=&\prod_{j=1}^{N} h[x-x_j] \np
\phi(x) &:=& \Bigl(\frac{h[x+(\frac{3}{2}-u)i] h[x-(\frac{3}{2}-u)i]}{h[2i] h[3i]}\Bigr)^{\frac{N}{2}},
\qquad h[x] := \theta_1(ix),  \nonumber
\end{eqnarray}
where  $w=\exp(i \frac{\pi \ell}{L+1})$  and $\ell=1$ for the largest eigenvalue
sector.
It is often referred to as the dressed vacuum form.

The parameters, $\{x_j \}$ are solutions to "Bethe ansatz equation" (BAE)\cite{BWN},
\begin{equation}
w \frac{\phi(x_j+i)}{ \phi(x_j-i)}=
-\frac{Q(x_j-i) Q(x_j+2 i)}{Q(x_j+i)Q(x_j-2 i)},
\quad j=1,\cdots, N.
\label{bae}
\end{equation}

We introduce
\begin{equation}
P=
\begin{cases}
\frac{4(L+1)}{ L}  ,&  \hbox{ for regime 1} \\
 \frac{4(L+1)} { L+2}   ,& \hbox{ for regime 2} .
\end{cases}
\label{period}
\end{equation}

Then  the periodicity  $T_1(x) = T_1(x+Pi)$  follows obviously from the above expression. 

We represent the three terms in
eigenvalue of the quantum transfer matrix by three boxes with
letter 1,2 and 3 suffixed by the spectral parameter .

\begin{eqnarray}
T_1(u,x) &=&  \framebox[0.4cm][c]{1}_{x} + \framebox[0.4cm][c]{2}_{x} +
\framebox[0.4cm][c]{3}_{x},  \nonumber  \\
\framebox[0.4cm][c]{1}_{x} &:=&  w \phi(x+\frac{3}{2}i)
      \phi(x+\frac{1}{2}i)\frac{Q(x-\frac{5}{2}i)}{Q(x-\frac{1}{2}i)}    \nonumber  \\
\framebox[0.4cm][c]{2}_{x} &:=&   \phi(x+\frac{3}{2}i) \phi(x-\frac{3}{2}i) 
\frac{Q(x-\frac{3}{2}i)\, Q(x+\frac{3}{2}i)}{Q(x-\frac{1}{2}i)\, Q(x+\frac{1}{2}i)}    \label{box_expr} \\
\framebox[0.4cm][c]{3}_{x} &=& w^{-1} \phi(x-\frac{3}{2}i) \phi(x-\frac{1}{2}i)  
 \frac{Q(x+\frac{5}{2}i)}{Q(x+\frac{1}{2}i)}.    \nonumber
\end{eqnarray}

These boxes are analogue to those for Young tableaux
and play a fundamental role in the following.

\subsection{The symmetric fusion structure}

In the bootstrap procedure of the $S-$ matrix, 
$\phi^3$ property is most essential.
The similar feature , referred to as the $a^{(2)}_2$ fusion structure,
as been discussed   in the dilute $A_L$ model  \cite{GPZ}.
This comes from the singularity of the RSOS weights at
$u=\pm 2$.

Due to the fusion procedure, 
the eigenvalue of a fusion QTM can be represented by
sum over products of "boxes" with different letters and spectral parameters.
The point is that 
the sum of such boxes can be identified with \underline{S}emi-\underline{S}tandard 
Young \underline{T}ableaux (SST) for $sl_3$.
Let us present a simple example.  See \cite{BR,SuzG2,KS,GPZ, YKZ}
for details.
By the fusion procedure, one can construct a transfer matrix 
of which auxiliary space acts on a symmetric subspace of 
$V \times V$. We associate this to the set of the 
SST ,
$\framebox[0.4cm][c]{$i_1$}\framebox[0.4cm][c]{$i_2$}, (i_1\le i_2)$.
The eigenvalue of the transfer matrix  is then represented by,
\begin{equation*}
\sum_{i_1\le i_2} \framebox[0.4cm][c]{$i_1$}_{\,x-i} \,
\framebox[0.4cm][c]{$i_2$}_{\,x+i}.
%\label{box-expr-rule}
\end{equation*}
In a same manner, one can introduce fusion models based on general Young diagrams,
of which eigenvalues are specified  by their shapes.
On each diagram, the spectral parameter changes by $+2i$ from left to right and
$-2i$ from top to the bottom. 
First, we argue  QTMs  associated  to rectangular shape diagrams.
Although they are simpler, some of the properties below 
are crucial for the discussion for  QTMs  associated  to complex Young  diagrams.
First, due to identities,
\begin{eqnarray}
%------------------------------------------------
\begin{tabular}{|l|}
\hline
1 \\
\hline
2 \\
\hline
\end{tabular}
\raise 2mm \vtop{  \hbox{$\phantom{}_{x+i}$} \hbox{$\phantom{}_{x-i}$}  }
&=& \phi_2(x)   \framebox[0.4cm][c]{$1$}_{x},
%
%----------------------------------------------------------------
\quad
%------------------------------------------------------------------
\begin{tabular}{|l|}
\hline
1 \\
\hline
3 \\
\hline
\end{tabular}
\raise 2mm \vtop{  \hbox{$\phantom{}_{x+i}$} \hbox{$\phantom{}_{x-i}$}   }
=\phi_2(x)   \framebox[0.4cm][c]{$2$}_{x}  \np
%
%--------------------------------------------------------------------
%
\begin{tabular}{|l|}
\hline
2 \\
\hline
3 \\
\hline
\end{tabular}
\raise 2mm \vtop{  \hbox{$\phantom{}_{x+i}$} \hbox{$\phantom{}_{x-i}$}   }
&=&\phi_2(x)   \framebox[0.4cm][c]{$3$}_{x}   ,
%------------------------------------------------------------------------
\qquad
\phi_2(x):=
  \phi(x\pm \frac{5}{2}i)
\label{twotoone}  \\
%
%--------------------------------------------------------------------------
%
\begin{tabular}{|l|l}
\hline
1\\
\hline
2 \\
\hline
3\\
\hline
\end{tabular}
\begin{array}{l}
 \phantom{}_{x+2i}  \\
 \phantom{}_{x}    \\
 \phantom{}_{x-2i}
\end{array}
&=& \phi_3(x) :=\prod_{j=1}^3 \phi(x\pm (\frac{9}{2}-j)i),   \label{threetoscalar}
\end{eqnarray}
the QTMs from  $2\times m$ ( $3\times m$)  Young diagram can be reduced
to those from $1\times m$ (or just scalars).
Hereafter we adopt abbreviations for any function $f$, 
$$f(x\pm i a):= f(x+i a) f(x-ia ).$$

The first three relations brings additional symmetry besides $sl_3$,
and make the $a^{(2)}_2 $  symmetry explicit.
We call this  the $a^{(2)}_2 $ structure hereafter.

Second, the eigenvalues of $1\times m$  fusion QTMs have the "duality" in
the following sense.
Let us denote a renormalized $1\times m$  fusion QTMs by
$T_m(x)$;
\begin{equation}
T_m(x)= \frac{1}{f_m(x)} 
\sum 
\begin{tabular}{|l|l|l|l|}
\hline
$i_1$& $i_2$& $\cdots$& $i_3$ \\
\hline
\end{tabular}
\label{deftm}
\end{equation}
where the semi-standard (SS) condition $i_1 \le i_2 \le \cdots \le i_m$ is
imposed on the summation. 
The spectral parameters 
are assigned $x-i(m-1) \cdots x+i(m-1)$ from left to right. 
The renormalization factor, which is the common factor of 
 the expressions from tableaux of length $m$, is given by
$$
f_m(x):= \prod_{j=1}^{m-1} \phi(x\pm i(\frac{2m-1}{2}-j)) ,
$$
such that the  resultant $T_m$'s are all degree $2N$ w.r.t. 
$h[x+\hbox{ shift }]$.
For later convenience, we also denote by $T^{\rm non}_m(x)$,
the un-renormalized $T_m(x)$, namely, the object without the factor 
$f_m(x)$ in (\ref{deftm}).
Obviously, we have a periodicity due to Boltzmann weights;
$$
T_m(x+P i) =T_m(x).  \qquad (T^{\rm non }_m(x+P i) =T^{\rm non }_m(x))
$$
where $P$ is defined in (\ref{period}). \pn
Hereafter,  we frequently use functions of which argument is shifted by
the half period in the imaginary direction, i.e., $x+i\frac{P}{2}$.
We shall denote them by a symbol $\phantom{}^{\vee}$. 
For example,
$$
T^{\vee}_m(x) = T_m(x+\frac{P}{2} i) .
$$
The following functional relations are direct consequence of 
 the $a^{(2)}_2$ structure. %, together with the above property,
\begin{eqnarray}
T_m(x-i) T_m(x+i) &=&g_m(x) T_m(x) + 
          T_{m+1}(x) T_{m-1}(x),  \quad m \ge 1 \label{sl3fusion} \\
g_m(x) &=& \phi(x\pm i(m+\frac{3}{2})) ,	 \np
T_{-1}(x) &:=& 0   \np
T_0(x) &:=& f_2(x).
\end{eqnarray}
The scalar function $g_m(x)$ satisfies,
$$
g_{2 L-1-m} (x) =
\begin{cases}
   g_{m} (x) ,&  \hbox {if $L$ is  even } \\
     g_{m}^{\vee} (x) ,&  \hbox {if $L$ is  odd} .
   \\\end{cases}
$$
The adjacency condition leads to $T_{2L} (x)=T_{2L+1}(x)=0$.
Thus one deduces the duality relations,
\begin{equation}
T_m(x)=
\begin{cases}
 T_{2L-1-m}(x), \,\, m=0,\cdots, 2L& \, \hbox {for $L$ even } \\
 %
%T_{2L-1-m}(x+p/2 i), \,\, m=0,\cdots, 2L&, \hbox {for $L$ even } \\
%  
T_{2L-1-m}^{\vee}(x), \,\, m=0,\cdots, 2L& \, \hbox {for $L$ odd } \\
 \end{cases}
\label{dualonerow}
\end{equation}
for the solutions to eq.(\ref{sl3fusion}). 
Note that this duality is not valid for $T^{\rm non }(x)$; obviously 
 $T^{\rm non }_{2L-1-m}(x)$ and $T^{\rm non }_{m}(x)$ have different
 order in $h[x]$.
These relations can be 
in principle proved by using explicit fusion weights.
We have at least checked this numerically and assume the validity in this report.

The duality brings about many relations. It thus  happens seemingly quite
different equations are identical; an identical  object can have (infinitely ) many equivalent expressions.
It sometimes makes the proof of the functional relations quite involved.

Before closing this section, we make a comment.

The simple relation (\ref{sl3fusion}) suggests a choice of $Y$ function,
\begin{equation}
Y_m(x)=\frac{T_{m+1}(x) T_{m-1}(x)}{g_m(x) T_m(x)} \: (m=1,\cdots, 2L-2).
\label{trialY1}
\end{equation}

This leads to a  set of simple equations (the $Y-$ system)\cite{YKZ}
\begin{equation}
{\cal Y}_m(x-i) {\cal Y}_m(x+i)=
 \frac{(1+{\cal Y}_{m+1}(x)) (1+{\cal Y}_{m-1}(x))}
              {1+{\cal Y}_m(x)}.
 \label{trialYsys}
\end{equation}

The relations seem to be tractable and they indeed work well 
in a nonunitary case \cite{EllenBaz2}.
This is  not the case for the unitary models.
The transformation of   the $Y-$ system  into a simple set of integral equations
requires the both sides to be Analytic and NonZero and to have Constant 
asymptotic behavior 
as $ |x| \rightarrow \infty $ (ANZC) in certain strips.

As is demonstrated numerically in appendix in the case of  $L=8$ 
in  appendix \ref{numericsA8},
this requirement is not satisfied for  (\ref{trialYsys}).
Therefore we must seek for alternatives, which may not be
found only among symmetric fusion transfer matrices.

In the next subsection,
we thus consider a much wider class of skew Young diagrams.
%

%########################################################
%
%
\subsection{The quantum Jacobi-Trudi formula }
%
%##########################################################
%
Let $\mu$ and $\lambda$ be a pair of Young diagrams satisfying 
 $\mu_i \ge \lambda_i, \forall i$.
We subtract  a diagram  $\lambda$ from $\mu$. 
The resultant "narrower" one, consisted of
$(\mu_1-\lambda_1,\mu_2-\lambda_2,\cdots )$ boxes is called 
a skew Young diagram $\mu-\lambda$.
(The usual Young diagram is the special case that $\lambda $ is empty, and
we will omit $\lambda$ in the case hereafter.)

Consider a set of semi-standard skew Young table of the shape $\mu-\lambda$.
The spectral parameters are assigned to boxes in a table
such  that they change by $+2i$ from left to right and
$-2i$ from top to the bottom. 
The spectral parameters of the head (left-top)
and the tail (right-bottom) are fixed to be
$x+i(\mu_1+\mu'_1-2) $ and  $x-i(\mu_1+\mu'_1-2) $
where $\mu'_1$ denotes the depth of the diagram.
One identifies each box in a table with an expression 
under the rule (\ref{box_expr}) with
a shift of the spectral parameter.
Then the product over all constituting boxes yields
the desired expression for a table.
We then have the following theorem which is a quantum analogue to
Jacobi-Trudi formula for the Schur function.
\begin{theorem}
Let ${\cal T}_{\mu/\lambda}(x)$ be the sum 
over the resultant expressions divided by a common factor,
$
\prod_{j=1}^{\mu'_1} f_{\mu_j-\lambda_j} 
  (x+i(\mu_1' -\mu_1+\mu_j+\lambda_j-2j+1)).
$
Then the following  equality holds\cite{BR, KS}.
\begin{equation}
{\cal T}_{\mu/\lambda}(x) =
  \hbox{det }_{1\le j,k\le \mu_1'} 
         ( T_{\mu_j-\lambda_k-j+k} 
            (x+i(\mu_1' -\mu_1+\mu_j+\lambda_k-j-k+1)) )
\label{qJT}
\end{equation}
where $T_{m<0}:=0$.
\end{theorem}
${\cal T}_{\mu/\lambda}(x)$ may be naturally identified with 
the eigenvalue of QTM corresponding to fusion $\mu-\lambda$,
although this is yet to be proved.
We note that 
the pole-free property of ${\cal T}_{\mu/\lambda}(x)$,
obvious from the quantum Jacobi-Trudi formula.

Due to the semi-standard condition,  products of ${\cal T}_{\mu/\lambda}(x)$
enjoy the  decomposition similar to that of the products of Young diagrams.
This observation is crucial in the proving the functional relations below.
One, should , however note that each box can only stay at positions with
identical spectral parameter.
Therefore, the product of two ${\cal T}$ shows nontrivial
decomposition only when corresponding  diagrams are situated 
at proper relative positions specified by their spectral  parameters.

\begin{figure}[hbtp]
\centering
\includegraphics[width=6cm]{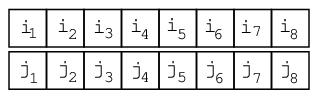}
\caption{ 
A typical table in the product  $T^{\rm non}_m(x+i)  T^{\rm non}_m(x-i)$.  
Hereafter, disjoint diagrams represent their product.
  }
\label {TmTm}
\end{figure}

For illustration, let us prove (\ref{sl3fusion}) diagrammatically.
We consider $T^{\rm non}_m(x+i)  T^{\rm non}_m(x-i)$. 
The spectral parameters are assigned for each box, $x-(m-2)i, \cdots , x+mi$ from the
left to the right for $T^{\rm non}_m(x+i)$
, and  $x-mi, \cdots , x+(m-2)i$  for  $T^{\rm non}_m(x-i)$.
Thus  two diagrams are situated so that these 
 two width $m$ rectangles piled vertically (fig. \ref{TmTm}). @

We consider a particular assignment of letters, $i_1 , \cdots , i_m$ for the upper
and  $j_1 , \cdots , j_m$ for the lower diagram.
$T^{\rm non}_m(x+i)  T^{\rm non}_m(x-i)$ is obtained by the sum over tableaux under the condition,
$i_1\le  i_2 \le \cdots \le i_m$ and   $j_1\le  j_2 \le \cdots \le j_m$, respectively.

\begin{figure}[hbtp]
\centering
\includegraphics[width=14cm]{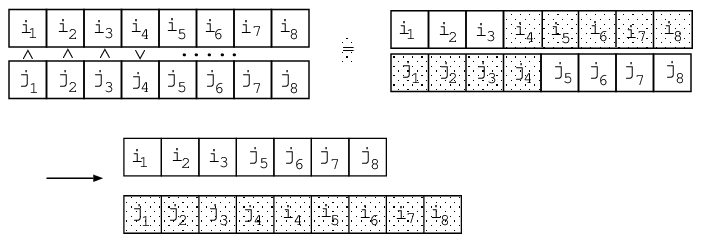}
\caption{ The pair $(i_4, j_4)$  is the first from the left which breaks
the semi-standard condition.   Then the recombination of tableaux occurs there, into
hatched and unhatched ones.      }
\label {TmTm2}
\end{figure}

The first term in the lhs of  (\ref{sl3fusion})  comes from the configurations, $i_{k}<j_{k}, (k=1, \cdots, m)$.
The resultant table  is  of the shape $2 \times m$ and satisfy the semi-standard condition.
Due to the reduction (\ref{twotoone}), it is a member of 
$T^{\rm non}_m(x)$ times $\prod_{j=1}^m \phi_2(x-i(m+1)+2 ij )$.
The second term comes from the configurations which breaks the condition (fig. \ref{TmTm2}).
By  $i_{\alpha}\ge j_{\alpha}$, 
we mean  the first pairs breaking the condition from the left end  .
Then the ordering within $i$'s and $j'$s concludes that
the new sequences $j_1, \cdots, j_{\alpha}, i_{\alpha}, i_{\alpha+1}, \cdots, i_m$ 
and  $i_1, \cdots, i_{\alpha-1}, j_{\alpha+1}, \cdots, j_m$ satisfies the semi-standard condition.
One can also check that the new  sequences satisfy the proper spectral parameter assignment,
namely increment of the spectral parameter by 2i from the left to right.
Thus after the recombination of sequences , we obtain  $T^{\rm non}_{m+1}(x)  T^{\rm non}_{m-1}(x)$.
Finally, by dividing the both sides by the factor $f_m(x\pm i)$ , one arrives at  (\ref{sl3fusion}).

The above procedure is formally interpreted as follows.
Place two diagrams  properly according to their spectral parameters.
We just join them or recombine them such that the total set of spectral parameters  is
conserved. 

The case studies for $L=3,4, 6$, however,  suggest that 
${\cal T}_{\mu/\lambda}(x)$ is not the right object in the discussion of 
functional relations.
Instead, we should introduce  $\Lambda_{\mu/\lambda}(x)$, 
which is analytic under BAE,
  from ${\cal T}_{\mu/\lambda}(x)$ by putting
$T_{m\ge 2L}(x)=0$ in the latter,
\begin{equation}
 \Lambda_{\mu/\lambda}(x):={\cal T}_{\mu/\lambda}(x) /.\{ T_{m\ge 2L}(x)=0 \}.
 \label{LamcalT}
\end{equation}
The pole-free property of $\Lambda_{\mu/\lambda}(x)$ is obvious from 
(\ref{qJT}).

The diagrammatic decomposition applies to
${\cal T}_{\mu/\lambda}(x)$  but not to $\Lambda_{\mu/\lambda}(x)$ in a strict sense.
%
%On the other hand,  the quantities which appear in  our functional relations  are $\Lambda_{\mu/\lambda}(x)$.
%
Therefore the above decomposition rules, demonstrated above,
 may not  seem to be helpful in our discussion.
For most of case which we treat below, however, 
the skew diagram $\mu-\lambda$ is no so huge  and $\Lambda$ and ${\cal T}$ coincide.
Thus the above rules are  quite often applicable and play significant roles ;
as already remarked,  there are plenty of expressions for an identical object.
Therefore it is sometimes crucial to choose "the most" relevant expression among
many. 
The above graphical decomposition provides
hints on the optimal candidate in proving the functional relations .

Before closing the section, we mention an important diagrammatic
symmetry of $\Lambda_{\mu/\lambda}$ .

\begin{corollary} \label{diagramsymm}
Let $Y$ and  $\widehat{Y}$ be skew diagrams. $\widehat{Y}$
is obtained from $Y$ after 180$^\circ$ rotation.
The following relation is a consequence of the eq (\ref{qJT})
\begin{equation}
{\cal T}_Y (x)  = ( {\cal T}_{\widehat{Y}} (x^*) )^*
\label{diagramsym}
\end{equation}
where the asterisk stands for complex conjugations.
\end{corollary}

The proof is direct; one substitutes 
$\widehat{Y}_i =(\mu_1-\lambda_{d+1-i} )-  (\mu_1-\mu_{d+1-i})$
for $Y=\mu-\lambda$ and $\mu'_1=d$.

This corollary is crucial in establishing a property of the
transfer matrices (see  section \ref{reality}).
It will be also used  implicitly in the proof of functional relations in the following.

After these preparations, 
we introduce the first  of  our three ingredients in 
the closed functional relations, the Breather  $T-$ system in the next section.

%########################################################
%
%
\section{Breather $T-$ system }
%
%##########################################################

The RSOS weights possesses a singularity at $u=\pm 3$ besides the
 $u=\pm 2$ singularity related to $a^{(2)}_2$ structure \cite{GPZ}.
It is shown that the singularity at $u=\pm 3$ signifies the
``hidden'' $sl_2$ structure behind the model.
The breather type relation originates from this  $sl_2$ structure.

We define the "breather" transfer  matrices for the dilute $A_L$ model in regime  2.
\begin{eqnarray*} 
T_{B_1}(x) &= &T_1(x)   %\label{deftb1} 
\\
T_{B_3}(x) &= &
\begin{cases} 
\frac{1}{ \phi^{\vee}(x)   }  \Lambda_{2L-2,1} (x-\frac{5}{2} i),&  \hbox{ for $L$ odd} \\
\frac{1}{ \phi^{\vee}(x)   }  \Lambda_{2L-2,1} ^{\vee} (x-\frac{5}{2} i) ,&  \hbox{ for $L$ even} \\
\end{cases}
%\label{deftb3} 
\\
T_{B_5}(x) &= &
\frac{1}{\phi(x\pm \frac{3}{2} i) } \Lambda_{(4L-5,2L-2,2L-2)/(2L-3, 2L-3)} (x) % \label{deftb5}  
\end{eqnarray*} 
One immediately verifies the validity of the following functional relations,
\begin{eqnarray} 
T_{B_1}(x \pm i \frac{L-2}{2 (L+2)} ) &=& T_0(x\pm i\frac{L+6}{2(L+2)}) + \phi^{\vee}(x) T_{B_3}(x)  \label{tb1tb1}\\
T_{B_3}(x \pm i \frac{L-2}{2 (L+2)} ) &=& T_0(x) T_0(x\pm i\frac{4}{(L+2)}) + T_{B_1}(x) T_{B_5}(x) . \label{tb3tb3}
\end{eqnarray} 
\begin{proof}
Proof utilizes the duality (\ref{dualonerow}). 
 For (\ref{tb1tb1}) , we first consider the multiplication of
  $T_1(x) (=T_{B_1}) $  by  its dual $ T_{2L-2}(x+(2L-1)i) $ .  
  See  fig. \ref{figbreather_tb1tb1} for a case $L=5$.

    \begin{figure}[hbtp]
\centering
\includegraphics[width=8cm]{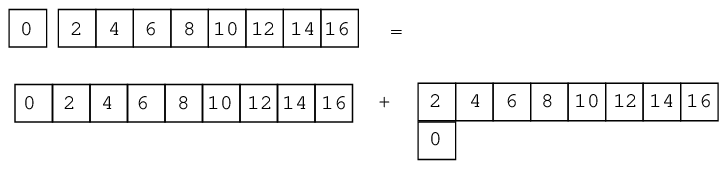}
\caption{ The multiplication of
  $T_1(x) (=T_{B_1}) $  by  $ T_{2L-2}(x+(2L-1)i) $.
  The letter $a$ in a box implies its spectral parameter $x+ia$.
   }
\label {figbreather_tb1tb1}
\end{figure}
  The spectral parameters are assigned so that they align horizontally.  
 For $L$  even (odd)  $ T_{2L-2}(x+(2L-1)i)$ is  equal to 
  $T_{B_1}(x+(2L-1)i)$ (  $T_{B_1}(x+(2L-1)i+\frac{P}{2} i)$   ). 
  Thus the product is equivalent to that of two $T_{B_1}$.
  It is decomposed into  two terms;
  the first is $T_{2L-1}$ and the second is  $ T_{B_3}$ (with some shifts in spectral parameters
  and with coefficients.)   By the duality,  $T_{2L-1}$ is replaced by $T_0$.
  Then  one arrives  (\ref{tb1tb1}) after the shift in
  the spectral parameter and taking account of proper coefficients.

 The second relation is shown similarly.  One only should notice that dual of the diagram $(2L-1, 1)$ is
 given by a skew diagram$(2L-1, 2L-1)-(2L-2)$.
 The case  $L=5$ is depicted in   fig. \ref{figbreather_tb3tb3} .
  \begin{figure}[hbtp]
\centering
\includegraphics[width=14cm]{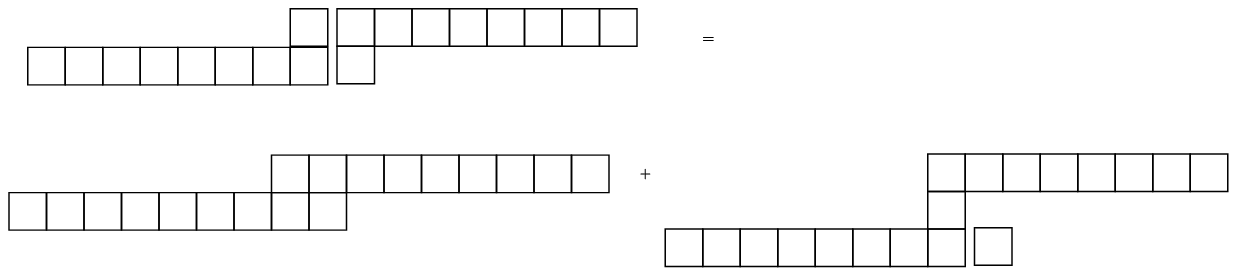}
\caption{ 
   }
\label {figbreather_tb3tb3}
\end{figure}

 Despite its disguise,  the first diagram in the rhs corresponds to the products of $T_0$ due to
 (\ref{qJT}).  Then it is straightforward to verify  (\ref{tb3tb3}).
 \end{proof}
 
By noticing 3$= -\frac{(L-2)}{(L+2)}  $(mod $\frac{4(L+1)}{L+2}$ ),
we understand the above relation stems from the singularity of
the RSOS weight at $u=\pm 3$ related to $sl_2$ structure.
%
%The above functional  relations imply the singularity of intertwinners at $u=\pm 
% \frac{L-2}{2 (L+2)}$ . ( Note that the normalizations of $x$ and $u$ differ by factor $i$.)
%t is the  aforementioned singularity related to $sl_2$ by
%as $3= -\frac{(L-2)}{(L+2)}  $(mod $\frac{4(L+1)}{L+2})$ )$.

The following representation of  $T_{B_3}(x)$ and    $T_{B_5}(x)$ will be useful in the following argument for the regime 2
\begin{eqnarray} 
T_{B_3}(x-\frac{L-2}{2(L+2)}i) &=& \frac{1}{\phi(x-\frac{5}{2}i) } \Lambda_{(2,1)}(x+ i  )    \label{anotherdeftb3} \\
T_{B_5}(x) &=& \Lambda_{(3,2)/(1)}(x).  \label{anotherdeftb5} 
\end{eqnarray} 

\begin{proof}
The first equation utilizes the representation of $T_{B_1}(x)$ by the height 2 diagram.
\begin{equation}
T_{B_1}(x)  T_{B_1}(x+3 i)  =\frac{ f_2(x+2 i)}{\phi_2(x)} \Lambda_{(2,1)}(x+i) +\frac{ \phi_3(x+i)}{\phi_2(x)}
\label{anothert1t1}
\end{equation}
See  (\ref{twotoone}) and (\ref{threetoscalar}) for $\phi_{2,3}$.
  \begin{figure}[hbtp]
\centering
\includegraphics[width=5cm]{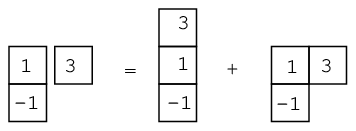}
\caption{ 
A graphical  representation of (\ref{anothert1t1}).
 The letter $a$ in a box implies its spectral parameter $x+ia$
   }
\label{foranotherdeftb3fig}
\end{figure}

We present  graphically  eq.(\ref{anothert1t1}) in  fig. \ref{foranotherdeftb3fig}.
By noticing 3$= -\frac{(L-2)}{(L+2)}  $(mod $\frac{4(L+1)}{L+2}$ ),
 and comparing the above with (\ref{tb1tb1}) (after trivial shift in $x$),
one concludes  (\ref{anotherdeftb3}).  \par\noindent
To prove the second, we consider  ${\cal T}^{\rm non}_{(4L-5,2L-2,2L-2)/(2L-3, 2L-3)} (x)$ instead of  
$\Lambda_{(4L-5,2L-2,2L-2)/(2L-3, 2L-3)} (x)$ in the definition of $T_{B_5}$. 
Note their difference, (\ref{LamcalT}).
From theorem 1, ${\cal T}^{\rm non}_{(4L-5,2L-2,2L-2)/(2L-3, 2L-3)} (x)$ decomposed into 3 pieces
as in the case of usual Young diagrams. 
\begin{equation*}
{\cal T}^{\rm non}_{(4L-5,2L-2,2L-2)/(2L-3, 2L-3)} (x) = T^{\rm non}_{2L-3} (x\pm 2iL)  \phi_3(x) = 
T_{2}(x \pm 2i)  \phi_3(x) f_{2L-3} (x\pm 2iL)  
\label{dec1zero}
\end{equation*}
where the duality relation  (\ref{dualonerow}) is used in the second equality.   $\phi_3(x)$ is defined in 
(\ref{threetoscalar}).  After proper normalizations, it reads,

\begin{equation}
{\cal T}_{(4L-5,2L-2,2L-2)/(2L-3, 2L-3)} (x) =T_{2}(x \pm 2i) .
\label{dec1}
\end{equation}

On the other hand,  the formula  (\ref{qJT}) leads to
\begin{equation}
{\cal T}_{(4L-5,2L-2,2L-2)/(2L-3, 2L-3)} (x) =\Lambda_{(4L-5,2L-2,2L-2)/(2L-3, 2L-3)} (x)+T_0(x\pm i) T_{4L-3}(x).
\label{dec2}
\end{equation}
Note  $T_{4L-3}(x)=T_4(x)$ , which is derived in parallel to the duality.
Then  the comparison (\ref{dec1}) and (\ref{dec2}) leads to
$$
\Lambda_{(4L-5,2L-2,2L-2)/(2L-3, 2L-3)} (x)=T_{2}(x \pm 2i) - T_0(x\pm i) T_{4}(x)=\Lambda_{(3,2)/(1)}(x),
$$
where (\ref{qJT}) is applied in the last equality.
In this manner, one verifies  (\ref{anotherdeftb5} ) . 
\end{proof}

The transformation of the $T$ system to the "breather" $Y-$ system is standard\cite{KluemperPearce92, KNS1, KSS98}.
We define,
\begin{eqnarray*}
Y_{B_1}(x)&:=& \frac{ \phi^{\vee}(x) T_{B_3}(x)  }{T_0(x\pm i\frac{L+6}{2(L+2)})  }  %\label{defyb1} 
\\
Y_{B_3}(x)&:=& \frac{ T_{B_1}(x) T_{B_5}(x) }{ T_0(x) T_0(x\pm i\frac{4}{(L+2)}) }.    %\label{defyb3} 
\end{eqnarray*}

Then the  following  "breather" $Y-$ system is directly derived with the help of  the "breather "
functional relation  (\ref{tb3tb3}).
\begin{equation}
Y_{B_1}(x \pm i \frac{L-2}{2 (L+2)} ) =(1+Y_{B_3})(x).
\label{breatherY}
\end{equation}

The main goal in this report is to evaluate the largest eigenvalue of $T_1(u, 0)$ or $T_{B_1}(0)$ in the
Trotter limit.
For this, our strategy is to  utilize a set of closed  functional relations, including  $T_{B_1}(x)$.
The $sl_2$ originated functional relations are indeed enough to obtain the desired relations
in the case of $M_{3,4}+\phi_{1,2}$. 
This is no longer valid for general case.
In the next sections, we introduce further objects and relations 
to achieve the goal.
%
% ##############################################################
%
\section{The $sl_3$ type functional relations}
%
%################################################################

%As is mentioned  in the preceding
%section, there is another common structure for the four families,
%which is similar to the functional relations associated to the
%$sl_3$ model.
As is mentioned  in the preceding
section, there is another structure analogous to the
$sl_3$ models.
% 
%The latter takes the following form.
%
Let $W^{(a)}_m, ( a=1, 2. \quad  m \in Z_{\ge 0})$ 
be a Yangian module corresponding to
$m \Lambda_a$ as a classical $sl_3$ module.
We associate a transfer matrix $T^{(a)}_m(x)$ to $W^{(a)}_m$.
Then the following functional relations hold,
\begin{equation*}
T^{(a)}_m(x \pm i \gamma) = T^{(a+1)}_m(x) T^{(a-1)}_m(x)+
                            T^{(a)}_{m+1}(x) T^{(a)}_{m-1}(x)
\end{equation*}
where $\gamma$ is a parameter related to a singularity in $R$ matrix.
$T^{(a)}_0 , T^{(0)}_m$ and $T^{(3)}_m$ are given by 
trivial scalar factors times an identity operator.
Thus when multiplied on a common eigenspace of the commuting transfer
matrices, they amount to known trivial scalar functions.
The above relation can be graphically represented as follows.
Consider a product of two rectangles of the shape $a\times m$.
It decomposed into two subsets. The one is the products of
 two rectangles of the shape $(a+1) \times m$ and  the shape $(a-1) \times m$ 
and  the other is the products of rectangles , $a \times (m+1)$ and  $a \times (m-1)$ .

We turn back to the  dilute $A_L$  models in regime 2.
The above relations among rectangular can not hold ; 
 due to the $a^{(2)}_2$ symmetry, the $2\times m$ rectangle is always
reduced to the rectangle of    $1\times m$ .
Nevertheless, an analogue can be still found .
Let us denote 
$$
T_{D_j}(x) = \Lambda_{(j+1, j)/(1)}(x).
$$

In this notation, which is adopted hereafter, 
 $T_{B_5}(x)$ in the preceding section is identical to
$T_{D_2}(x)$.
Note also that $T_{D_1}(x)= T_1(x\pm 2i)$ and 
$T_{D_0}(x)= T_0(x\pm 2i)$.

%
%Consider the $A_L$ model at regime 2, (corresp. to $M_{L, L+1}+\phi_{1,2}$)
%
\begin{lemma}\label{su3liketsys}
The following functional relations, similar to $sl_3$ type, hold.
\begin{eqnarray}
T_{2j}(x \pm 2i) &=& T_{2(j+1)}(x) T_{2(j-1)}(x) + T_{D_{2j}}(x) 
   \label{sl3one}   \\
T_{D_{2j}}(x \pm 2i) &=& 
 \begin{cases}
    T_{D_{2(j+1)}}(x) T_{D_{2(j-1)}}(x) + 
       T_{2j}(x) T_{B_3}( x \pm i\frac{L-2-4j}{2L+4 } )& j \quad {\rm even}\\
  T_{D_{2(j+1)}}(x) T_{D_{2(j-1)}}(x) + 
   T_{2j}(x)T^{\vee}_{B_3}( x \pm i\frac{L-2-4j}{2L+4 } )& j \quad {\rm odd}.\\
 \end{cases}
 \label{sl3two}   
\end{eqnarray}
\end{lemma}
The similarity is clearer by writing $T^{(1)}_j (x)=T_{2j}(x)$ and $T^{(2)}_j (x)=T_{D_{2j}}(x)$.
The difference lies in that fact that  $T^{(3)}_j (x)$,  which amounts to  a known scalar  for  the $sl_3$ case,
remains non trivial; it is given by a product of $T_{B_3}$.

\begin{proof}
The first equality  (\ref{sl3one}) is trivial shown by the quantum Jacobi-Trudi formula.
The second one needs one step further.
First, one uses the following relation, proved by the quantum Jacobi-Trudi formula,
\begin{equation}
T_{D_{2j}}(x \pm 2i) -T_{D_{2(j+1)}}(x) T_{D_{2(j-1)}}(x) 
= T_{2j}(x) {\cal T}_{(2j+2, 2j+1, 2j)/(2,1)}(x).
\label{forproofsu31}
\end{equation}
From the tableaux rule, a hatched part of the diagram  in ${\cal T}_{(2j+2, 2j+1, 2j)/(2,1)}(x)$
reduces to product of  scalars, see fig. \ref{decmp}.

\begin{figure}[hbtp]
\centering
\includegraphics[width=5cm]{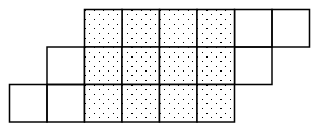}
\caption{ ${\cal T}_{(2j+2, 2j+1, 2j)/(2,1)}(x)$ for $j=3$.    }
\label {decmp}
\end{figure}
The resultant two pieces of diagrams represents the  products of  $\Lambda_{(2,1)}(x+(2j+1)i)$
and $\Lambda_{(2,2)/(1)}(x-(2j+1)i)$  apart from normalization factors.
Note that $ 2j  \equiv \frac{4j}{2L+4} \; ({\rm mod }\, P) $ for $j$ even and 
 $ 2j \equiv \frac{4j}{2L+4}+\frac{P}{2}  \; ({\rm mod } \, P) $ for $j$  odd .
Then from the expression  (\ref{anotherdeftb3} ),  one finds for the first piece,
$$
%T_{B_3}( x - \frac{L-2-4j}{2L+4}i) = \frac{1}{\phi(x+(2j-5/2)i}  \Lambda_{(2,1)}(x+(2j+1)i).
\Lambda_{(2,1)}(x+(2j+1)i) =\phi(x+(2j-\frac{5}{2})i )   T_{B_3}( x - \frac{L-2-4j}{2L+4}i) .
$$
The second piece, $\Lambda_{(2,2)/(1)}(x-(2j+1)i)$,  equals to $(\Lambda_{(2,1)/(1)}(x+(2j+1)i))^* $
by corollary \ref{diagramsymm} .
Therefore,  
after taking account of proper normalization factors and using the property $(T_{B_3}(x))^* = T_{B_3}(x^*)$, 
one finds,
\begin{equation}
T_{B_3}( x \pm  \frac{L-2-4j}{2L+4}i) =
\begin{cases}
      {\cal T}_{(2j+2, 2j+1, 2j)/(2,1)}(x)&   j \; {\rm even} , \\
     {\cal T}^{\vee} _{(2j+2, 2j+1, 2j)/(2,1)}(x)&   j \; {\rm odd} .
\end{cases}
\label{forproofsu32}
\end{equation}
Thus the second relation is  proved from  (\ref{forproofsu31}) and (\ref{forproofsu32}).
\end{proof}

For $L=4k-1, 4k$ and $4k+1$, we introduce
\begin{eqnarray}
Y^{(1)}_j (x) &:=& 
   \begin{cases}
       \frac{T_{D_{2(j+1)}}(x) }{T_{2j}(x)T_{2(j+2)}(x)} ,&  \quad j \; {\rm even}  \\
	    \frac{T^{\vee}_{D_{2(j+1)}}(x) }{T^{\vee}_{2j}(x)T^{\vee}_{2(j+2)}(x)} ,&  \quad  j \;  {\rm odd}  
   \end{cases} 
 \qquad   0\le j\le k-3     \label{sl3y1byt}  \\ 
Y^{(2)}_j (x) &:=& 
   \begin{cases}
       \frac{T^{\vee}_{2(j+1)}(x)  T_{B_3}(x\pm \frac{L-2-4(j+1)}{2L+4}) }
	           {T^{\vee}_{D_{2j}}(x)T^{\vee}_{D_{2(j+2)}} (x)} ,&  \qquad  j \;  {\rm even}   \\
      \frac{T_{2(j+1)}(x)  T_{B_3}(x\pm \frac{L-2-4(j+1)}{2L+4}) }
	          {T_{D_{2j}}(x)T_{D_{2(j+2)}} (x)} ,&  \qquad j \; {\rm odd}  
	 \end{cases}
    \qquad   0\le j\le k-3.        \label{sl3y2byt} 
  \end{eqnarray}
 while these do not show up for  $L=4k+2$.

By noticing   $2 \equiv \frac{4}{2L+4} + \frac{P}{2}$,  one immediately obtains
the following $Y-$ system,
\begin{eqnarray}
Y^{(1)}_j ( x \pm \frac{4}{2L+4}i) &=& 
   \frac{1+Y^{(2)}_j(x)   }   
           { (1+\frac{1}{Y^{(1)} _{j+1}(x)} )   (1+\frac{1}{Y^{(1)} _{j-1}(x) })  }
                                                                  \label{sl3Ysys1}  \\
Y^{(2)}_j ( x \pm \frac{4}{2L+4}i) &=& 
        \frac{1+Y^{(1)}_j(x)     }   
		        { (1+\frac{1}{Y^{(2)} _{j+1}(x)} )   (1+\frac{1}{Y^{(2)} _{j-1}(x)} )  }
  	 \qquad  1 \le  j	\le k-3.						        \label{sl3Ysys2}    			  
\end{eqnarray}
These are desired relations. See  the $Y-$ system in the appendix \ref{list_Y-system}.

The $sl_2$ $T-$ system is not itself closed.  Here we find it connected to the $sl_3$ like $T-$ system
through $T_{B_3}(x) = T_{D_2}(x)$.
The latter system is, however, neither  closed. 
In the next section,  other functional relations  are introduced, 
referred to as "magnon-like" , which turns out to "close" the total  $T-$ systems.

% ##############################################################
%
\section{The magnon-like $t-$ system and the Dynkin like diagrams}
%
%################################################################

The  above transfer matrices are not enough to obtain sets of closed functional relations,  which are
necessary to derive TBA.
We have to introduce further objects separately depending on four families.
They are related to the magnon-like $Y-$ system introduced in \cite{DPT2, DTT}.
A magnon-like $Y-$ system is associated to a diagram of the Dynkin type.
Although they can be classified into the four families, 
the  diagrams are all different depending on values of $L$.
Consequently, the  magnon-like $Y$ system are all distinct for different $L$.

Here the situation is slightly unified.
The  diagrams we have to deal with are "tails" of those in  \cite{DPT2, DTT}, independent of
$L$,  except for the dilute $A_{4k+2}$ models.
%
%We only have to deal with identical "  magnon-like system " for  the other three families .

%The  Dynkin like diagrams are explicitly given as follows.

Before starting discussions on individual cases, we shall make remarks.
A blank node  in each  Dynkin like diagram
indicates its different character from the rests.
To  node $a$ in the diagram, we associate $t^{(a)}_m (x)$, where $m$ takes 
an integer value from a finite set.
Our magnon-like $t-$ system  is a set of functional relations among  $t^{(a)}_m (x)$,
which should be described separately for four families.
The most crucial observation below will  be that $t^{(a)}_m$ associated to a blank node
is always expressed by (product of ) $T_{B_3}$.

The dilute $A_{4k-1}$  model and the dilute $A_{4k+1}$ model
share a same diagram but possess different $t-$ system.
The only "$L$" depending diagram  for   $A_{4k+2}$ ($L=4k+2$) remarkably coincides
with  the  Dynkin diagram for $D_{k+2}$.

Below we take $x$ to be real.
As usual, the symbol $a \sim b$ means  
that nodes $a$ and $b$ are adjacent on the diagram.
%
%------------------------------------------------------
\subsection{The magnon-like $t-$ system for $A_{4k-1}$ }
%------------------------------------------------------
%
\begin{figure}[hbtp]
\centering
\includegraphics[width=8cm]{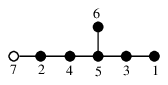}
\caption{Dynkin like diagram  for $A_{4k-1}$  }
\label{fig_a4kminus1}
\end{figure}

We define a magnon-like $t-$ system associated to fig. \ref{fig_a4kminus1},
\begin{eqnarray}
t^{(a)}_m(x \pm \frac{i}{2L+4})  &= &
t^{(a)}_{m-1}(x)  t^{(a)}_{m+1}(x)  + \prod_{b \sim a} t^{(b)}_m (x) , \quad a=1 \sim  6, \quad  m=1,2 \label{tsys11} \\
t^{(a)}_0&= &1 \quad (a=1 \sim  6), \quad 
t^{(a)}_3=0,  (a=3 \sim 6 )  \nonumber \\
t^{(7)}_1(x \pm \frac{i}{2L+4}) &=&  t^{(7)}_2 (x).   \label{tsys12} 
\end{eqnarray}
This is not a set of  closed functional relation. 
We, however, note its affinity to the $E_6$ related $T-$ system \cite{KNS1}.
Indeed if  $t^{(7)}_1(x)=1$ and   $t^{(1)}_3(x)= t^{(2)}_3(x)=0$, the above
functional relation coincides with the $T-$ system for level 3 RSOS model, 
corresponding to the coset 
$(E^{(1)}_6)_2 \times  (E^{(1)}_6)_1 /    (E^{(1)}_6)_3$. 
Our interpretation is as follows.
 The nontrivial $t^{(7)}_1(x)$  makes the
  $t^{(1)}_3(x)$ and $ t^{(2)}_3(x)$ non zero and connects the $E_6$ level 3 $T-$ system
  to others.
Thus  $t^{(7)}_1(x)$ plays a role of "glue".
This seems to be quite parallel to the $sl_3$ like $T$ system,  which is connected to
$sl_2$  $T$ system  by nontrivial   $T_{B_3}$.
Here the situation is indeed the same;  we will see below  $t^{(7)}_1(x)=T_{B_3}$. 
 $T_{B_3}$ thus plays a role in gluing  the $sl_3$ ,  the $E_6$   and the
 $sl_2$   $T-$ systems.
 
 We shall define "gauge invariant" functions  for later discussions.\cite{KNS1, KLWZ}
 \begin{eqnarray*}
 Y^{(a)}_{k-1} (x)  &:=&   \frac{ \prod_{b \sim a} t^{(b)}_1 (x)  }{  t^{(a)}_2  (x)  } , \quad a=1\sim 6   
 %\label{dA4km1ydef1} 
 \\
 Y^{(1)}_{k-2} (x)  &:=&   \frac{ t^{(3)}_2(x)  }{ t^{(1)}_1 (x) t^{(1)}_3(x)  } %\label{dA4km1ydef2} 
 \\
 Y^{(2)}_{k-2} (x)  &:=&   \frac{ t^{(4)}_2(x)  t^{(7)}_2(x)  }{ t^{(2)}_1 (x) t^{(2)}_3(x)  } .  
 %\label{dA4km1ydef3} 
 \end{eqnarray*}

%------------------------------------------------------
\subsection{The magnon-like $t-$ system for $A_{4k}$ }
%------------------------------------------------------
%
\begin{figure}[hbtp]
\centering
\includegraphics[width=8cm]{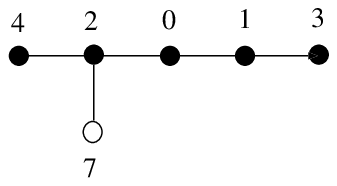}
\caption{Dynkin like diagram  for $A_{4k}$}
\label {fig_a4k}
\end{figure}

Associated to fig. \ref{fig_a4k}
we propose a magnon-like $t-$ system for this case,
\begin{eqnarray*}
t^{(a)}_m(x \pm \frac{i}{L+2})  &= &
t^{(a)}_{m-1} (x)  t^{(a)}_{m+1} (x)   + \prod_{b \sim a} t^{(b)}_m (x) , \quad a=1\sim 6 , \quad m=1,2 ,%\label{tsys31} 
\\
t^{(a)}_0&=& 1  \quad( a=1\sim 6 ),   \qquad 
 t^{(a)}_3(x)  =  0   \quad (a=0,3,4 ),     \\
t^{(7)}_1(x \pm \frac{i}{L+2}) &=&  t^{(7)}_2 (x).   %\label{tsys32} 
\end{eqnarray*}

Again, this is not a set of  closed functional relation. 
In this case we note its similarity to  the $T-$ system  for level 3 RSOS model, 
corresponding to the coset 
$(A^{(1)}_5)_2 \times  (A^{(1)}_5)_1 /    (A^{(1)}_5)_3$  \cite{KNS1}.
As in the preceding case, $t^{(7)}_1(x)$ will turn out to be equal to $T_{B_3}(x)$.
Thus the interpretation is the same;
 $ t^{(7)}_1 $ connects the $T-$ system for level 3 $A_5$ RSOS model ,
the $sl_3$ and the $sl_2$ $T-$ systems.

$Y$ functions  for this case read, 
 \begin{eqnarray}
 Y^{(a)}_{k-1} (x)  &:=&   \frac { \prod_{b \sim a} t^{(b)}_1 (x)  }
                           {  t^{(a)}_2  (x)  } , \quad a=1\sim 6     \label{dA4kydef1} 
						   \\
 Y^{(1)}_{k-2} (x)  &:=&   \frac{ t^{(0)}_2 (x)  t^{(3)}_2 (x)}
                                 { t^{(1)}_1(x)  t^{(1)}_3(x)  }   \label{dA4kydef2} 
								 \\
 Y^{(2)}_{k-2} (x)  &:=&   \frac{ t^{(0)}_2(x)  t^{(4)}_2(x) t^{(7)}_2(x)  }
                                                     { t^{(2)}_1 (x) t^{(2)}_3 (x)  } .  \label{dA4kydef3} 
 \end{eqnarray}

%------------------------------------------------------
\subsection{The magnon-like $t-$ system for $A_{4k+1}$ }
%------------------------------------------------------

\begin{figure}[hbtp]
\centering
\includegraphics[width=8cm]{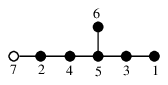}
\caption{Dynkin like diagram  for $A_{4k+1}$ }
\label{fig_a4kplus1}
\end{figure}

Although it shares the same diagram with the case $A_{4k-1}$ , 
 a magnon-like $t-$ system takes a slightly different form,
\begin{eqnarray*}
t^{(a)}_1 (x\pm \frac{i}{2L+4})   &=& 
t^{(a)}_2 (x) +  \prod_{a \sim b}  t^{(b)}_1(x)  , \quad a=3,4,5,6    %\label{tsys21} 
\\
t^{(1)}_1 (x\pm \frac{4 i}{2L+4})   &=&
     t^{(1)}_2(x) + t^{(2)}_1(x)    % \label{tsys22} 
	 \\
t^{(2)}_1 (x\pm \frac{4 i}{2L+4})   &=& 
      t^{(2)}_2(x) + t^{(1)}_1(x)  t^{(7)}_1(x)   %\label{tsys23}
	  \\
t^{(7)}_1 (x\pm \frac{i}{2L+4})   &=& 
      t^{(7)}_2(x)      %\label{tsys24} 
	  \\
t^{(a)}_2 (x\pm \frac{i}{2L+4})   &=&
   \prod_{b \sim a}  t^{(b)}_2 (x)    
      \quad  \quad a=5,6     %\label{tsys25}
	  \\
t^{(3)}_2 (x\pm \frac{i}{2L+4})   &=&
 t^{(3)}_1 (x)     t^{(3)}_3 (x)  +  
 t^{(5)}_2 (x)   t^{(1)}_1 (x\pm \frac{i}{2L+4})    %\label{tsys26}
 \\
t^{(4)}_2 (x\pm \frac{i}{2L+4})   &=& 
  t^{(4)}_1 (x)     t^{(4)}_3 (x)  + 
    t^{(5)}_2 (x)   t^{(2)}_1 (x\pm \frac{i}{2L+4}) .  %\label{tsys27}
\end{eqnarray*}
They are analogous to   the $T-$ system  for level 3 RSOS model, 
corresponding to the coset 
$(D^{(1)}_4)_2 \times  (D^{(1)}_4)_1 /    (D^{(1)}_4)_3$  \cite{KNS1}, 
which is clear by setting  $t^{(1)}_1=  t^{(2)}_1= t^{(7)}_1=1$ and $ t^{(3)}_3=  t^{(4)}_3=0$.
In this case,  $t^{(7)}_1$ will turn out to be a product of  $T_{B_3}$, 
$t^{(1)}_1$ and  $t^{(2)}_1$ are shown to be originated from the $sl_3$ structure.
Again, our interpretation is  that non-trivial $T_{B_3}$ glues 
the $D_4$, the $sl_3$ and the $sl_2$ $T-$ systems.

 $Y$ functions  are defined by,
 \begin{eqnarray*}
 Y^{(a)}_{k-1} (x)  &:=&   \frac{  t^{(a)}_2  (x)  }  { \prod_{b \sim a} t^{(b)}_1 (x)  }, \quad a=3\sim 6   
 %\label{dA4kp1ydef1} 
 \\
 Y^{(1)}_{k-1} (x)  &:=&   \frac{ t^{(3)}_1 (x) t^{(3)}_3(x)  }{ t^{(5)}_2(x)   t^{(1)}_1(x\pm \frac{1}{2L+4}i )     }     %\label{dA4kp1ydef2}
 \\
 Y^{(2)}_{k-1} (x)  &:=&   \frac{ t^{(4)}_1 (x) t^{(4)}_3(x)  }{ t^{(5)}_2(x)   t^{(2)}_1(x\pm \frac{1}{2L+4}i )     }           %\label{dA4kp1ydef3} 
 \\
 Y^{(1)}_{k-2} (x)  &:=&   \frac{ t^{(2)}_1(x)  }{ t^{(1)}_2 (x) }     %\label{dA4kp1ydef4} 
 \\
 Y^{(2)}_{k-2} (x)  &:=&   \frac{ t^{(1)}_1(x)  t^{(7)}_1(x)  }{ t^{(2)}_2 (x)  } .    %\label{dA4kp1ydef5} 
 \end{eqnarray*}

%------------------------------------------------------
\subsection{The magnon-like $t-$ system for $A_{4k+2}$ }
%------------------------------------------------------
%

\begin{figure}[hbtp]
\centering
\includegraphics[width=8cm]{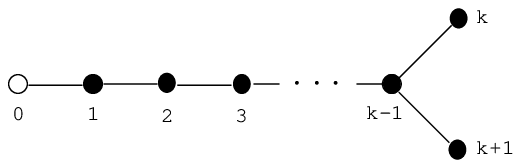}
\caption{Dynkin like diagram  for $A_{4k+2}$}
\label{fig_a4kplus2}
\end{figure}

Only in this case, the diagram depends on $k$.
A magnon-like $t-$ system for the fig. \ref{fig_a4kplus2}
is defined as ,
\begin{eqnarray}
t^{(a)}_m(x \pm \frac{2 i}{L+2})  &= &
t^{(a)}_{m-1} (x)  t^{(a)}_{m+1} (x)   + \prod_{b \sim a} t^{(b)}_m (x) , 
 \quad a=1\sim k+1  \quad  m=1,2,3       \label{tsys41}    \\
t^{(a)}_0&=& 1 \quad  (a=1\sim k+1),   \quad 
 t^{(a)}_4(x)  = 0   \quad (a =2   \sim k+1  )   \nonumber \\
t^{(0)}_1(x \pm \frac{2 i}{L+2}) &=&  t^{(0)}_2 (x),  \quad
t^{(0)}_2(x \pm \frac{2 i}{L+2}) =  t^{(0)}_1 (x) t^{(0)}_3 (x).
  % \label{tsys42} 
 %
\end{eqnarray}
In this case, the functional relations looks like the level 4 RSOS $T-$ system for
$D_{k+1}$, instead of level 3.  
% 
%By examining its solution, however,we find  that  $t^{(a)}_3 (x)$ is given by a product
%of $T_{B_3}$. 
%
%Thus again our interpretation is again that non trivial $ T_{B_3} $ connects the
% $T-$ system for level 2(!)  $D_{k+1}$ RSOS model ,
%the $sl_3$ and the $sl_2$ $T-$ systems.

%Only for this case,  none of $Y^{(a)}_j$ are defined yet.
%They take the most simplest form  among 4 categories,
The description of  $Y^{(a)}_j$  can de done in the most systematic way  among 4 categories,
 \begin{equation*}
Y^{(m) }_{a-1}(x) =\frac{t^{(a)}_{m+1}(x)  t^{(a)}_{m-1}(x)  }{  \prod_{b \sim a}  t^{(b)}_{m}(x) }
\quad m=1,2 \quad  1 \le a \le k +1.
\end{equation*}

%------------------------------------------------------
\subsection{The solutions to magnon-like $t-$ system }
%------------------------------------------------------

The main finding in this report is that the solutions to the above magnon-like $t$ system
are expressible in terms of fusion transfer matrices appearing in dilute $A_L$ models.
We summarize the explicit solutions in appendix \ref{solutiontsys}.
The proof requires several steps and rather lengthy, part of which will be given in section \ref{prooftext}.
Before going into such details, we would like draw attention
to a general property of the solution and also its implication for the $Y-$ system .
Firstly, we need to explain  $T_K(x)$   appearing in the solutions for $L=$  even.

%
% ##############################################################
%
\section{Kink transfer matrix }
%
%################################################################
%
The case studies on $M_{4,5}, M_{6,7}$ \cite{SuzE7} imply the necessity of the introduction of 
objects  which are not produced from boxes  by (\ref{qJT}) .
Generally, we find it possible to introduce the desired objects for the $A_L$ model ,
$L$ even.

The explicit procedure  procedure to construct is as follows. Let $L=2\ell$.
and consider the model in regime 2, corresponding the perturbation $\phi_{1,2}$.
Take 
$$
\omega^{\ell} \phi(x+\frac{\ell}{\ell+1} i) \frac{Q(x+\frac{\ell}{\ell+1} i  )}{Q(x-\frac{1}{\ell+1} i)}
$$
as the top term (or the highest term).  Next, one adds a lower term
$$
\omega^{\ell-1} \phi(x-\frac{\ell+2}{\ell+1} i) 
\frac{Q(x-\frac{\ell+2}{\ell+1} i  ) Q(x-  \frac{2\ell+1}{\ell+1} i   )}
         {Q(x-\frac{1}{\ell+1} i)       Q(x+  \frac{2\ell-1}{\ell+1} i   )          }
$$
so that the singularity at $x= x_j + \frac{1}{\ell+1} i$ cancels due to the Bethe ansatz equation (\ref{bae}).
Then we introduce another counter term so as to kill 
 the singularity at $x= x_j - \frac{2\ell-1}{\ell+1} i$, and so on.
  
One observes that this procedure closes by finite steps and a pole-free object, 
consists of $2 \ell+1$  terms, results.
We refer to this as an eigenvalue of  "kink  transfer matrix" , $T_K(x)$,  although the corresponding
transfer matrix is not yet found in the lattice model.
For example,   
\begin{eqnarray*}
T_K(x) &=& 
    w^4 \phi (x+\frac{4}{5}i) \frac{Q(x+\frac{4}{5}i)} {Q(x-\frac{1}{5}i)}  +
    w^3 \phi (x-\frac{6}{5}i) \frac{Q(x-\frac{6}{5}i) Q(x-\frac{9}{5}i)} {Q(x-\frac{1}{5}i)Q(x+\frac{7}{5}i)} \\
 &  &
      + w^2 \phi (x+\frac{2}{5}i) \frac{Q(x+\frac{2}{5}i) Q(x-\frac{9}{5}i)} {Q(x-\frac{3}{5}i)Q(x+\frac{7}{5}i)} 
    +w \phi (x-\frac{8}{5}i) \frac{Q(x-\frac{8}{5}i) Q(x-\frac{9}{5}i)} {Q(x-\frac{3}{5}i)Q(x+i)}  \\
&  &
  +  \phi (x) \frac{Q(x) Q(x-\frac{9}{5}i)} {Q(x+i)Q(x-i)}  
  +w^{-1}  \phi (x+\frac{8}{5}i) \frac{Q(x+\frac{8}{5}i) Q(x+\frac{9}{5}i)} {Q(x+\frac{3}{5}i)Q(x-i)} \\
 &  &
 + w^{-2 }\phi (x-\frac{2}{5}i) \frac{Q(x-\frac{2}{5}i) Q(x+\frac{9}{5}i)} {Q(x+\frac{3}{5}i)Q(x-\frac{7}{5}i)} 
 +w^{-3}  \phi (x+\frac{6}{5}i) \frac{Q(x+\frac{6}{5}i) Q(x+\frac{9}{5}i)} {Q(x+\frac{1}{5}i)Q(x-\frac{7}{5}i)}  \\
&   &
 +w^{-4} \phi (x-\frac{4}{5}i) \frac{Q(x-\frac{4}{5}i)} {Q(x+\frac{1}{5}i)}
\end{eqnarray*}
for the dilute $A_8$ model. 

The numerical investigation on the largest eigenvalue
indicates  $(Q(x^*))^*= Q(x)$, which 
 leads to an important property  $(T_K(x^*))^*= T_K(x)$.

For our argument, it is not important whether it is a "real " transfer matrix or not,
but significant are its pole-free property as well as functional relations  among
the products of kink transfer matrices and $T_m(x)$.
One finds, for instance,   
$$
T_K(x \pm \frac{2k+1}{4k+2}i)  =  2 T_{4k-1}(x)  
$$
in the case of  the dilute $A_{4k}$ model.

These relations are summarized as lemma  \ref{kinkfunctionalrel} in appendix \ref{kinkfr}.

As mentioned there,  although the explicit proof has been done only for smaller values
of $k$, we assume the following. 

\begin{conjecture}
\label{conjecturemain1}
Lemma \ref{kinkfunctionalrel}  holds for general $k$.
\end{conjecture}

The functional relations there  play an important role in the proof of the magnon-like $t-$ system
for $L=$ even.

%#####################################################
 %
\section{The reality property of magnon-like $t-$  functions  }\label{reality}
%
%#####################################################

We remark the following  significant property of our solutions to the $T$ system.

\begin{lemma} \label{realproperty}
Let $T(x)$ be an either solution of  the magnon-like $t-$ system ,
the $sl_3$ like $T-$ system, or breather $T$ system.
Then,  in "the largest eigenvalue sector", 
\begin{equation}
T(x) = (T(x^*))^*
\label{realT}
\end{equation}
which we call the reality property. 
Namely it assures the reality of
$T(x)$ on the real axis.
\end{lemma}

As is commented in the previous section, $T_K(x)$ satisfies the reality property.
One also verifies 
the property for most of $t^{(a)}_m (x)$ listed in the appendix  \ref{solutiontsys}
using the diagrammatic symmetry  (\ref{diagramsym}), immediately.
It can be  shown for the remaining   $t^{(a)}_m (x)$, which takes "asymmetric" shape , 
with utilizing the magnon-like "t-" system which can be established without use of
the  reality property.
We take the dilute $A_{4k}$ case for example.
Using the diagrammatic symmetry, $t^{(a)}_m (x)$ are shown to satisfy (\ref{realT}) 
except for $(a,m)=(0,1)$ and  $(a,m)=(4,1)$.
The one of the  magnon-like $t-$ system,
$$
t^{(1)}_1(x \pm i\frac{1}{2L+4})- t^{(1)}_2(x)  = t^{(0)}_1(x)  t^{(0)}_3(x) 
$$
can be established independent of the property.
Since all other entries other than $t^{(0)}_1(x)$ satisfy   (\ref{realT}) ,
it follows from the above relation that   $t^{(0)}_1(x)$  also should satisfy  (\ref{realT}) .
The case  $(a,m)=(4,1)$ can be treated similarly.
There are, however,  two exceptions,
$t^{(3)}_1(x)$ and  $t^{(4)}_1(x)$  for the dilute $A_{4k+1}$ model.
One still can devise other functional relations which 
explain the reality property of them.
Since the proof is rather technical, we omit here.

%#####################################################
 %
\section{The $Y-$ system and the $T-$ system  }
%
%#####################################################

Several $Y-$ functions are already introduced for 4 categories by 
appropriate ratios of $t$ or $T$ functions.
Thanks to  the $T-$ system or  the  $t-$ system, one can easily check that 
the $Y-$ system is satisfied for the case in which the lhs are products of 
$Y^{(1)}_j $ or $ Y^{(2)}_j ( x )  $   except for $j=k-2$.
One can verify the rest of the $Y-$ system by utilizing the
explicit expressions of $t$ in  appendix \ref{solutiontsys}.
% 
%This is a little bit different from the situation in $\Phi_{1,3}$ perturbation
%theory in which the $Y-$ system is a direct consequence of the $T-$ system.

Our main message in this report is summarized as follows.
\begin{proposition}\label{YT}
 The four categories of  the $Y-$ system proposed in \cite{DPT2} can be
 solved in terms of transfer matrices appeared in dilute $A_L$ model, if
 $x$ is real.
\end{proposition}
Due to the different forms of $Y-$ system,
the proof must be done  separately for four categories.
Below we shall concentrate on the case  $L=4k$ .
The other cases can be established with minor modifications.
\begin{proof} Proposition \ref{YT} for  $L=4k$\\
Thanks to explicit relations between  $Y$ and $t$ in  (\ref{dA4kydef1}) -(\ref{dA4kydef3})
the "tail " part of the $Y-$ system   (\ref{y4k_4}  ) is shown to be a consequence of 
the magnon-like $t$ system.
Eq. (\ref{y4k_3}  ) also follows from the $sl_3$ structure  (\ref{sl3Ysys1}  )  except for $j=k-2$  . 
Consider the case $\alpha=1$ and $j=k-2$.
By replacing the $Y$ functions by  $t^{(a)}_m$ ,
one finds the lhs,
$$
\frac{ t^{(0)}_2(x\pm \frac{2}{L+2} i )     t^{(3)}_2(x\pm \frac{2}{L+2} i )     }
        {    t^{(1)}_3 (x\pm \frac{2}{L+2} i )      t^{(1)}_1(x\pm \frac{2}{L+2} i ) }
$$
while the rhs reads
$$
\frac{t^{(2)}_2(x \pm \frac{1}{L+2} i) }
        {t^{(2)}_3(x)  t^{(1)}_1(x \pm \frac{2}{L+2} i) (1+\frac{1}{Y^{(1)}_{k-3}(x)  } ) }.
$$
At this stage,  the equality is not obvious.
By the explicit forms of $Y^{(1)}_{k-3} (x) $   in (\ref{sl3y1byt} ), 
we rewrite
$$
1+\frac{1}{Y^{(1)}_{k-3}(x)  } =
\begin{cases}
  \frac{T_{2(k-2)}(x\pm 2i)}{T_{D_{2(k-2)}}(x) },& k \quad {\rm odd}  \\
  \frac{T^{\vee}_{2(k-2)}(x\pm 2i) }{T^{\vee}_{D_{2(k-2)}}(x)}=
     \frac{T_{2(k-2)}(x\pm \frac{2}{L+2}i) }{T^{\vee}_{D_{2(k-2)}}(x)}  ,& k \quad {\rm even}  .
\end{cases}
$$
By substituting further   the explicit forms of $t^{(a)}_m$ in  appendix
 \ref{tsol_A4k} ,   one verifies the case $a=1$ and $j=k-2$; for example
 both sides reduce to
 $$
 \frac{T_{D_{2(k-1)}}(x\pm \frac{2}{L+2}i) }
        { T_{2(k-2)}(x  \pm \frac{2}{L+2}i)   T_{2k}(x  \pm \frac{2}{L+2}i) }
 $$
 in case $k$ even.
 The case  $\alpha=2$ and $j=k-2$  can be shown in the same manner.
 
 Since  eq. (\ref{y4k_1}  ) is  already proved in  (\ref{breatherY}), it remains to 
 show eq. (\ref{y4k_2}  ) . We must treat the cases $k$ even and odd separately.
 For $k= 2 m+1$ ,  
 the product 
$$
  \Xi^{(2)}_{k-1}(x)         
\prod_{\ell=0}^{k-2}  \Xi^{(2)}_{\ell}(x\pm \frac{4k-4\ell-6}{2L+4}i)   =
  \Xi^{(2)}_{k-1}(x)         
\prod_{\ell=1}^{k-1}  \Xi^{(2)}_{k-\ell-1}(x\pm \frac{4\ell-2}{2L+4}i) 
 $$
 is written in terms of fusion transfer matrices  using (\ref{sl3y2byt}), 
 \begin{eqnarray*}
&  &   \frac{T_{D_{2k}} (x\pm \frac{1}{L+2}i ) }
          {T^{\vee}_{D_{2k-2}} (x\pm \frac{1}{L+2}i )}       
\prod_{\ell=1, \ell={\rm odd} }^{k-1}  
  \frac{T^{\vee}_{D_{2k-2\ell }} (x\pm \frac{2\ell+1}{L+2}i )
                       T^{\vee}_{D_{2k-2\ell }} (x\pm \frac{2\ell-3}{L+2}i )   }
          {T_{D_{2k-2\ell-2 }} (x\pm \frac{2\ell-1}{L+2}i )  
		               T_{D_{2k-2\ell+2 }} (x\pm \frac{2\ell-1}{L+2}i )         }  
					                                                     \nonumber  \\
 &  & 
 \prod_{\ell=1, \ell={\rm even} }^{k-1}  
  \frac{T_{D_{2k-2\ell }} (x\pm \frac{2\ell+1}{L+2}i ) 
                    T_{D_{2k-2\ell }} (x\pm \frac{2\ell-3}{L+2}i )   }
          {T^{\vee}_{D_{2k-2\ell-2 }} (x\pm \frac{2\ell-1}{L+2}i ) 
		            T^{\vee}_{D_{2k-2\ell+2 }} (x\pm \frac{2\ell-1}{L+2}i )       
					        }  
							                                               \nonumber  \\
&  &= \frac{T_{D_{2k}} (x\pm \frac{1}{L+2}i ) }
          {T^{\vee}_{D_{2k-2}} (x\pm \frac{1}{L+2}i )}    
\prod_{\ell'=1}^{m}
\frac{ T^{\vee}_{D_{2k-4(\ell'-1)-2}}(x\pm \frac{4(\ell'-1)-1}{L+2} i )  
                    T_{D_{2k-4\ell'}}(x\pm \frac{4\ell' +1}{L+2} i )      }
       {  T^{\vee}_{D_{2k-4\ell'-2}}(x\pm \frac{4\ell'-1}{L+2} i )     
	                T_{D_{2k-4(\ell'-1) }}(x\pm \frac{4(\ell'-1) +1}{L+2} i ) } 
					                                                     \nonumber  \\
& & 
= \frac{T_{D_2}(x \pm \frac{2k-1}{L+2} i)}  
            {T_0(x\pm \frac{2k-1}{L+2}i)   T_0(x\pm \frac{2k-5}{L+2}i)    }
 \end{eqnarray*}
 
 where we have used  $T^{\vee}_{D_0}(x)=T^{\vee}_0(x\pm 2i)=T_0(x\pm \frac{2}{4k+2}i)$.
Multiplying  this by the remaining term  in the rhs, 
$(1+Y_{B_3}(x)) =T_{B_1}(x \pm \frac{2k-1}{L+2}) / T_0( x\pm \frac{2k+3}{L+2} i) $,
one reaches the expression identical to the lhs. Note $T_{D_2}(x)=T_{B_5}(x)$.
The case $k$ even can be treated analogously.
\end{proof}

Once the solutions to magnon-like $t$ system are proved to be given by those
 in the appendix \ref{solutiontsys}, the solutions to $Y$ system
immediately follow.
It is thus vital to verify the solutions.
The proofs for the $L=$ even cases and the odd cases are similar, respectively.
 We then present the proof  for   the case  $L=4k+2$ in the next section, where it can be
 done most systematically.
 As a representative of the odd case,   we supplement the  outline of 
 the proof for $L=4k-1$ case
in the  appendix \ref{appproofL4kp1}.

%###########################################
%
\section{ Proof of the magnon-like $t-$ system  for  $L=4k+2$}\label{prooftext}
%
%#############################################
%
We first prepare a few lemmas
%
%-----------------------------------------
%
\begin{lemma} \label{tate3tate2}
$$
\Lambda_{(m,m,1)/(m-1)}(x)= T_{D_{m-1}}(x) \phi(x\pm \frac{2m+5}{2}i)
$$
\end{lemma}
This lemma holds not only for  $L=4k+2$, but all $L$.
The proof is direct by the $a^{(2)}_2$ property in (\ref{twotoone} ).  
Although it is not used in this section,  this lemma is frequently used
in the proof of the magnon-like $t-$ system. See appendix \ref{prooflemmas}
for example.
%
%
%------------------------------------------------------
%
\begin{lemma} \label{modrelations}
The following equality holds by mod $P$
\begin{eqnarray}
-\frac{k}{2k+2}  & \equiv&     
  \begin{cases}
               2k+3,&              k: \quad {\rm even}   \\
			    2k+3+\frac{P}{2} ,&      k: \quad {\rm odd}   
	\end{cases}
	            \label{modrel2}
\\
\frac{j}{2k+2}  & \equiv&  
  \begin{cases}
               2j,&               j: \quad {\rm even}   \\
			    2j+\frac{P}{2},&       j: \quad {\rm odd} .
	\end{cases}
	            \label{modrel3}
\end{eqnarray}
\end{lemma}
They can be checked trivially.
%
%------------------------------------------------------
%
\begin{lemma}  \label{tksquare}
$$
(t^{(k)}_2(x))^2 =  
 \begin{cases} 
                      T^{\vee}_{D_{2k}}(x)   ,&               k: \quad {\rm even}   \\
					   T_{D_{2k}} (x)  ,&               k: \quad {\rm odd}   \\
  \end{cases}
$$
\end{lemma}
%
%
%------------------------------------------------------
%
\begin{lemma} \label{tk1tk3}
\begin{equation}
t^{(k)}_1(x) t^{(k)}_3(x) =
   \begin{cases}
        \frac{1}{2}( T_{D_{2k+2}}(x)-  T_{D_{2k-2}}(x)  ) ,&  k:\,  {\rm even}  \\
        \frac{1}{2}( T^{\vee}_{D_{2k+2}}(x)-  T^{\vee}_{D_{2k-2}}(x)  ) ,&  k:\,  {\rm odd}  \\   
	\end{cases}
	\label{tk1tk3prod}
\end{equation}
 \end{lemma}
 The proofs of the above two lemmas are given in the appendix \ref{prooflemmas}.
 
With these preparations, we shall  prove the magnon-like $t$ system equation by equation.
Consider first the case $ m=1$ in (\ref{tsys41} ).
For $a=1,\cdots, k-2$, by paying attention to (\ref{modrel3} ) with $j=1$, one immediately sees
that  the assertion is actually equivalent to  (\ref{sl3one} ).
Thus only the cases $a=k, k-1$ need proofs.
Using  $t^{(k)}_1(x) =T_{K}(x)/2$ and the functional relation  (\ref{kinkt5} ) with $j=k$,
we find
$$
t^{(k)}_1(x\pm \frac{1}{2k+2}i)  =
\begin{cases}
      T^{\vee}_{2k-2}(x)  + \frac{1}{2}( T^{\vee}_{2k+2}(x) -   T^{\vee}_{2k-2}(x) ) & k:\,{\rm even}   \\
      T_{2k-2}(x)  + \frac{1}{2}( T_{2k+2}(x) -   T_{2k-2}(x) ) & k:\,{\rm odd}. 
	  \end{cases}
$$
The first and the second term in the rhs  coincides with $t^{(k-1)}_1(x)$ and  $t^{(k)}_2(x)$, respectively, which 
proves   the case $a=k$ of (\ref{tsys41} ).
The $a=k-1$ case is less trivial. 
Substituting $a=k-1$ in (\ref{dA4kp2_s2})   and using  (\ref{sl3one} ),  one finds 

\begin{equation}
t^{(k-1)}_1(x\pm \frac{1}{2k+2}i)  =
\begin{cases}
T_{2k-2}(x \pm 2i)   =T_{2k}(x)  T_{2k-4}(x)  +T_{D_{2k-2}}(x) &   k\,{\rm even} \\
T^{\vee} _{2k-2}(x \pm 2i)=   T^{\vee}_{2k}(x)  T^{\vee}_{2k-4}(x)  +T^{\vee}_{D_{2k-2}}(x) & k\,{\rm odd}.
\end{cases}
\label{tkm1tkm1}
\end{equation}
The second term in the rhs is identified with $t^{(k-1)}_2(x)$.

We note  
$$
t^{(k)}_1(x) t^{(k+1)}_1(x) =(T_K(x)/2)^2 = 
\begin{cases}
   T_{2k}(x)    & k\,{\rm even} \\ 
    T^{\vee}_{2k}(x)  & k\,{\rm odd} 
\end{cases}
$$
where (\ref{kinkt5} ) with $j=k+1$ is used in the second equality.  
Since  $T_{2k-4}(x)=t^{(k-2)}_1(x)$,  the first term in  (\ref{tkm1tkm1}) can be
written as   $t^{(k-2)}_1(x) t^{(k)}_1(x) t^{(k+1)}_1(x)$.
Combining these observations,  we find that 
 the  rhs of (\ref{tkm1tkm1})  coincides with 
$t^{(k-1)}_2(x)+  t^{(k-2)}_1(x) t^{(k)}_1(x) t^{(k+1)}_1(x)$ , which proves  (\ref{tsys41} ) for  $a=k-1, m=1$ .

Next, consider the case   $ m=2$ in (\ref{tsys41} ).
As in the case of   $ m=1$ ,  the relation is  equivalent to  (\ref{sl3two} ) for $a=1 \sim k-2$.
One only has to pay an attention,
$$
T^{\vee}_{D_0}(x)= T_0(x \pm \frac{1}{2k+2}i) =t^{(0)}_2(x)
$$ 
for $a=1$.
In the case of $a=k-1$, the product of $t^{(k-1)}_2(x)$ reads,
$$
t^{(k-1)}_2(x\pm \frac{1}{2k+2}i)  =
\begin{cases}
T^{\vee}_{D_{2k-2}}(x \pm 2i)   =T^{\vee}_{2k-2}(x)T_{B_3}(x \pm \frac{1}{2k+2}i)
       +T^{\vee}_{D_{2k}}(x)T^{\vee}_{D_{2k-4}}(x) &   k\,{\rm even} \\
T_{D_{2k-2}}(x \pm 2i)=  T_{2k-2}(x)T_{B_3}(x \pm \frac{1}{2k+2}i)  +T_{D_{2k}}(x)T_{D_{2k-4}}(x)& k\,{\rm odd} 
\end{cases}
$$
where  (\ref{sl3two} )  is again applied.
The first term in the rhs is the product  $t^{(k-1)}_1(x)  t^{(k-1)}_3(x) $, while
the second is  $t^{(k)}_2(x)  t^{(k+1)}_2(x)   t^{(k-2)}_2(x) $.
This comes from Lemma \ref{tksquare},  as   $t^{(k)}_2(x)  t^{(k+1)}_2(x)= (t^{(k)}_2(x) )^2$.
Thus the  case $a=k-1, m=2$ is proved.

The   case $a=k, m=2$ is the most non-trivial.  We treat  only the case  $k$ even for brevity.
When taking the product $t^{(k)}_2(x \pm i\frac{1}{2k+2})$,
we substitute two different expressions for  $t^{(k)}_2(x)$.
The first comes from  the already proved relation, $t^{(k)}_2(x)=t^{(k)}_1(x\pm i\frac{1}{2k+2}) -t^{(k-1)}_1(x)$.
The second is as listed in  (\ref{dA4kp2_s3}).

The first expression results, after  applying (\ref{kinkt5}), 

\begin{eqnarray}
& &t^{(k)}_2(x \pm i\frac{1}{2k+2})   \nonumber  \\
&  & \quad  =
 \frac{1}{2}  \Bigr( T_{2k}(x)   T_{2k-4}(x) +  T_{2k}(x)   T_{2k+4}(x)     \nonumber \\
    &  &  \qquad - T^{\vee}_{2k-2}(x-\frac{1}{2k+2}i)   T^{\vee}_{2k+2}(x+\frac{1}{2k+2}i)
					 - T^{\vee}_{2k-2}(x+\frac{1}{2k+2}i)   T^{\vee}_{2k+2}(x-\frac{1}{2k+2}i) \Bigl)    \nonumber \\
& & \quad =    \frac{1}{2}  \Bigr( T_{2k}(x)   T_{2k-4}(x) +  T_{2k}(x)   T_{2k+4}(x)     \nonumber \\
 &  & \qquad 
                     - T_{2k-2}(x-2 i)   T_{2k+2}(x+2i) 
					 - T_{2k-2}(x+2i)   T_{2k+2}(x-2 i) \Bigl) .
 \label{tk2exp1}
 \end{eqnarray}
 From the second expression, $t^{(k)}_2(x)=\bigl (T^{\vee}_{2k+2}(x)-T^{\vee}_{2k-2}(x) \bigr )/2 $,  we find 
 \begin{eqnarray}
t^{(k)}_2(x \pm i\frac{1}{2k+2})&=&
 \frac{1}{4}  \Bigr( T_{2k}(x)   T_{2k-4}(x) +  T_{2k}(x)   T_{2k+4}(x)  + T_{D_{2k-2}}(x) +   T_{D_{2k+2}}(x)  \nonumber \\
 &  &
                       - T_{2k-2}(x-2 i)   T_{2k+2}(x+2i)) 
					 - T_{2k-2}(x+2i)   T_{2k+2}(x-2 i)  \Bigl)  
 \label{tk2exp2}
  \end{eqnarray}
 where (\ref{sl3one}) is used.
Twice  (\ref{tk2exp2}) and subtract  (\ref{tk2exp1}), one arrives at,
$$
t^{(k)}_2(x \pm i\frac{1}{2k+2})=\frac{1}{2}(  T_{D_{2k-2}}(x) +   T_{D_{2k+2}}(x) )
=T_{D_{2k-2}}(x)+ \frac{1}{2}(  T_{D_{2k+2}}(x) -   T_{D_{2k-2}}(x) ).
$$
The first term in the rhs coincides with $t^{(k-1)}_2(x)$, while the second agrees with
$t^{(k)}_1(x)  t^{(k)}_3(x) $ due to Lemma  \ref{tk1tk3}.
Thus the case $a=k, m=2$  is established.

Finally consider the case $m=3$.
Except for  the case $a=1$ , the relation is satisfied trivially.
When $a=1$ , the lhs reads

\begin{align*}
t^{(1)}_3(x\pm \frac{1}{2k+2}i) &=  T_{B_3}(x \pm \frac{k-2}{2k+2}i) T_{B_3}(x \pm \frac{k}{2k+2}i)  \\
&= T_{B_3}(x \pm \frac{k-2}{2k+2}i) ( T_0(x) T_0(x  \pm \frac{2}{2k+2}i)+T_{B_1}(x) T_{B_5}(x) ) \\
&=T_{B_3}(x \pm \frac{k-2}{2k+2}i)   T_0(x) T_0(x  \pm \frac{2}{2k+2}i)+
 T_{D_2}(x)  T_{B_1}(x) T_{B_3}(x \pm \frac{k-2}{2k+2}i)  
\end{align*}

where (\ref{tb3tb3}) is used in the second quality and $T_{D_2}(x)= T_{B_5}(x) $ is 
used in the last.

By noting $t^{(0)}_3(x)=T_0(x) T_0(x  \pm \frac{2}{2k+2}i)$, one find the resultant
expression coincides with
$t^{(2)}_3(x)  t^{(0)}_3(x) + t^{(1)}_2(x)   t^{(1)}_4(x) $.

We thereby verified the magnon-like $t$ system for $L=4k+2$.

%%%%%%%%%%%%%%%%%%%%%%%%%%%%%%%%%%%%%%%%%%%%%%
                                                                                                                                       %
\section{Thermodynamic Bethe Ansatz in the scaling limit }     \label{derTBA}                  %
                                                                                                                                       %
%%%%%%%%%%%%%%%%%%%%%%%%%%%%%%%%%%%%%%%%%%%%%%
\subsection{Analytic properties}\label{Yanaly}

The $Y-$ systems are obtained in \cite{DPT2} by the simply Fourier transformation of  the
thermodynamic Bethe ansatz equation (TBA, for short), which is proposed in advance.
In the present report, the $Y-$system has been proved first, in the preceding sections.
The (inverse) transformation of the  $Y-$system to TBA seems thus to be direct.
This is, however, not necessary true.  One needs further analytic properties of 
of $Y-$ functions as we shall see below.

The $Y-$ system typically assumes the following form. (See  appendix \ref {list_Y-system} for notations.)
\begin{equation}
Y^{(a)}_m(x\pm i \alpha) = \prod_{b,j } \Xi^{(b)}_j (x \pm i  \gamma^{(b)}_j)   
           \prod_{c,k } {\cal L}^{(c)}_k(x \pm i  \gamma^{(c)}_k)     \label{typical}
\end{equation}
We try to take the logarithm of both sides and take the Fourier transformation.
For the necessary condition that  the last transformation can be done simply,
one must ensure that the lhs (rhs) is analytic and nonzero in the strip 
$\Im x \in [-\alpha, \alpha]$   ($\Im x \in [ -\gamma^{(b)}_j, \gamma^{(b)}_j]$ ).
Otherwise, one must take account of contributions from zeros in the strips. 

As is already shown, complex combinations of "transfer matrices" solve the both sides. 
The  explicit expressions for eigenvalues of transfer matrices thus enable the investigations 
of the analytic properties of the  $Y-$ systems, once parameters, $N, q, \beta$ are fixed.
We have carried out numerical investigation for ranges of parameters, $L=5,7,8,9,10$, 
$ 4 \le N \le 10, 0\le q \le 0.5,  \beta =0.04, 0,05, \cdots, 0.1 $.
Although our main interest is focus on the limit $N \rightarrow  \infty$, 
our data suggests weak dependency on $N$.
For illustration, appendix \ref{numericsA8} presents
some numerical data and graph showing locations of zeros
for various transfer matrices for the dilute $A_8$ model in regime 2 .
The analytic property of $Y$ can be easily read off from these zeros.

To state our conclusion,  we remark that the $Y-$ system is invariant , for even $N$,  if
 $Y$ is replaced by $\widetilde{Y}$,  defined by
\begin{equation*}
\widetilde {Y}_{B_1}(x) =
  \begin{cases}
    \frac{Y_{B_1}(x) }{\kappa(x \pm i(1+u') \frac{L-2}{2(L+2)})}&  {\rm for }   \quad u<0 \\
   Y_{B_1}(x)  \kappa(x \pm i(1-u') \frac{L-2}{2(L+2)})   &  {\rm for }  \quad  u>0 \\
  \end{cases}
\end{equation*}
and all other cases,   $\widetilde{Y}=Y$.  The parameter $u'$ stands for $\frac{2(L+2)}{L-2} u$.
This is due to the definition of $\kappa$,
$$
\kappa(x) =\Bigl (
 i \frac{ \vartheta_1 ( i \frac{(L+2)}{2(L-2)} \pi x, \tau') }
           {   \vartheta_2 ( i \frac{(L+2)}{2(L-2)} \pi x, \tau') }
  \Bigr )^N
$$
which satisfies  $ \kappa(x\pm i \frac{L-2}{2(L+2)})=1$.
The elliptic nome $q'=\exp(-\tau'), \tau'=\frac{2(L+1)}{L-2} \tau$ is introduced so as to 
respect the periodicity of the $Y$ function on the real direction of $x$. 
We refer the resultant functional relation to $\widetilde{Y}$- system. 

Our conjecture is as follows.
\begin{conjecture}\label{analy}
The both sides of the $\widetilde{Y}$-system
are analytic and nonzero for appropriate strips, for any $L, N, q, \beta$.
\end{conjecture}
\subsection{The coupled Integral equations}

Once the conjecture  \ref{analy} is taken for granted, the transformation of 
the $\widetilde{Y}$-system into a set of integral equations  is straightforward.
When all $\widetilde{Y}$ functions is replaced by  $Y$, only the integral equation
for $\log Y_{B_1}(x)$ possesses a drive term.
That is, for $u>0$, with appropriate kernel functions $K_{B_1, a, m}$, it takes of the form,
\begin{eqnarray}
\log Y_{B_1}(x) &=&  - \log  \kappa(x \pm i(1-u') \frac{L-2}{2(L+2)})  +  
       \sum_{(a,m)} \int_{-\tau"}^{\tau"}  K_{B_1, a, m }(x-x') \log \Xi^{(a)}_m (x')  dx'   \nonumber \\
 &  & + \sum_{(a,m)} \int_{-\tau"}^{\tau"}K_{B_1, a, m }(x-x') \log {\cal L}^{(a)}_m (x')  dx'    \label{nlin1}
\end{eqnarray} 
where $\tau" = \frac{(L-2)   }{(L+2)} \frac{\tau'}{\pi} $.
For $u<0$, the driving term should be  $ \log  \kappa(x \pm i(1+u') \frac{L-2}{2(L+2)}) $.

The free-energy is evaluated from
\begin{eqnarray}
\log T_{B_1}(x)&= & \log r_N(x) + \int_{-\tau"}^{\tau"} s(x-x') \log \varepsilon_N(x') dx'   \nonumber \\
& &+ \int_{-\tau"}^{\tau"} s(x-x') \log(1+Y_{B_1}(x'))dx'   \label{free1} \\
 r_N(x)&=&  \Bigl (  
                              \frac{h(x \pm(\frac{2L-4}{2L+4}+u)i)   h(x \pm(\frac{4L}{2L+4}+u)i)       }
                                      {h( \pm(\frac{2L-4}{2L+4})i)   h( \pm(\frac{4L}{2L+4})i)   }
                         \Bigr )^{\frac{N}{2}}    \nonumber  \\
\varepsilon_N(x)&=&  \Bigl (  
                              \frac{h(x \pm(\frac{L-2}{2L+4}-u)i)   h(x \pm(\frac{3L+2}{2L+4}-u)i)       }
                                      {h(x \pm(\frac{L-2}{2L+4}+u)i)   h( x\pm(\frac{3L+2}{2L+4}+u)i)   }
                         \Bigr )^{\frac{N}{2}}      \nonumber \\
 \widehat{s(k)}&=&\frac{1}{2 \cosh \frac{L-2}{2(L+2)} k } . \nonumber
\end{eqnarray}
Here we denote the Fourier transformation of $F(x)$ by $\widehat{F}(k)$,
\begin{eqnarray*}
\widehat{F}(k) &=& \int_{-\tau"}^{\tau"} F(x) e^{-ik x} dx \\
F(x) &=& \frac{\delta}{2 \pi}  \sum_{n}  \widehat{F}(k_n) e^{ik_n x}, \quad k_n= n \delta 
\end{eqnarray*}
  and $\delta=\pi/\tau"$.
The free energy is determined by the value of $T_{B_1}(x)$ at the origin,
$\beta f = -\log T_{B_1}(0)$, thus we are only interested in $x \sim 0$ in  (\ref{free1} ).

 All the above results are valid for any $N$.
 Now the Trotter limit can be performed analytically.
The drive term in  (\ref{nlin1}) becomes,
$$
 - \log  \kappa(x \pm i(1-u') \frac{L-2}{2(L+2)})  \rightarrow 4 \beta \pi s(x)
$$
  while
  \begin{eqnarray*}
  \log  r_N(0)  &\rightarrow&  \lambda \beta 
     \Bigl(  (\log \vartheta_1(\frac{L-2}{L+2}\lambda))'+  (\log \vartheta_1(\frac{2L}{L+2}\lambda))' \Bigr ) \\
	\log  \varepsilon_N(x)  &\rightarrow&  \lambda \beta  
     \Bigl(  \bigl(\log   \frac{  \vartheta_1(\lambda(ix -\frac{L-2}{2L+4})  }
	                                       { \vartheta_1(\lambda(ix+\frac{L-2}{2L+4})  }
                	       \bigr )'+  
			\bigl(\log   \frac{  \vartheta_1(\lambda(ix -\frac{3L+2}{2L+4})  }
	                                       { \vartheta_1(\lambda(ix+\frac{3L+2}{2L+4})  }
                	       \bigr )'
						   \Bigr )
  \end{eqnarray*}
  for $u=\beta/N$.  The crossing parameter $\lambda$ is set to be $\frac{\pi(L+2)}{4(L+1)}$.
  
\subsection{Thermodynamic Bethe Ansatz equations}  

We consider the scaling limit $q \rightarrow 0, \beta \rightarrow \infty$ \cite{BWN}.
The precise tuning of these two parameters will be specified below.
In order to take this limit. 
we  conveniently rewrite the integrals in the preceding subsection
in such a form that "kernel function times small functions" .
The numerical calculation indicates,  in the vicinity of the origin, 
$$
\begin{cases}
 |Y| >>1,  &   u>0  \\
 |Y| << 1,  &   u<0 . \\
\end{cases}
$$
if $\beta >>1$.
In the connection with $M_{L, L+1}+\Phi_{1,2}$ theory, we are interested in 
the regime 2 ($u>0$) case.
Thus we  first rewrite the $Y$system such that the relations involves only $(1+\frac{1}{Y})$.
Now the important contribution for the integral comes from "Fermi" surface $x' \sim \tau"$.
Thus we are interested in the scaling function,  $\lim_{q \rightarrow 0} y_{B_1}(x+\tau")$
etc and  integral equations among them, TBA equation.
To derive them, we 
take the logarithm of both sides  of the transformed $Y$ system above, 
take also the Fourier transform and find
\begin{equation}
\widehat{M }
\begin{pmatrix}   
                        \widehat{\log} Y_{B_1} \\
                            \widehat{\log} Y_{B_3} \\
                           \vdots \\
\end{pmatrix} 
=4 \pi \beta 
 \begin{pmatrix}   1 \\
                           0 \\
                           \vdots \\
\end{pmatrix} 
+
\widehat{K_0 }
 \begin{pmatrix} 
                             \widehat{L}_{B_1}  \\
                            \widehat{L}_{B_3}   \\
                           \vdots \\
\end{pmatrix} 
\label{fouriernlie}
\end{equation}
where $\widehat{L}_{B_1} = \widehat{\log}(1+ \frac{1}{Y_{B_1}})$ and similarly for others.
Note that $M$ and $K_0$ are asymmetric matrices. See  appendix \ref{kernelexample} for the example for $L=5, 7$.
By multiplying $M^{-1}$ from the left,  @the kernel matrix for TBA, 
 $M^{-1} K_0$  turns out to be symmetric, remarkably.
This property is crucial in applying the dilogarithm technique to evaluate the central charge.
Its elements agree with the expression  described in \cite{DPT2} in terms of $S$ matrices,
under identification $x =3 \theta/\pi$  in the limit $q \rightarrow 0$.
Finally, we have checked that nonzero drive terms exist 
only for $Y$ functions  corresponding to Breathers and Kinks. 
For example, we consider $L=5$.
 The explicit forms of the drive terms in the Fourier space read,
 \begin{eqnarray*}
\widehat{ d}_{B_1}&=&\frac{ 8 \pi \beta  \cosh\frac{11}{14}k }{(2\cosh\frac{2}{14}k -1)D(k)}   \\
 \widehat{ d}_{B_3}&=&\frac{4 \pi \beta  (2 \cosh\frac{2}{14}k +1)  (2\cosh\frac{4}{14}k -1)   }{D(k)}   \\
 \widehat{ d}_{B_5}&=&\frac{ 16 \pi \beta \cosh\frac{1}{14}k \cosh\frac{4}{14}k  }{(2\cosh\frac{2}{14}k -1) D(k)}   \\
 \widehat{ d}_{B_2}&=&\frac{ 8 \pi \beta  \cosh\frac{1}{14}k  }{(2\cosh\frac{2}{14}k -1) D(k)}   \\
 \widehat{ d}_{K_1}&=&\frac{ 4 \pi \beta }{(2\cosh\frac{2}{14}k -1) D(k)}   \\
\widehat{ d}_{K_2}&=&\frac{ 8 \pi \beta \cosh\frac{4}{14}k }{(2\cosh\frac{2}{14}k -1) D(k)} ,
 \end{eqnarray*}
where we denote by $\widehat{d}_{B_1}$ for the drive term associated to $\widehat{\log} Y_{B_1}$ and so on.
A common denominator $D(k)$ denotes
$$
D(k)=2 \cosh \frac{12}{14}k+2  \cosh \frac{10}{14}k -2  \cosh \frac{6}{14}k-2  \cosh\frac {4}{14}k+1.
$$
When taking the inverse Fourier transformation, the nearest zero to the real axis of the denominator is 
relevant in the "scaling" limit, $q \rightarrow 0$. 
We have checked for various $L$ that the zero of  the common denominator $D(k)$
brings the nearest zero, and always, $k=\pm i\frac{\pi}{3}$.

Let us be more explicit for $L=5$.  %
As announced above, we are interested in the integral equation
for $y(x)=\lim_{q \rightarrow 0}  Y(x+\tau")$.
Thus the associated drive term reads,
$$
d^{\rm scaling}(x):= \lim_{q \rightarrow 0} d(x+\tau")=
\lim_{q \rightarrow 0} \frac{1}{2 \pi} \sum_m \int_{-\infty}^{\infty}
 e^{ 2 \pi i m \frac{k'}{\delta} +i(x+\tau")  k'} \widehat{d}(k') d k'
$$
where the Poisson's summation formula is applied.
As $q \rightarrow 0$,   $\delta, \tau"$ tends to be infinity, the only $m=0, m=-1$ terms contribute,
$$
d^{\rm scaling}(x)  \sim  \lim_{q \rightarrow 0} 
\Bigl ( 
  \frac{1}{2 \pi} \int_{-\infty}^{\infty} e^{i(x+\tau")  k'}  \widehat{d}(k') d k' +
  \frac{1}{2 \pi} \int_{-\infty}^{\infty} e^{-i(\tau"-x)} \widehat{d}(k') d k' 
  \Bigr )   .
$$
We enclose the complex contour of the integration  in the upper half plane for the first term 
and  in the lower half plane for the second. For   $\tau" \rightarrow \infty $, the nearest poles 
$k=\pm i \pi/3$ bring the dominant contributions.
We denote 
$$
\frac{1}{D(k)} \sim \frac{r}{(k-i\frac{\pi}{3})i},
$$
then 
$$
d^{\rm scaling}_{K_1}(x) = \frac{8 \pi r}{2 \cos \frac{\pi}{21} -1}  e^{-\frac{4}{7}\tau}  \cosh \frac{\pi}{3}x 
$$
for example.
Let $R$ be radius of compactification. 
Then  we take $q \rightarrow 0, \beta \rightarrow \infty$ while $m_k$ finite;
$m_k R=\frac{8 \pi \beta r}{2 \cos \frac{\pi}{21} -1} q^{\frac{4}{7}}$.
This is the precise meaning of our scaling limit.
We write  $\frac{\pi}{3}x  =\theta$.
Immediately, one verifies that all other drive terms also take the form $m R \cosh \theta$
and their mass ratio agree with those in \cite{DPT2}.
\begin{align*}
m_{B_1} &= 2 m_K \cos\frac{11}{42} \pi &   m_{B_3} &=4 m_K  \cos\frac{11}{42}\pi  \cos\frac{3}{42}\pi  \\
m_{B_5}&=4 m_K  \cos\frac{1}{42}\pi  \cos\frac{4}{42}\pi & 
m_{B_2} &= 2 m_K \cos\frac{1}{42} \pi   \\
m_{K_2} &= 2 m_K \cos\frac{4}{42} \pi .    &  \\     
\end{align*}
We then write  $y_a(\theta) = \exp(\epsilon(\theta))$, and verify
that the TBA is recovered in the scaling limit.

We comment that for general $L$, the straightforward calculation shows
$m_K \propto q^{\frac{2(L+1)}{3(L+2)}}$ . This is consistent with scaling relation,
$\xi=\frac{1}{2-2\triangle_{1,2}}$ for  $m  \propto q^{\xi}$.
Indeed comparison of the result leads to proper scaling dimension,
$\triangle = \frac{L-2}{4(L+1)}$.

It remains to check the expression for the scaling free energy.

The scaling limit of free energy comes from the contribution from
the last term in (\ref{free1} ) at $x=0$.
$$
\int_{-\infty}^{\infty} s(x') \log (1+Y_{B_1}(x')) dx' =
\int_{-\infty}^{\infty} s(x') \log Y_{B_1}(x') dx' + \int_{-\infty}^{\infty}  s(x') \log L_{B_1}dx' .
$$
We substitute  the inverse transformed (\ref{fouriernlie}) into
the first term in the rhs above to find,
$$
\int_{-\infty}^{\infty}s(x') \log (1+Y_{B_1}(x')) dx' =
\int_{-\infty}^{\infty} s(x') d_{B_1}(x') dx' +
 \sum_{\ell} \int_{-\infty}^{\infty} {\cal K}_{\ell} (x') \log L_{\ell}dx' .
$$
where  ${\cal K}_{\ell}$ reads in the Fourier space, 
$$
\widehat{ {\cal K} }_{\ell} = \widehat{s} (\widehat{M}^{-1})_{1,j} \widehat{D}_{j,\ell}.
$$
and
$$
\widehat{D}_{i,j} :=\widehat{ M}_{i,j}+ (\widehat{K}_0)_{i,j}.
$$
We adopt indices $y_{B_1}=y_1, y_{B_3}=y_2, $ and so on.

The scaling free energy $f_s$, the contribution of excitations near the 
Fermi surface,   is thus identified with

$$
f_s = -  \lim_{q \rightarrow 0}
\frac{1}{\beta}   \sum_{\ell}  \int_{-\infty}^{\infty}  {\cal K}_{\ell} (x'+\tau") \log (1+\frac{1}{y_\ell}(x')) dx'.
$$
To perform the limit, we use a remarkable relation,
 $$
 \widehat{ {\cal K} }_{\ell} =\widehat{d}_{\ell} 
 $$
 which we  checked case by case.
 Then one repeats the same argument in deriving the drive terms in TBA equation, 
 and reaches
 $$
f_s= -\frac{1}{2 \pi} \sum_{\ell} \int_{-\infty}^{\infty}  m_{\ell} \cosh \theta   \log (1+\frac{1}{y}_{\ell})  d\theta,
= -\frac{1}{2 \pi} \sum_{\ell} \int_{-\infty}^{\infty}
      m_{\ell} \cosh \theta   \log (1+{\rm e}^{-\epsilon(\theta)})  d\theta
$$
under tuning  $\beta =\sqrt{\frac{2 \pi}{3}} R$.
Therefore  the field theoretical  expression of the free energy 
is  also reproduced.
%
%
%%%%%%%%%%%%%%%%%%%%%%%%%%%%%%%%%%%%%%%%%%%%%
%%%%%%%%%%%%%%%%%%%%%%%%%%%%%%%%%%%%%%%%%%%%%
%%%%%%%%%%%%%%%%%%%%%%%%%%%%%%%%%%%%%%%%%%%%%
%%%%%%%%%%%%%%%%%%%%%%%%%%%%%%%%%%%%%%%%%%%%%
%######################################################

\section{Summary and discussions}

In this report,  we establish relevant functional relations 
of the dilute $A_L$ models for {\itshape arbitrary} $L$, 
to analyze finite temperature property 
we demonstrate explicitly that  TBA equations for $M_{L,L+1}+\phi_{1,2}$
, conjectured by Dorey, Pocklington and Tateo \cite{DPT2}, is realized in the scaling limit.
The crucial idea is to introduce fusion transfer matrices associated to skew Young
tableaux and to investigate the functional relations among them.

There are still many open problems. 
The explicit identification of string solutions would be
definitely one of the most important.
The complete study on this will shed some light on the 
way how to proceed  for TBA in the case of  perturbed nonunitary minimal
 models.
We mention the first step in this direction in \cite{EllenBaz2}.
The minimal  unitary theory perturbed by $\phi_{2,1}$ may be treated within the
same framework. One only has to deal with the regime 1 instead of the regime 2.
We already have  some results about the "exceptional" cases in terminology of \cite{DPT2}
, however,  they are omitted in this report for brevity.
We hope to complete the program on the $\phi_{2,1}$ theory for arbitrary $L$ in a subsequent
report.

Before closing this paper, we comment on a possible link to Fermionic formulae of the Virasoro characters.
In \cite{KKMM, DKKMM, berkovich94, Melzer94, BM96}, remarkable results  are reported on 
  new representations of  the Virasoro characters which
  originate form  fermionic particle bases of the Bethe ansatz equations.
See also combinatorial arguments in view of TBA\cite{Warnaar96I, Warnaar96II}
and mathematical generalizations based on the combinatorial objects, the 
rigged configurations\cite{KKR88}-\cite{Schilling02}.
The Fermionic characters obtained in  \cite{KKMM}
-\cite{Warnaar96II} seem to be connected to $\phi_{1,3}$ perturbation.  
In \cite{BMP, BM98}, 
they also explicitly showed the different  realization of   Fermionic formulae 
which are related to  $\phi_{2,1}$ or   $\phi_{1,5}$ perturbations.

Since the TBA  in \cite{DPT2}
 offers a description of the associated Hilbert space, 
 it might be natural to expect that a new  fermionic representation
is possible based on the $\phi_{1,2}$   perturbation.  
There exists already  positive results based on the simplest case,
 the dilute $A_3$ model \cite{WarnaarPearce}. 
 It may be natural to expect the generalizations.

Indeed, we check up to $O(q^{200})$ that the following formula is valid,
$$
 \chi^{5,6}_{1,1} = \sum 
      \frac{ q^{\frac{1}{2} {\bf n} g {\bf n}   } }
              {\prod_{i=1}^6 (q)_{n_i}}
 \begin{bmatrix}
 \frac{n_5+n_6+n_8}{2} \\
 n_7\\
 \end{bmatrix}
 \begin{bmatrix}
 \frac{n_7}{2} \\
 n_8\\
 \end{bmatrix}
 $$
where 
$$
 g=
 \begin{pmatrix}
 4, &  6,&   8,&  4,&  2,& 4,& 0, &0\\
 6, &12, &16,&  8, & 4,& 8, &0, &0\\
 8, &16, &24, &12,& 6, &12,& 0,& 0\\
 4, & 8,&12, &8, &4,& 6,& 0,& 0\\   
 2, &4, &6, &4, &3, &3,& -\frac{1}{2},& 0\\
 4,& 8, &12, &6, &3, &7,& -\frac{1}{2}, &0\\
 0, &0, &0,& 0,& -\frac{1}{2}, & -\frac{1}{2}, & 1,& -\frac{1}{2}\\
 0, &0, &0, &0,& 0, &  0,& -\frac{1}{2}, & 1\\
 \end{pmatrix}.
 $$
 The summation is taken under the condition,  
 $n_7 \le \frac{n_5+n_6+n_8}{2}\in Z_{\ge 0} ,  n_8 \le \frac{n_7}{2}\in Z_{\ge 0}$.
  Note that  $g$ has its origin in the kernel functions of the TBA equation.
 
 A similar formula for $\chi^{7,8}_{1,1}$  is checked up to  $O(q^{100})$, which may  imply the existence of 
  Fermionic formulae for  $L$ odd case in general .  We are not yet able to find such a simple expression for
 the $L$ even case.
 We hope to clarify this  in a future publication \footnote{ We have learnt that S.O. Warnaar has independently
 found polynomial version of these formulas corresponding to both $\Phi_{1,2}$ and $\Phi_{2,1}$ perturbations.  }.
 
\noindent
{\bf Acknowledgements}
\noindent
 The author would like to thank P. Dorey and R. Tateo for many valuable comments,
 discussions, cirtical reading of the manuscript
 and collaborations at early stage. 
 He also thanks S.O. Warnaar for correspondence,
 B.M. McCoy, P.A. Pearce and J. Shiraishi for conversations.
 A part of the present work has been reported at the APCTP focus program,
 'Finite size technology in low dimensional quantum field theory'.
  The author thanks organizers for the kind invitation. 
This work  has been supported by a Grant-in-Aid
for Scientific Research from the Ministry of Education, Culture,
Sports and Technology of Japan, no.~14540376.

\appendix
%
%####################################################

%
% ##########################################
%
 \section{Y-system }\label{list_Y-system}
 %
 %############################################
 %
 As obtained in \cite{DPT2}, $Y$ system for the minimal unitary 
 theories perturbed by $\Phi_{1,2}$ falls into 4 categories.
 Below we shall summarize their result with  slight changes in 
 notations.
 Instead of their rapidity variable $\theta$, 
 we use $x$ as in the text.
 They are simply related by
 $
 x= \frac{3 \theta}{\pi}.
 $
 By $\CL$ we mean
 $
 \CL = \frac{1}{1+ \frac{1}{Y}}
 $
 for a $Y$ function.  In the section \ref{derTBA},  an inverse function
 $
 L ={1+ \frac{1}{Y}}
 $
is also introduced but it will not be utilized here.
  We understand indices associated with $\CL$ specify 
 these for $Y$ functions, e.g.,
 $$
 \CL^{(a)}_m(x) = \frac{1}{1+ \frac{1}{Y^{(a)}_m  (x)  }}
 $$
 etc.
 Similarly $\Xi^{(a)}_m$ denotes
 $$
 \Xi^{(a)}_m (x)= 1+Y^{(a)}_m(x)
 $$
 %
 %-----------------------------------------------------
 %
 \subsection {\bf Case, ($L=4k-1, k \ge$ 2):}
 The Y-system is
\begin{eqnarray*}
& &Y_{B_1}(x \pm \frac{L-2}{2L+4 }i ) = \Xi_{B_3}(x)       %\label{y4km1_1} 
  \\
& &Y_{B_3}(x \pm \frac{L-2}{2L+4 }i ) = \nt \Xi_{B_1}(x)     \Xi^{(2)}_{k-1}(x)
     \prod_{\ell=0}^{k-2}  \Xi^{(2)}_{\ell} (x\pm \frac{ 4k-4\ell-7}{2L+4}i)      %  \label{y4km1_2}  
	 \\
& & Y_{j}^{(\alpha)}(x \pm \frac{4}{2L+4 }i  =  [\Xi_{B_3}(x) ]^{\delta_{j 0}     
\delta_{\alpha 2}}
    \Xi_{j}^{(\bar \alpha )}(x)
        \prod_{\ell=0}^{k-2} \left(  \CL_{\ell}^{(\alpha)}(x) \right)^{I_{\ell, j}^{[A_{k-1}]}}   \nonumber \\
&  & \phantom{  } \times   \left[
                          \CL^{(4)}_{k-1} (x) \CL^{(6)}_{k-1} (x)
                         \CL^{(5)}_{k-1}(x\pm \frac{1}{2L+4 }i  )\CL^{(3)}_{k-1}(x \pm \frac{2}{2L+4 }i )
                          \CL^{(1)}_{k-1} (x \pm \frac{3}{2L+4 }i )
               \right]^{\delta_{j,k-2} \delta_{\alpha 1}}                        \nonumber  \\
&  & \phantom{  } \times   \left[
    \CL^{(3)}_{k-1} (x)  \CL^{(6)}_{k-1}(x)  \CL^{(5)}_{k-1} (x \pm \frac{1}{2L+4 }i )
	\CL^{(4)}_{k-1}(x \pm \frac{2}{2L+4 }i)  \CL^{(2)}_{k-1} (x \pm \frac{3}{2L+4 }i )
                   \right]^{\delta_{j,k-2} \delta_{\alpha 2}}  \nonumber \\
&&\big(\mbox{with } j=0, \dots,k-2;~ \alpha=1,2\,\mbox{ and } 
\bar{\alpha}=3-\alpha \big )       %  \label{y4km1_3}   
  \\ 
& &Y^{(\gamma)}_{k-1} (x  \pm \frac{1}{2L+4 }i ) =
\left[ \CL_{k-2}^{(\gamma)}(x) \right]^{ \delta_{\gamma 1} +\delta_{\gamma 2} }
\prod_{\beta \sim \gamma} 
\left[    \Xi^{(\beta)}_{k-1}(x) \right]
       ~~  (\gamma=1,\dots 6)
\end{eqnarray*}
 where the last product is taken over  nodes on the $E_6$ Dynkin diagram, which is
 obtained from  fig. \ref{fig_a4kplus1} by deleting  the node 7.
 $I^{[A_{k-1}]}$ stands for the incidence matrix for $A_{k-1} $, where nodes are indexed by  0 to $k-2$.
 %
 %----------------------------------------------------
 %
 \subsection {\bf Case, ($L=4k, k \ge$ 2):}
%
%--------------------------------------
%
The Y-system is
\begin{eqnarray}
Y_{B_1}(x\pm \frac{L-2}{2L+4} i) &=& \Xi_{B_3}(x)    \label{y4k_1}  \\
Y_{B_3}(x\pm \frac{L-2}{2L+4} i)  &=& 
   \Xi_{B_1}(x)   \Xi^{(2)}_{k-1}(x)         
\prod_{\ell=0}^{k-2}  \Xi^{(2)}_{\ell}(x\pm \frac{4k-4\ell-6}{2L+4} i)     \label{y4k_2}   \\
Y_{j}^{(\alpha)}(x \pm \frac{4}{2L+4} i)) &=&  
  [\Xi_{B_3}(x)]^{\delta_{j 0} \delta_{\alpha 2}}
    \Xi_{j}^{(\bar \alpha)}(x) 
    \prod_{\ell=0}^{k-2} \left(  \CL_{\ell }^{(\alpha)}(x) \right)^{I_{\ell j}^{[A_{k-1}]}} 
                                      \nonumber  \\
    &\times&  \left[\CL^{(0)}_{k-1} (x)\CL^{(3)}_{k-1} (x)\CL^{(1)}_{k-1} (x \pm \frac{2}{2L+4} i)
                      \right ]^{\delta_{j,k-2} \delta_{\alpha 1}}       
                                       \nonumber \\
&\times&  \left[\CL^{(0)}_{k-1}(x)\CL^{(4)}_{k-1}(x)\CL^{(2)}_{k-1}(x \pm \frac{2}{2L+4} i)
\right ]^{\delta_{j,k-2} \delta_{\alpha 2}}   \nonumber \\
&&\big(\mbox{with } j=0, \dots,k-2;~ \alpha=1,2\,\mbox{ and } 
\bar{\alpha}=3-\alpha \big )   \label{y4k_3}   \\
Y^{(\gamma)}_{k-1} (x \pm  \frac{2}{2L+4} i  )&=& \left[ \CL_{k-2}^{(\gamma)}(s) \right]^{
\delta_{\gamma 1} +\delta_{\gamma 2}}
\prod_{\beta \sim \gamma} 
\left[ \Xi^{(\beta)}_{k-1}(x) \right]
~~ (\gamma=0,\dots 4)     \label{y4k_4}  
\end{eqnarray}
 where the last product is taken over  nodes on the $A_5$ Dynkin diagram, which is
 obtained from  fig. \ref{fig_a4k} by deleting  the node 7.

 %----------------------------------------------------
 %
 \subsection {\bf Case, ($L=4k+1, k \ge$ 2):}
%
%--------------------------------------
%
The Y-system is
\begin{eqnarray*}
Y_{B_1}(x \pm  \frac{(L-2)}{2L+4}i ) \nt &=&\nt \Xi_{B_3}(x)     %\label{y4kp1_1}    
 \\
Y_{B_3}(x \pm  \frac{(L-2)}{2L+4} i) \nt &=& \nt\Xi_{B_1}(x)         
\Xi^{(2)}_{k-1} (x)
\prod_{\ell=0}^{k-2}  \Xi^{(2)}_i(x\pm \frac{ 4k-4\ell-5}{2L+4}i )          %\label{y4kp1_2}  
\\
Y_{j}^{(\alpha)}(x  \pm  \frac{4}{2L+4} i) \nt&=&\nt  [\Xi_{B_3}(x) ]^{\delta_{j 0} 
\delta_{\alpha 2}}
\Xi_{j}^{(\bar \alpha)}(x) 
\prod_{\ell=0}^{k-2} \left(  \CL_{\ell}^{(\alpha)}(x) \right)^{I_{\ell ,j}^{[A_{k-1}]}}   \nonumber \\
       &\times&  \left[
	           \CL^{(3)}_{k-1} (x) \CL^{(1)}_{k-1}  (x\pm  \frac{1}{2L+4} i)
                   \right ]^{\delta_{j,k-2} \delta_{\alpha 1}}        \nonumber \\
       &\times&  \left[\CL^{(4)}_{k-1}(x)\CL^{(2)}_{k-1}(x \pm \frac{1}{2L+4} i) 
	              \right ]^{\delta_{j,k-2}  \delta_{\alpha 2}}    \nonumber \\
&&\big(\mbox{with } j=0, \dots,k-2 ;~ \alpha=1,2\,\mbox{ and } 
\bar{\alpha}=3-\alpha \big )      % \label{y4kp1_3}
 \\ 
Y^{(1)}_{k-1} (x \pm \frac{2}{2L+4} i)\nt&=& \nt
    \CL_{k-2}^{(1)}(x)   
	   \Xi^{(4)}_{k-1} (x)\Xi^{(6)}_{k-1}(x)\Xi^{(5)}_{k-1}(x \pm \frac{1}{2L+4} i) 
          \Xi^{(3)}_{k-1} (x \pm \frac{2}{2L+4} i)       % \label{y4kp1_4} 
		  \\
Y^{(2)}_{k-1} (x \pm  \frac{3}{2L+4} i)\nt&=&\nt  \CL_{k-2}^{(2)}(0) 
\Xi^{(3)}_{k-1} (x)\Xi^{(6)}_{k-1} (x)\Xi^{(5)}_{k-1} (x \pm  \frac{1}{2L+4} i)    
\Xi^{(4)}_{k-1} (x\pm \frac{2}{2L+4} i)   % \label{y4kp1_5} 
\\
Y^{(3)}_{k-1} (x\pm \frac{1}{2L+4} i)\nt&=&\nt  \Xi^{(1)}_{k-1} (x)\CL^{(5)}_{k-1} (x) 
% \label{y4kp1_6}  
\\
Y^{(4)}_{k-1} (x\pm \frac{1}{2L+4} i)\nt&=&\nt  \Xi^{(2)}_{k-1} (x)\CL^{(5)}_{k-1} (x)  
%\label{y4kp1_7}  
\\
Y^{(5)}_{k-1} (x\pm \frac{1}{2L+4} i)\nt&=&\nt  \CL^{(6)}_{k-1} (x)\CL^{(3)}_{k-1} (x) \CL^{(4)}_{k-1} (x)  % \label{y4kp1_8} 
 \\
Y^{(6)}_{k-1} (x \pm \frac{1}{2L+4} i)\nt&=& \nt \CL^{(5)}_{k-1} (x).
\end{eqnarray*}

 %----------------------------------------------------
 %
 \subsection {\bf Case, ($L=4k+2, k \ge$ 2):}
%
%--------------------------------------
\begin{eqnarray*}
Y_{B_1}(x\pm   \frac{L-2}{2L+4}i ) &=& \Xi_{B_3}(x) %\label{y4kp2_1}   
 \\
Y_{B_3}(x \pm   \frac{L-2}{2L+4}i ) &=& \Xi_{B_1}(x) 
\Xi^{(2)}_{k}(x)
\prod_{\ell=0}^{k-1}  \Xi^{(2)}_{\ell} (x\pm \frac{4k-4\ell-4}{2L+4} i)   % \label{y4kp2_2}  
\\
Y_j^{(\alpha)}(x \pm  \frac{4}{2L+4}i ) &=&
  \left [
       \Xi_{B_3}(x) \right]^{\delta_{j 0} \delta_{\alpha 2}}
        \Xi_j^{(\bar \alpha)}(x) 
		\prod_{\ell \sim j } 
\left[ 
          \CL_{\ell}^{(\alpha)}(x) \right]    \nonumber \\
&&\big(\mbox{with } j=0, \dots,k ;~ \alpha=1,2\,\mbox{ and } 
\bar{\alpha}=3-\alpha \big )
\end{eqnarray*}

where the last product is taken over  nodes on the $D_{k+1} $ Dynkin diagram, which is
 obtained from  fig. \ref{fig_a4kplus2} by deleting  the node 0.

 %#####################################################
 %
\section{explicit solutions to magnon-like $t-$ system  } \label{solutiontsys}
%
%#####################################################

Below,  sets of explicit solutions to magnon-like $t-$ system
are  listed.
 There are, however, many equivalent expressions as remarked in the text.
We write just representatives among them.
When other expressions are convenient, some remarked will be made
then. 
 %
%------------------------------------------------------
\subsection{solutions for $A_{L=4k-1}$ }
%------------------------------------------------------
%
\begin{eqnarray}
t^{(1)}_1(x) &= &  
\begin{cases}
      T_{2k}(x)   , \qquad k \; {\rm even }    \\
      T^{\vee} _{2k}(x)    , \qquad k \; {\rm odd }  
  \end{cases} 
         \label{t11A4kminus1} \\
t^{(2)}_1(x) &= &  
\begin{cases}
      T^{\vee}_{D_{2k}}(x)   , \qquad k \; {\rm even }    \\
      T_{D_{2k}}(x)    , \qquad k \; {\rm odd }  
  \end{cases} 
   \label{t21A4kminus1} \\
t^{(3)}_1 (x) &=& 
\begin{cases}
      \Lambda_{(6k-2, 2k)/(1)}(x-\frac{3}{2L+4}i)   , \qquad k \; {\rm even }    \\
       \Lambda^{\vee}_{(6k-2, 2k)/(1)}(x-\frac{3}{2L+4}i)  , \qquad k \; {\rm odd }  
  \end{cases} 
   \label{t31A4kminus1}\\
t^{(4)}_1 (x) &=& 
\begin{cases}
      \frac{1}{\phi^{\vee}(x+2ki+\frac{7}{2} i -\frac{i}{2L+4})} 
	  \Lambda^{\vee}_{(4k, 4k , 2k)/(4k-1,1) }(x-2ki +i -\frac{i}{2L+4})   , 
	          \qquad k \; {\rm even }    \\
     \frac{1}{\phi(x+2ki+\frac{7}{2} i -\frac{i}{2L+4})}   
	     \Lambda_{(4k, 4k , 2k)/(4k-1,1) }(x-2ki +i -\frac{i}{2L+4}) , 
	      \qquad k \; {\rm odd }  
  \end{cases} 
 \label{t41A4kminus1}\\
t^{(5)}_1 (x) &=& 
\begin{cases}
      \Lambda_{(10k-4,6k-2,6k-3)/(4k-1,4k-2)}(x)  , 
	          \qquad k \; {\rm even }    \\
	  \Lambda^{\vee}_{(10k-4,6k-2,6k-3)/(4k-1,4k-2)}(x)   , 
	      \qquad k \; {\rm odd }  
  \end{cases} 
 \label{t51A4kminus1} \\
t^{(6)}_1(x) &=&
   \frac{1}{\phi(x-(L+2)i-\frac{2i}{2L+4})} 
     \Lambda_{(L,1)}(x-\frac{L+4}{2L+4}i)  
	  \label{t61A4kminus1}\\
t^{(7)}_1(x) &=& T_{B_3}(x) . 
 \label{t71A4kminus1}
 \end{eqnarray}
 
\begin{figure}[hbtp]
\centering
\includegraphics[width=14cm]{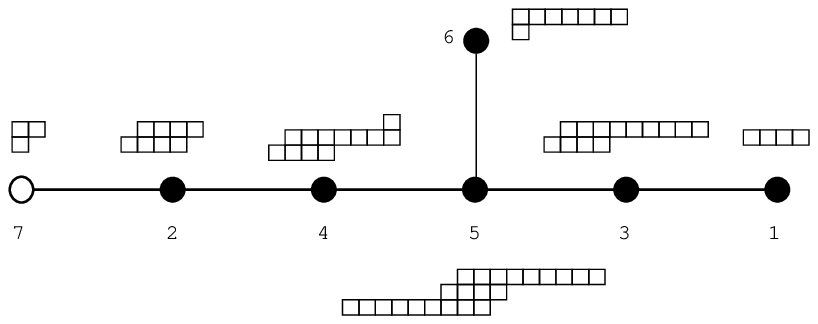}
\caption{ Young diagrams associated to $t^{(a)}_1(x)$ for $k=2$.    }
\label {dA4km1YD}
\end{figure}

 For a guide to eye, corresponding diagrams 
 are illustrated in  fig. \ref{dA4km1YD} for $k=2$.
 
 For $t^{(a)}_2$ , 
 \begin{eqnarray}
t^{(1)}_2(x) &= &  
 \begin{cases}
  T^{\vee}_{2k-2}(x\pm \frac{3}{2L+4}i)    , \qquad k \; {\rm even } \\
 T_{2k-2}(x\pm \frac{3}{2L+4}i)    , \qquad k \; {\rm odd } 
\end{cases} 
   \label{t12A4kminus1}  \\
t^{(2)}_2(x) &= &  
 \begin{cases}
 T_{D_{2k-2}}(x\pm \frac{3}{2L+4}i)    , \qquad k \; {\rm even } \\
   T^{\vee}_{D_{2k-2}}(x\pm \frac{3}{2L+4}i)    , \qquad k \; {\rm odd } 
\end{cases} 
  \label{t22A4kminus1}   \\
t^{(3)}_2(x) &= &  
 \begin{cases}
   T_{D_{2k-2}}(x)  T^{\vee}_{2k-2}(x\pm \frac{2}{2L+4}i)     , \qquad k \; {\rm even } \\
   T^{\vee}_{D_{2k-2}}(x)   T_{2k-2}(x\pm \frac{2}{2L+4}i)    , \qquad k \; {\rm odd } 
\end{cases} 
  \label{t32A4kminus1}   \\
t^{(4)}_2(x) &= &  
 \begin{cases}
   T_{D_{2k-2}}(x\pm \frac{2}{2L+4}i)  T^{\vee}_{2k-2}(x)     , \qquad k \; {\rm even } \\
   T^{\vee}_{D_{2k-2}}(x\pm \frac{2}{2L+4}i)   T_{2k-2}(x)    , \qquad k \; {\rm odd } 
\end{cases} 
  \label{t42A4kminus1}   \\
t^{(5)}_2(x) &= &    t^{(6)}_2(x\pm \frac{1}{2L+4}i)       \label{t52A4kminus1}    \\
t^{(6)}_2(x) &=&
 \begin{cases}
   T_{D_{2k-2}}(x)  T^{\vee}_{2k-2}(x)     , \qquad k \; {\rm even } \\
     T^{\vee}_{D_{2k-2}}(x)  T_{2k-2}(x)    , \qquad k \; {\rm odd } 
\end{cases} 
  \label{t62A4kminus1}   \\
\end{eqnarray}
and
\begin{eqnarray}
t^{(1)}_3(x)   &=& 
\begin{cases}
   T^{\vee}_{2k-2}(x \pm \frac{2}{2L+4}i)  T_{2k-4}(x)     , \qquad k \; {\rm even } \\
    T_{2k-2}(x\pm \frac{2}{2L+4}i)  T^{\vee}_{2k-4}(x)      , \qquad k \; {\rm odd } 
\end{cases} 
\\
t^{(2)}_3(x)   &=& 
\begin{cases}
   T_{D_{2k-2}}(x\pm \frac{2}{2L+4}i)  T^{\vee}_{D_{2k-4}}(x)     , \qquad k \; {\rm even } \\
  T^{\vee}_{D_{2k-2}}(x\pm \frac{2}{2L+4}i)  T_{D_{2k-4}}(x)    , \qquad k \; {\rm odd } .
\end{cases} 
\\
\end{eqnarray}

%------------------------------------------------------
\subsection{solutions for $A_{L=4k}$ } \label{tsol_A4k}
%------------------------------------------------------
\begin{eqnarray*}
t^{(0)}_1 (x)& =& 
 \begin{cases}
    t^{(3)}_1(x-\frac{2 i}{L+2}) T_{2k}(x+ \frac{i}{L+2})  -  t^{(3)}_1(x+\frac{2i}{L+2}) T^{\vee} _{2k-2}(x- \frac{i}{L+2}), k \;{\rm even}  \\
    t^{(3)}_1(x-\frac{2 i}{L+2}) T^{\vee}_{2k}(x+ \frac{i}{L+2})  -  t^{(3)}_1(x+\frac{2i}{L+2})
	T_{2k-2}(x- \frac{i}{L+2}), k \;{\rm odd}  
\end{cases}       % \label{dA4k_tsol_1}
 \\
t^{(1)}_1(x)&=&
\begin{cases}
   T_{2k}(x)    ,\quad    k \;{\rm even}  \\
  T^{\vee}_{2k}(x)    , \quad   k \;{\rm odd}  
 \end{cases}        \label{dA4k_tsol_2} 
 \\
t^{(2)}_1(x)&=&
\begin{cases}
   T^{\vee} _{D_{2k}}(x)    ,\quad  k \;{\rm even}  \\
  T_{D_{2k}}(x)                  , \quad k \;{\rm odd}  
 \end{cases}        \ %label{dA4k_tsol_3}  
 \\
 t^{(3)}_1(x) &=& \frac{T_K(x)}{\sqrt{2}}     %\label{dA4k_tsol_4}   
 \\
 t^{(4)}_1(x)&=& 
 \frac{1}{\phi(x-\frac{2L}{L+2}i)}
     ( t^{(3)}_1(x-\frac{L }{L+2}i) T_1(x+ \frac{i}{2})  - 
	       t^{(3)}_1(x+\frac{L+4}{L+2}i) T_0(x-\frac{1}{2}i)  )   %  \label{dA4k_tsol_5} 
		   \\
t^{(7)}_1(x) &=& T_{B_3}(x)    % \label{dA4k_tsol_6}
\end{eqnarray*}

Among  $t^{(a)}_2(x)$,  non-trivial are the cases $a=3, 4$.
\begin{eqnarray*}
 t^{(3)}_2(x) &=&
 \begin{cases}
     T^{\vee} _{2k-2}(x)    ,\quad  k \;{\rm even}   \\
    T_{2k-2}(x)                  , \quad k \;{\rm odd}  
 \end{cases}     
 %\label{dA4k_tsol_7}   
 \\
 t^{(4)}_2(x)&=& 
 \begin{cases}
     T _{D_{2k-2}}(x)    ,\quad  k \;{\rm even}  \\
    T^{\vee} _{D_{2k-2}}(x)                  , \quad k \;{\rm odd}  
 \end{cases}   
 % \label{dA4k_tsol_8}
%
\end{eqnarray*}
and others are easily obtained from them due to the $t-$ system, 
\begin{eqnarray*}
t^{(0)}_2 (x)& =& t^{(3)}_2 (x) t^{(4)}_2(x)      %\label{dA4k_tsol_9}
\\
t^{(1)}_2(x)&=&    t^{(3)}_2 (x\pm \frac{i}{L+2} )     %\label{dA4k_tsol_10}  
\\
t^{(2)}_2(x)&=&    t^{(4)}_2 (x\pm \frac{i}{L+2} )  .
 \end{eqnarray*}   
Finally, 
\begin{eqnarray*}
 t^{(1)}_3(x) &=&
 \begin{cases}
   \bigl ( T^{\vee} _{2k-2}(x) \bigr )^2  T_{2k-4}(x)    ,\quad  k \;{\rm even}  \\
   \bigl ( T _{2k-2}(x) \bigr)^2  T^{\vee}_{2k-4}(x)                 , \quad k \;{\rm odd}  
 \end{cases}     %\label{dA4k_tsol_11}
 \\
 t^{(2)}_3(x)&=& 
 \begin{cases}
  \bigl (  T _{D_{2k-2}}(x) \bigr )^2    T^{\vee} _{D_{2k-4}}(x)    ,\quad  k \;{\rm even}  \\
   \bigl( T^{\vee} _{D_{2k-2}}(x)  \bigr )^2          T _{D_{2k-4}}(x)         , \quad k \;{\rm odd}  .
 \end{cases}    % \label{dA4k_tsol_12} 
\end{eqnarray*}

%------------------------------------------------------
\subsection{solutions for $A_{L=4k+1}$ }
%------------------------------------------------------
\begin{eqnarray}
t^{(1)}_1(x)   &=& 
\begin{cases}
   T^{\vee}_{2k-2}(x)    , \qquad k \; {\rm even } \\
    T_{2k-2}(x)       , \qquad k \; {\rm odd } 
\end{cases} 
\label{dA4kp1_tsol_11}     \\
t^{(2)}_1(x) &= &  
\begin{cases}
      T_{D_{2k-2}}(x)   , \qquad k \; {\rm even }    \\
      T^{\vee}_{D_{2k-2}}(x)    , \qquad k \; {\rm odd }  
  \end{cases} 
\label{dA4kp1_tsol_21}  \\
t^{(3)}_1 (x) &=& 
\begin{cases}
      \Lambda_{(6k+2, 2k)/(1)}(x-\frac{1}{2L+4}i)   , \qquad k \; {\rm even }    \\
       \Lambda^{\vee}_{(6k+2, 2k)/(1)}(x-\frac{1}{2L+4}i) , \qquad k \; {\rm odd }  
  \end{cases} 
\label{dA4kp1_tsol_31} \\
t^{(4)}_1 (x) &=& 
      \frac{1}{\phi^{\vee}(x+(4k+3) i)} 
	  \Lambda^{\vee}_{(4k, 4k , 2k)/(4k-1,1) }(x+\frac{i}{2})   , 
\label{dA4kp1_tsol_41} \\
t^{(5)}_1 (x) &=& 
\begin{cases}
      \Lambda^{\vee}_{(10k+2,6k+2,6k+1)/(4k+1,4k)}(x)  , 
	          \qquad k \; {\rm even }    \\
	     \Lambda_{(10k+2,6k+2,6k+1)/(4k+1,4k)}(x)    , 
	      \qquad k \; {\rm odd }  
  \end{cases} 
\label{dA4kp1_tsol_51} \\
t^{(6)}_1(x) &=&
   \frac{1}{\phi(x-(L+\frac{3}{2})i+\frac{3 L }{2L+4}i)} 
     \Lambda_{(L,1)}(x+\frac{3 L }{2L+4}i)
\label{dA4kp1_tsol_61}  \\
t^{(7)}_1(x) &=&
  T_{B_3}(x \pm \frac{3i}{2L+4})
\label{dA4kp1_tsol_71}
 \end{eqnarray}
 
\begin{figure}[hbtp]
\centering
\includegraphics[width=14cm]{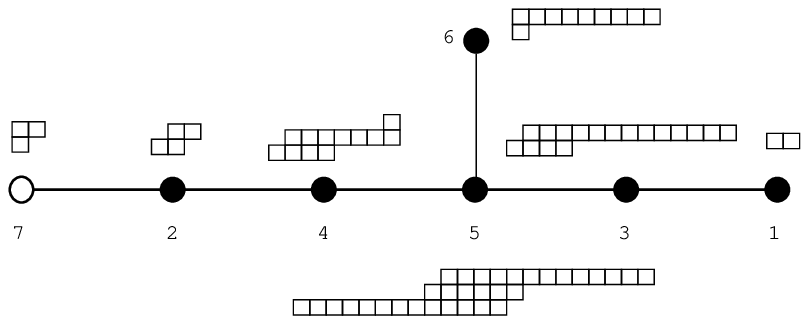}
\caption{ Young diagrams associated to $t^{(a)}_1(x)$ for $k=2$.    }
\label {dA4kp1YD}
\end{figure}

 The corresponding diagrams 
 are illustrated in  fig. \ref{dA4kp1YD} for $k=2$.
 Notice  their similarity to those in  fig. \ref {dA4km1YD}.
 
 For $t^{(a)}_2$ , 
 \begin{eqnarray}
t^{(1)}_2(x) &= &  
 \begin{cases}
  T_{2k}(x) T_{2(k-2)}(x)   , \qquad k \; {\rm even } \\
  T^{\vee}_{2k}(x) T^{\vee}_{2(k-2)}(x)    , \qquad k \; {\rm odd } 
\end{cases} 
 \label{dA4kp1_tsol_12} \\
t^{(2)}_2(x) &= &  
 \begin{cases}
 T^{\vee}_{D_{2k}}(x) T^{\vee}_{D_{2(k-2)}}(x), \qquad k \; {\rm even } \\
   T_{D_{2k}}(x) T_{D_{2(k-2)}}(x)   , \qquad k \; {\rm odd } 
\end{cases} 
\label{dA4kp1_tsol_22} \\
t^{(3)}_2(x) &= &  
 \begin{cases}
   T^{\vee}_{D_{2k}}(x)  T_{2k}(x\pm \frac{2}{2L+4}i)     , \qquad k \; {\rm even } \\
   T_{D_{2k}}(x)  T^{\vee}_{2k}(x\pm \frac{2}{2L+4}i)    , \qquad k \; {\rm odd } 
\end{cases} 
\label{dA4kp1_tsol_32}  \\
t^{(4)}_2(x) &= &  
 \begin{cases}
    T^{\vee}_{D_{2k}}(x\pm \frac{2}{2L+4}i)  T_{2k}(x)  , \qquad k \; {\rm even } \\
     T_{D_{2k}}(x\pm \frac{2}{2L+4}i)  T^{\vee}_{2k}(x)   , \qquad k \; {\rm odd } 
\end{cases} 
\label{dA4kp1_tsol_42} \\
t^{(5)}_2(x) &= &    t^{(6)}_2(x\pm \frac{1}{2L+4}i)     \label{dA4kp1_tsol_52}   \\
t^{(6)}_2(x) &=&
 \begin{cases}
   T^{\vee}_{D_{2k}}(x)  T_{2k}(x)     , \qquad k \; {\rm even } \\
    T_{D_{2k}}(x)  T^{\vee}_{2k}(x)   , \qquad k \; {\rm odd } 
\end{cases}  
\label{dA4kp1_tsol_62}
\end{eqnarray}
 and the rests are found to be
 \begin{eqnarray}
 t^{(3)}_3(x) &=& t^{(5)}_2(x)    \label{dA4kp1_tsol_33}  \\
 t^{(4)}_3(x) &=& t^{(5)}_2(x)   T_{B_3}(x).  \label{dA4kp1_tsol_43} 
 \end{eqnarray}

%------------------------------------------------------
\subsection{solutions for $A_{L=4k+2}$ }
%------------------------------------------------------
In this case,  the eigenvalues of $t$ associated to the tails,
$k$ and $k+1$ , are identical.
\begin{eqnarray}
t^{(k)}_1(x) &= &  t^{(k+1)}_1(x) = \frac{T_K(x)}{2}    \label{dA4kp2_s1}  \\
t^{(a)}_1 &=& 
\begin{cases}
    T_{2a}(x)   , \qquad a \; {\rm even }  \qquad  0\le a \le k-1   \\
      T^{\vee} _{2a}(x)    , \qquad a \; {\rm odd }   % \qquad  0\le a \le k-1 
  \end{cases} 
    \label{dA4kp2_s2}
 \end{eqnarray}
 \begin{eqnarray}
t^{(k)}_2(x) &= &  t^{(k+1)}_2(x) =
 \begin{cases}
  \frac{T^{\vee} _{2k+2}(x)-T^{\vee} _{2k-2}(x)   }{2}    , \qquad k \; {\rm even} \\
       \frac{T_{2k+2}(x)-T_{2k-2}(x)}{2}  , \qquad k \; {\rm odd } 
  \end{cases} 
                   \label{dA4kp2_s3} \\
t^{(a)}_2(x) &= & 
 \begin{cases}
 T^{\vee}_{D_{2a}} (x)  , \qquad a \; {\rm even }  \qquad 0\le a \le k-1 \\
  T_{D_{2a}} (x)    , \qquad a\; {\rm odd } 
  \end{cases} 
                         \label{dA4kp2_s4} \\
\end{eqnarray}
\begin{eqnarray}
t^{(k)}_3(x)&=& t^{(k+1)}_3(x)=T_{B_3}(x) \\
t^{(a)}_3(x) &= &    T_{B_3}(x \pm \frac{k-a}{2k+2}i)  , \quad  1\le a \le k-1   \\
t^{(1)}_4 (x) &=&T_{B_1}(x)   T_{B_3}(x \pm \frac{k-2}{2k+2}i) .
\end{eqnarray}

%%%%%%%%%%%%%%%%%%%%%%%%%%%%%%%%%%%%%%
%
\section{Functional relations for the kink transfer matrix} \label{kinkfr}
%
%%%%%%%%%%%%%%%%%%%%%%%%%%%%%%%%%%%%%%

We verify the following lemma for relatively small values of $k$

\begin{lemma}
\label{kinkfunctionalrel}
The following functional relations among  kink transfer matrices
and $T_m$ are valid for the dilute $A_{4k}$ and $A_{4k+2}$ 
models ($k=1,2,3$ for the former and $k=0 \sim 3$ for the latter) in regime 2. 
(corresponding to $M_{L,L+1}+\phi_{1,2}$)  \pn
%
%-----------------------------------------------------
%
$A_{4k}$ :
\begin{eqnarray*}
T_K(x \pm \frac{2k+1}{4k+2}i)  &=&  2 T_{4k-1}(x)   % \label{kinkt1} 
 \\
T_K(x \pm \frac{2(k-j)+1}{4k+2}i)  &=& 
   \begin{cases} 
          2(T_{2j-2}(x)+ T_{4k-2j}^{\vee} (x) ) , &   j:\quad {\rm odd} \\
%       2(T_{2j-2}(x)+ T_{4k-2j}(x+i\frac{P}{2}) ) , &   j:\quad {\rm odd} \\
%   
        2(T_{2j-2}^{\vee}(x)+ T_{4k-2j}(x) ) , &   j:\quad {\rm even} \\
 %       2(T_{2j-2}(x+i\frac{P}{2} )+ T_{4k-2j}(x) ) , &   j:\quad {\rm even} \\
%
   \end{cases}, \qquad  j=1, \cdots, k       % \label{kinkt2}  
   \\
T_K(x \pm \frac{2(k+j)+1}{4k+2}i)  &=& 
   \begin{cases} 
      2(T_{2j-1}(x)+ T_{4k-2j-1}^{\vee}(x) ) , &   j:\quad {\rm odd} \\
 %     2(T_{2j-1}(x)+ T_{4k-2j-1}(x+i\frac{P}{2}) ) , &   j:\quad {\rm odd} \\
%   
     2(T_{2j-1}^{\vee}(x)+ T_{4k-2j-1}(x) ) , &   j:\quad {\rm even} \\
%	    2(T_{2j-1}(x+i\frac{P}{2} )+ T_{4k-2j-1}(x) ) , &   j:\quad {\rm even} \\
%
   \end{cases}
     \qquad  j=1, \cdots, k     % \label{kinkt3}   
\end{eqnarray*}
%
%--------------------------------------------------------
%
$A_{4k+2}$ :
\begin{align}
T_K(x \pm \frac{k+1}{2k+2}i)  &=  2 T_{4k+1}^{\vee}(x)    \label{kinkt4}  \\
%T_K(x \pm \frac{k+1}{2k+2}i)  &=&  2 T_{4k+1}(x+i\frac{P}{2} )    \label{kinkt4}  \\
%
%
T_K(x \pm \frac{k+1-j}{2k+2}i)  &=
   \begin{cases} 
      2(T_{2j-2}(x)+ T_{4k-2j+2}(x) ) , &   j:\, {\rm odd} \\
      2(T_{2j-2}^{\vee}(x)+ T_{4k-2j+2}^{\vee}(x) ) , &   j:\, {\rm even} \\
%    2(T_{2j-2}(x+i\frac{P}{2} )+ T_{4k-2j+2}(x+i\frac{P}{2}) ) , &   j:\quad {\rm even} \\
%
   \end{cases}, \, j=1, \cdots, k+1      \label{kinkt5}    \\
T_K(x \pm \frac{k+1+j}{2k+2}i)  &= 
   \begin{cases} 
    2(T_{2j-1}(x)+ T_{4k-2j+1}(x) ) , &   j:\, {\rm odd} \\
    2(T_{2j-1}^{\vee}(x)+ T_{4k-2j+1}^{\vee}(x) ) , &   j:\,{\rm even} \\
%	    2(T_{2j-1}(x+i\frac{P}{2} )+ T_{4k-2j+1}(x+i\frac{P}{2}) ) , &   j:\quad {\rm even} \\
%
   \end{cases}, \,  j=1, \cdots, k+1      \label{kinkt6}   
\end{align}
We also note another functional relation for $T_K(x \pm i) $ of the $A_{4k+2}$ case,  
indicating the    $\phi^3$ property
\begin{equation}
T_K(x \pm i) =
   \begin{cases}
     2(\phi^{\vee}(x)  T_K(x) +  2 T^{\vee}_{2k-1}(x) )   , &   k:\quad {\rm odd} \\
    2(\phi^{\vee}(x)   T_K(x)  +  2 T_{2k-1}(x) )   , &   k:\quad {\rm even }. \\
    \end{cases}
\label{kinkt7}
\end{equation}
There are also relations obtained from the above by
using the duality and the periodicity,  omitted for brevity.
\end{lemma}

\begin{proof}
The proof is done by comparing the explicit dressed vacuum forms in the both sides.
In most cases, this procedure , although simple enough in disguise ,  is not straightforward .
For example, for the dilute $A_{10} $ model,  the product of two $T_K$ with identical spectral parameters,
decomposes as 
$$
T_K^2(x) = 3 T_4(x) + T_{15}(x).
$$
By the duality, $T_{15}(x)=T_4(x)$, one concludes (\ref{kinkt5}), with $k=2$ and $j=3$.
Similarly, consider 
$$
T_K^2(x\pm \frac{i}{6}) = 2T_2(x+\frac{11}{6}i) + T_6(x+\frac{11}{6}i)+T_{13}(x+\frac{11}{6}i) .
$$
The rhs is identical to $2(T_2(x+\frac{11}{6}i) + T_6(x+\frac{11}{6}i))$ due to the duality, which
proves  (\ref{kinkt5}), with $k=2$ and $j=2$.
The rests are also shown by using this kind of trick.
\end{proof}

\begin{corollary} \label{tkbytmregime2}
For $A_{4k+2}$ models, the kink transfer matrix is represented as
  \begin{equation}
         T_K(x)=
		 \begin{cases}
		   \frac{1}{\phi^{\vee}(x) }(T^{\vee}_{2k+1}(x)-T_{2k-1}^{\vee}(x))      , &   k:\quad {\rm odd} \\
		   \frac{1}{\phi^{\vee}(x) }(T_{2k+1}(x)-T_{2k-1}(x))      , &   k:\quad {\rm even}. \\
		 \end{cases}
		 \label{kinkphi3}
   \end{equation}
\end{corollary}

This is shown by  comparing (\ref{kinkt6}) with $ j=k+1$ and  (\ref{kinkt7}).

For the dilute $A_L$ models with odd $L$,  we can also construct $T_K(x)$ following the above manner.
The resulting pole-free object  consists of $2L+2$ terms.
It, however,  proves to be identically zero  under some assumption on analyticity.
 At the present, we do not know what is the  implication of this.

%
%------------------------------------------------------
%
%###################################
%
\section {Proofs of Lemma  \ref{tksquare} and \ref{tk1tk3} } \label{prooflemmas}
%
%###################################
\begin{proof}: Lemma  \ref{tksquare}

We use one of the magnon-like $t$ system 
$$
t^{(k)}_1(x \pm i\frac{1}{2k+2})  =t^{(k)}_2(x)+ t^{(k-1)}_1
$$
which has been shown in the main text.
Instead of using the solution for $t^{(k)}_2(x)$  in  (\ref{dA4kp2_s3} ),
we  solve the above for $t^{(k)}_2(x)$ and take its  square, 
\begin{eqnarray*}
\bigl (t^{(k)}_2(x) \bigr )^2  &=&
    \bigl( t^{(k)}_1(x \pm  i\frac{1}{2k+2} ) \bigr )^2   -
        2 t^{(k)}_1(x \pm   i\frac{1}{2k+2}) t^{(k-1)}_1(x) + (t^{(k-1)}_1(x))^2  \\
&=&  \frac{1}{4} \bigl( T_K(x + i\frac{1}{2k+2}  )  \bigr )^2  \bigl( T_K(x -i\frac{1}{2k+2}  ) \bigr )^2   -
  \frac{1}{2} T_K (x \pm   i\frac{1}{2k+2}) t^{(k-1)}_1(x)  \\
  & & + \bigl ( t^{(k-1)}_1(x)  \bigr )^2
\end{eqnarray*}
where the solutions for $t^{(k)}_1$ is substituted.
By further substituting the solutions for $t^{(k-1)}_1$ and making use of the functional relations
for $T_K$ (   the cases $ j=k$ and $j=k+1$   in eq.(\ref{kinkt5} )), 
one arrives
$$
\bigl ( t^{(k)}_2(x)  \bigr )^2  =
    \begin{cases} 
       T_{2k} (x \pm \frac{1}{2k+2} i) - T^{\vee}_{2(k+1)}(x)  T^{\vee}_{2(k-1)}(x)  ,& 
	         k:\, {\rm even}     \\			
      T^{\vee}_{2k} (x \pm \frac{1}{2k+2} i) - T_{2(k+1)}(x)  T_{2(k-1)}(x)  ,& 
	         k:\, {\rm odd}.     \\
     \end{cases}
$$

By noting the mod $P$ relation (\ref{modrel3}) with $j=1$ and   using the $sl_3$ type
relation (\ref{sl3one}),  one concludes
$\bigl ( t^{(k)}_2(x)  \bigr )^2=T^{\vee}_{D_{2k}}(x) $ for $k $ even and 
$\bigl ( t^{(k)}_2(x)  \bigr )^2=T_{D_{2k}}(x) $ for $k $ odd, which proves
  lemma \ref{tksquare}.
\end{proof}

\begin{proof}: Lemma  \ref{tk1tk3}

Thanks to  the $\phi^3$ property (\ref{kinkphi3}), $t^{(k)}_1(x)$ has an expression,
$$
t^{(k)}_1(x) = \frac{T_K(x)}{2}= 
   \begin{cases}
      \frac{1}{2 \phi^{\vee}(x)} ( T_{2k+1}(x)- T_{2k-1}(x)), & k:\, {\rm even} \\
	  \frac{1}{2 \phi^{\vee}(x)} ( T^{\vee}_{2k+1}(x)- T^{\vee}_{2k-1}(x)), & k:\, {\rm odd}. \\
   \end{cases}
$$

We solve (\ref{tb1tb1}) for   $t^{(k)}_3(x)=T_{B_3}(x)$ and substitute the result into the product ,
\begin{equation}
t^{(k)}_1(x)  t^{(k)}_3(x) =
\begin{cases}
       \frac{1}{2 (\phi^{\vee}(x))^2}  
	   ( T_1(x\pm \frac{k}{2k+2}i)T_{2k+1}(x) - T_1(x\pm \frac{k}{2k+2}i)T_{2k-1}(x) &  \\
	    -T_0(x \pm i  \frac{k+2}{2k+2}i)  T_{2k+1}(x) +  T_0(x \pm i  \frac{k+2}{2k+2}i)  T_{2k-1}(x))  & k:\, {\rm even} \\
		 \phantom{ } &  \\
       \frac{1}{2 (\phi^{\vee}(x))^2} 
	   ( T_1(x\pm \frac{k}{2k+2}i)T^{\vee}_{2k+1}(x) - T_1(x\pm \frac{k}{2k+2}i)T^{\vee}_{2k-1}(x) &  \\
	    -T_0(x \pm i  \frac{k+2}{2k+2}i)  T^{\vee}_{2k+1}(x) +  T_0(x \pm i  \frac{k+2}{2k+2}i)  T^{\vee}_{2k-1}(x)) & k:\, {\rm odd}. 
\end{cases}
\label{prodtk1tk3}
\end{equation}
We will show that the above expression coincide with the rhs of (\ref{tk1tk3prod}).
For simplicity,   only the case $k$ even is considered below. 
Consider the second term in the rhs of (\ref{prodtk1tk3}) first.
Noting the expression  $T_1(x\pm \frac{k}{2k+2}i)= T_1(x\pm 2k i)$  due to  (\ref{modrel3}) with $m=k$, we have
\begin{align}
 T_1(x\pm \frac{k}{2k+2}i)T_{2k-1}(x)  &=
T_0(x \pm \frac{k+2}{2k+2} i) T_{2k+1}(x) +   \Lambda_{(2k-1,2k-1,1)/(2k-2)}(x)  \nonumber \\
 &  + 
 \frac{f_{2k}(x-i)}{f_{2k-1}(x)} \Lambda_{(2k,2k)/(2k-1)}(x)+     \frac{f_{2k}(x+i)}{f_{2k-1}(x)} \Lambda_{(2k,1)}(x).
\label{secondterm}
\end{align} 
The graphical representation in fig. \ref{tk1tk3sgl} will be intuitively helpful.
\begin{figure}[hbtp]
\centering
\includegraphics[width=12cm]{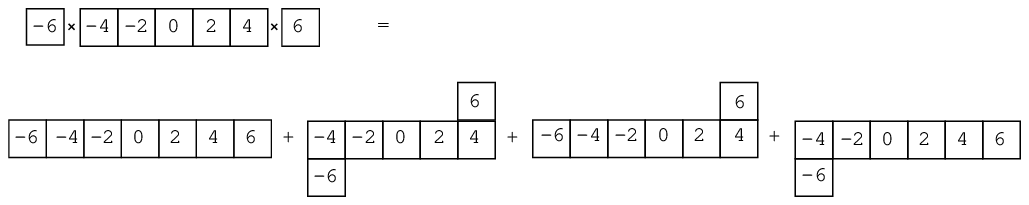}
\caption{ The graphical representation  of eq.(\ref{secondterm}) for $k=3$ .   
A  figure $a$ in a box specifies its spectral parameter $x+ia$. }
\label {tk1tk3sgl}
\end{figure}
The second term in  the rhs of (\ref{secondterm}) is identical to  $(\phi^{\vee}(x))^2  T_{D_{2k-2}}(x)$ due to Lemma \ref{tate3tate2} with $m=2k-1$.
The third and the fourth term in the rhs of  (\ref{secondterm} ) can de further decomposed with the aid of the quantum Jacobi-Trudi formula and mod P relations in Lemma \ref{modrelations}
\begin{eqnarray*}
& & 
T_0(x- (2k-1) i) T_{2k}(x-i) T_1(x+2k i) -T_0(x\pm (2k-1) i) T_{2k+1}(x) 
\\
&   &
=T_0(x+\frac{k+2}{2k+2} i) T_{2k}(x-i) T_1(x-(2k+3) i) -T_0(x\pm \frac{k+2}{2k+2} i) T_{2k+1}(x) 
\end{eqnarray*}
for the third, and 
\begin{eqnarray*}
& &  T_0(x+(2k+1) i) T_{2k}(x+i) T_1(x-2ki) -T_0(x\pm  (2k-1)  i) T_{2k+1}(x)   \\
& &= 
  T_0(x-\frac{k+2}{2k+2}  i) T_{2k}(x+i) T_1(x+(2k+3) i) -T_0(x\pm  \frac{k+2}{2k+2} i) T_{2k+1}(x) \\
\end{eqnarray*}
for the fourth  in  (\ref{secondterm}).
Next we consider the first term in the rhs of (\ref{prodtk1tk3}).
To obtain a non trivial decomposition,  we remark 
the following expression for  $T_{2k+1}(x) $ due to the $a^{(2)}_2$ property,
$$
T_{2k+1}(x) = \frac{1}{\phi(x \pm \frac{4k+5}{2}i)}  \Lambda_{(2k+1, 2k+1)}(x).
$$ 
We also use the expression  $T_1(x\pm \frac{k}{2k+2}i)= T_1(x\pm (2k+3) i)$  due to  (\ref{modrel2}).
By these tricks, three rectangles now set in the right position so that non trivial decomposition
occurs.
\begin{eqnarray} 
T_1(x\pm \frac{k}{2k+2}i) T_{2k+1}(x) &= &
   \phi(x\pm \frac{4k+3}{2}i) T_{D_{2k+2}}(x) + T_0(x \pm \frac{k+2}{2k+2} i) T_{2k-1}(x)    \nonumber \\
& &+ n_c  \Lambda^{\rm non}_{(2k+1, 2k)}(x+2 i)    +  
 n_c^*  \Lambda^{\rm non}_{(2k+1, 2k+1)/(1)}(x-2 i) .
 \label{firstterm}
\end{eqnarray}
The scalar function of the first term  in the rhs can be also written as
$\phi(x\pm \frac{4k+3}{2}i) =(\phi^{\vee}(x))^2$.

The graphical representation in fig . \ref{tk1tk3dbl} will be again helpful.
\begin{figure}[hbtp]
\centering
\includegraphics[width=12cm]{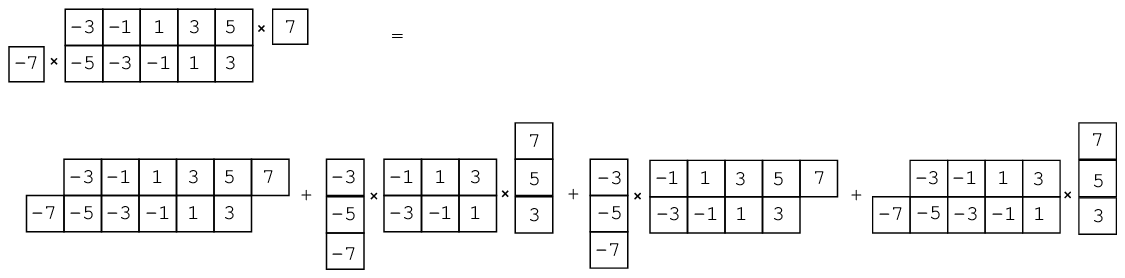}
\caption{ The graphical representation  of eq.(\ref{firstterm} ) for $k=2$.   
A  figure $a$ in a box specifies its spectral parameter $x+ia$. }
\label {tk1tk3dbl}
\end{figure}
The normalization factor $n_c$  in the third and the fourth term in  (\ref{firstterm} ) reads,
$$
n_c=\frac{\phi _3(x-i(2k+1)) }{f_{2k+1}(x)   \prod_{j=1}^{2k+1} \phi_2(x-2(k+1)i+2 ji) } .
$$
The upper index "non"  in the third and the fourth term  indicates the lack of 
their normalization factors.   See (\ref{twotoone}) for $\phi_2$.
The third term  in  (\ref{firstterm}) is further transformed with aid of the relation,
\begin{eqnarray*}
\Lambda^{\rm non}_{(2k+1, 2k)}(x+2 i)   
  &=  &
    \Lambda^{\rm non}_{(2k, 2k)}(x+ i)  T_1(x+(2k+3)i) -
      \Lambda^{\rm non}_{(2k-1, 2k-1,2k-1)/(2k-2)}(x+ 3i)  \\
   &=& 
   f_{2k}(x+i) \prod_{j=1}^{2k}\phi_2(x-2ki+2ji)  T_1(x+(2k+3)i) T_{2k}(x+i)  \\
   &  & -
   f_{2k-1}(x) \phi_3(x+(2k+1)i) \prod_{j=1}^{2k-1}\phi_2(x-2ki+2ji) T_{2k-1}(x)
\end{eqnarray*}
where we have used tableaux rule in the second equality.
After taking account of normalization factors and mod $P$ relations in Lemma \ref{modrelations} ,
the third term in   (\ref{firstterm}) is given by
$$
T_0(x- \frac{k+2}{2k+2} i) T_{2k}(x+i) T_1(x+(2k+3)i) -T_0(x\pm \frac{k+2}{2k+2} i) T_{2k-1}(x).
$$
Similarly the fourth term reads,
$$
T_0(x+\frac{k+2}{2k+2} i ) T_{2k}(x-i) T_1(x-(2k+3)i) -T_0(x\pm \frac{k+2}{2k+2} i) T_{2k-1}(x).
$$

Therefore the first two terms in the rhs of (\ref{prodtk1tk3} ) add up to
\begin{eqnarray*}
& &T_1(x\pm \frac{k}{2k+2}i)T_{2k+1}(x) - T_1(x\pm \frac{k}{2k+2}i)T_{2k-1}(x)  \\
& &=
\bigl(\phi^{\vee}(x) \bigr )^2   \bigl( T_{D_{2k+2}}(x)-T_{D_{2k-2}}(x) \bigr )     
 +  T_0(x\pm \frac{k+2}{2k+2} i ) \bigl (T_{2k+1}(x)-T_{2k-1}(x) \bigr )
\end{eqnarray*}
Obviously the last two terms in the above cancel the third and the fourth term  in   (\ref{prodtk1tk3}).
Thus we arrive 
$$
t^{(k)}_1(x)  t^{(k)}_3(x)   = \frac{1}{2}\bigl ( T_{D_{2k+2}}(x)-  T_{D_{2k-2}}(x) \bigr).
$$
The case for odd $k$ is  done in parallel.

\end{proof}

%##################################################
%
\section{Proof of the magnon-like $t$ system for $L=4k+1$   }\label{appproofL4kp1}
%
%###################################################
We summarize the proofs for  fort he  magnon-like $t$ system, $L=4k+1$.
Most of relations in  (\ref{tsys11}), of which lhs takes the form $t^{(a)}_2( x \pm \frac{1}{2L+4}i)$, are
shown without difficulty, thus they are omitted except for $a=3,4$ .

Note the useful mod $P$ relations
\begin{eqnarray}
-\frac{1}{4k+3} &\equiv& (4k+1) +\frac{P}{2}    \label{mod4kp11} \\
\frac{1}{4k+3} &\equiv& (4k+3)     \label{mod4kp12} \\
\frac{4}{2L+4}&=&\frac{2}{4k+3} \equiv 2 -\frac{P}{2}  \label{mod4kp13}  
\end{eqnarray}
which will be used frequently below.

%
%---------------------------------------------
%
 \subsection{ the proof for  $t^{(1)}_1(x \pm \frac{4}{2L+4}i) =t^{(2)}_1(x)+t^{(1)}_2(x) $ } 
 Thanks to  (\ref{mod4kp13}), the lhs equals to $T_{2(k-1)}(x\pm 2i)$ or  $T^{\vee}_{2(k-1)}(x\pm 2i)$,
 depending on $k$ even or odd.  Then the $sl_3$ relation  (\ref{sl3one})  proves the assertion.
%
%---------------------------------------------
%
 \subsection{ the proof for  $t^{(2)}_1(x \pm \frac{4}{2L+4}i) =t^{(1)}_1(x) t^{(7)}_1(x) +t^{(2)}_2(x) $ } 
 The proof is similar.   One only has to apply   (\ref{sl3two}).
 %
%---------------------------------------------
%
 \subsection{ the proof for  $t^{(3)}_1(x \pm \frac{1}{2L+4}i) =t^{(1)}_1(x) t^{(5)}_1(x) +t^{(3)}_2(x) $ } 
 We consider the case $k$ odd.  
 Using the realness of $t^{(3)}_1(x)$,  the mod $P$ relation (\ref{mod4kp11}) and the diagrammatic symmetry, we rewrite the lhs as
 $$
 \Lambda_{(6k+2, 2k)/(1)}(x+(4k+1) i)  \Lambda_{(6k+2, 6k+1)/(4k+2)}(x-(4k+1) i) ,
 $$
 while , thank to  the duality, the second term in the rhs is given by 
 $$
 t^{(3)}_2(x) = T_{D_{2k}}(x) T_{6k+1}(x\pm (4k+3)i).
 $$
 Then the quantum Jacobi Trudi formula leads to
 \begin{align*}
 &t^{(3)}_1(x \pm \frac{1}{2L+4}i) - t^{(3)}_2(x) =\\
 & \quad  T_{2(k-1)}(x) 
  \Bigl(   T_{2k+2}(x)T_{6k+1}(x\pm(4k+3)i)   
   + T_{2k-2}(x)T_{6k+3}(x\pm(4k+1)i)     \\
  & \qquad \qquad  -    T_{2k}(x-2i) T_{6k+1}(x+(4k+3)i)  T_{6k+3}(x-(4k+1)i)  \\
  & \qquad \qquad- T_{2k}(x+2i) T_{6k+1}(x-(4k+3)i)  T_{6k+3}(x+(4k+1)i) \Bigr )\\
&= T_{2(k-1)}(x) \Lambda_{(10k+2,6k+2,6k+1)/(4k+1,4k)}(x).
 \end{align*}
One can check the rhs of the above equation equals to  $t^{(1)}_1(x) t^{(5)}_1(x)$.

%
%---------------------------------------------
%
 \subsection{ the proof for  $t^{(3)}_2(x \pm \frac{1}{2L+4}i) =
 t^{(1)}_1(x\pm \frac{i}{2L+4}) t^{(5)}_2(x) +t^{(3)}_1(x) t^{(3)}_3(x) $ } 
 %
 %------------------------------------------
 
 Again we treat the case $k$ odd.  
 Since  $t^{(5)}_2(x)=t^{(3)}_3(x) $,  the equality to prove is,
 \begin{equation}
 t^{(3)}_2(x \pm \frac{1}{2L+4}i) = t^{(3)}_3(x) \Bigl(t^{(3)}_1(x) + t^{(1)}_1(x\pm \frac{i}{2L+4})  \Bigr ) .
 \label{redu}
 \end{equation}
Substitute  (\ref{dA4kp1_tsol_32}) and use (\ref{dA4kp1_tsol_33}) for    $t^{(3)}_3(x)$,
the lhs is given by

\begin{eqnarray*} 
t^{(3)}_2(x \pm \frac{1}{2L+4}i) &=&
t^{(3)}_3(x)
\Bigl (  T_{2k}(x+(4k+1)i) T_{2k}(x+2i)  \Bigr )|_{x \rightarrow x-\frac{i}{2L+4}}    \\
&=&t^{(3)}_3(x)
\Bigl (  T_{2k}(x+(4k+1)i) T_{6k+1}(x+(8k+6)i)  \Bigr )|_{x \rightarrow x-\frac{i}{2L+4}}  
\end{eqnarray*}
where the duality is used  in the second equality.
By the (\ref{qJT}), it is readily shown that the rhs of the bracket above coincide with
$$
t^{(3)}_3(x)
\Bigl( \Lambda^{\vee}_{(6k+2, 2k)/(1)}(x-\frac{i}{2L+4}) +T_{2k-2}(x\pm  \frac{i}{2L+4})   \Bigr ).
$$
It is immediate to check that the above equation coincides with the rhs of (\ref{redu})
%---------------------------------------------
%
\subsection{  the proof for    $ t^{(6)}_1(x\pm \frac{1}{2L+4}i) = t^{(5)}_1(x)+  t^{(6)}_2(x)$ } \label{subt6t6}
%
%------------------------------------------
 
\begin{figure}[hbtp]
\centering
\includegraphics[width=8cm]{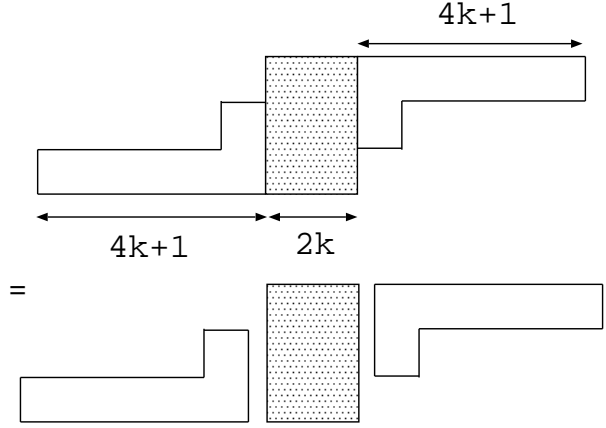}
\caption{ 
The decomposition of ${\cal T}_{(10k+2,6k+2,6k+1)/(4k+1,4k)}(x)$ into 
3 parts.  The hatched part   reduces to a product of  $\phi_3(x)$ }
\label {dcmp4kp1t61t61}
\end{figure}

 The first term in the rhs, $t^{(5)}_1(x)$ is identified with 
 $\Lambda_{(10k+2,6k+2,6k+1)/(4k+1,4k)}(x)$ for $k$ odd
 or $\Lambda^{\vee}_{(10k+2,6k+2,6k+1)/(4k+1,4k)}(x)$ for $k$ even.
We instead  consider ${\cal T}_{(10k+2,6k+2,6k+1)/(4k+1,4k)}(x)$.
See (\ref{LamcalT}) for their difference.
Due to the tableaux rule, it clearly decomposes into three parts.  
The "height-three" part reduces
to the shifted product of $\phi_3(x)$ in (\ref{threetoscalar}).
The remaining two parts are identified with the diagram for $t^{(6)}_1$ and
its 180$^\circ$   rotation, referring to (\ref{dA4kp1_tsol_61}) .  See fig. \ref{dcmp4kp1t61t61}.
By the reality property of  $t^{(6)}_1(x)$ and the diagrammatic symmetry
 (\ref{diagramsym}), one finds 
 $$
 {\cal T}_{(10k+2,6k+2,6k+1)/(4k+1,4k)}(x) =
 \begin{cases}
       t^{(6)}_1(x\pm \frac{1}{2L+4}i)&         k:\,{\rm odd}  \\
	   ( t^{(6)}_1)^{\vee} (x\pm \frac{1}{2L+4}i)      & k:\,{\rm even} 
\end{cases}
 $$
 where the mod relation has been applied.
 On the other hand,  the quantum Jacobi Trudi formula concludes
 
 \begin{eqnarray*}
 & & {\cal T}_{(10k+2,6k+2,6k+1)/(4k+1,4k)}(x)-
     \Lambda_{(10k+2,6k+2,6k+1)/(4k+1,4k)}(x)   \\
	 &   &\phantom{ } \qquad =T_{10k+4}(x) T_{D_{2k}}(x)
 =T^{\vee}_{2k}(x) T_{D_{2k}}(x),
 \end{eqnarray*}
 which coincides with $t^{(6)}_2(x)  \bigl(( t^{(6)}_2)^{\vee}(x)  \bigr  )$
 for $k$ odd (even) in  (\ref{dA4kp1_tsol_62}).
Thus by comparing above due expressions,  one concludes 

\begin{equation}
t^{(6)}_1(x\pm \frac{1}{2L+4}i) = t^{(5)}_1(x)+  t^{(6)}_2(x)
\label{t61t614kp1}
\end{equation}
irrespective of $k$ even or odd.

%
%---------------------------------------------
%
\subsection{  the proof for    $ t^{(4)}_1(x\pm \frac{1}{2L+4}i) = t^{(2)}_1(x)t^{(5)}_1(x)+  t^{(4)}_2(x)$ } 
We treat the case $k$ even.
First note the following representation of   $ t^{(4)}_1(x+\frac{1}{2L+4}i) $,

\begin{eqnarray*}
& & t^{(4)}_1(x+\frac{1}{2L+4}i) \phi(x+(6k+\frac{9}{2})i) =  \\
&  &\Lambda_{(4k-1,4k-1)/(4k-2)}(x+2i) T_{2k}(x-(2k+2)i) -
\Lambda_{(4k+1,4k+1)/(4k)}(x) T_{2k-2}(x-2 ki).
\end{eqnarray*}

This relation is natural by its diagrammatic representation, and can be 
verified easily by (\ref{qJT}).
Again  we use the reality property of  $t^{(4)}_1(x)$ and  the diagrammatic symmetry
 (\ref{diagramsym}) to evaluate  $t^{(4)}_1(x\pm \frac{1}{2L+4}i) \phi(x\pm (6k+\frac{9}{2})i) $.

After recombining the products of terms,   it is found ,
  \begin{eqnarray}
&  & t^{(4)}_1(x \pm \frac{1}{2L+4}i) \phi(x\pm (6k+\frac{9}{2})i)  =    \nonumber \\
& &\quad  \{ 
   T_{2k-2}(x\pm 2i) \Lambda_{(6k+1,6k+1,4k+1,1)/(6k,2k)}(x)+ 
  ( T_{2k}(x))^2  \Lambda_{(6k+1,6k+1,4k-1,1)/(6k,2k+2)}(x)      \nonumber     \\
 &   & \qquad - T_{2k}(x) T_{2k-2}(x- 2i) \Lambda_{(6k+1,6k+1,4k-1,1)/(6k,2k)}(x)  \nonumber   \\
  &   & \qquad   - T_{2k}(x) T_{2k-2}(x+ 2i) \Lambda_{(6k+1,6k+1,4k+1,1)/(6k,2k+2)}(x)  
   \}   \nonumber  \\
& & \quad + \{  
     T_{2k-4}(x)T_{2k}(x)-T_{2k-2}(x\pm 2i)
  \}  T_{2k}(x)  \Lambda_{(6k+2,6k+2,1)/(6k+1)}(x).
  \label{t41t41L4kp1} 
\end{eqnarray}

Thanks to (\ref{qJT}), the content of the first curly bracket is equal to

\begin{eqnarray}
&  &T_{2k}(x) {\cal T}_{(6k,6k,4k,4k-1,1)/(6k-1,2k+1,2k)}(x)  \nonumber  \\
&   &\qquad + T_{D_{2k-2}}(x) \Lambda_{(L,1)}(x-(2k+2) i) \Lambda_{(L,L)/(L-1)}(x+(2k+2)i).
\label{L4kp1bracket1}
\end{eqnarray}

The first term in the above is decomposed into three parts due to the tableaux rule as in the
case of   $ t^{(6)}_1(x\pm \frac{1}{2L+4}i) $.
An analogous argument leads to its expression,
\begin{eqnarray*}
& &\frac{1}{\phi(x \pm (2k-\frac{5}{2})i)} T_{2k}(x) 
   \Lambda_{(2k+1,2k+1,1)/(2k)  }(x\pm i(4k+1))    \\
&  &= \phi(x\pm(6k+\frac{9}{2})i) T_{2k}(x)  T_{D_{2k}} (x \pm (4k+1)i)  
=  \phi(x\pm(6k+\frac{9}{2})i) t^{(4)}_2(x)
\end{eqnarray*}
where Lemma \ref{tate3tate2} and the mod $P$ relation are used in the second equality.

The second term in  (\ref{L4kp1bracket1}) is identified with
$$
\phi(x\pm i(6k+\frac{9}{2})i) T_{D_{2k-2}}(x)  t^{(6)}_1(x \pm \frac{1}{2L+4}i)
$$
by the reality property of $t^{(6)}_1(x)$ and (\ref{diagramsym}) .
Substituting the established relation,  $t^{(6)}_1(x \pm \frac{1}{2L+4}i)=t^{(6)}_2(x)+ t^{(5)}_1(x)$,
we obtain the following expression for the first curly bracket term in   (\ref{t41t41L4kp1} ) ,
\begin{eqnarray*}
&  &\phi(x\pm i(6k+\frac{9}{2})i)
   \Bigl(t^{(4)}_2(x) + T_{D_{2k-2}}(x) \bigl( t^{(5)}_1(x) + t^{(6)}_2(x)  \bigr) \Bigr) \\
&    & =    \phi(x\pm i(6k+\frac{9}{2})i)  (t^{(4)}_2(x)+ t^{(2)}_1(x)  t^{(5)}_1(x)    + t^{(2)}_1(x) t^{(6)}_2(x)  ).
\end{eqnarray*}

The second curly bracket term in   (\ref{t41t41L4kp1} ) is easily found to be

 \begin{eqnarray*}
 & &- T_{D_{2k-2}}(x) T_{2k}(x) \Lambda_{(6k+2,6k+2,1)/(6k+1)}(x)
 =-\phi(x\pm i(6k+\frac{9}{2})i) T_{D_{2k-2}}(x) T_{2k}(x) T^{\vee}_{D_{2k}}(x) \\
 & &\qquad =-\phi(x\pm i(6k+\frac{9}{2})i) t^{(2)}_1(x)  t^{(6)}_2(x).
 \end{eqnarray*}
 
Lemma \ref{tate3tate2} is applied  first,  then the definitions in (\ref{dA4kp1_tsol_21}) and 
 (\ref{dA4kp1_tsol_62})   is utilized.
 Combining these results, we obtain
  $ t^{(4)}_1(x\pm \frac{1}{2L+4}i) =t^{(2)}_1(x)  t^{(5)}_1(x)+  t^{(4)}_2(x)$.
%
%---------------------------------------------
%
\subsection{  the proof for    $ t^{(4)}_2(x\pm \frac{1}{2L+4}i) = t^{(2)}_1(x\pm \frac{1}{2L+4}i)t^{(5)}_2(x)+  t^{(4)}_1(x) t^{(4)}_3(x)$ } 
Here we set $k=$ odd.
Since  $t^{(4)}_3(x)=T_{B_3}(x)t^{(5)}_2(x)$, the assertion is equivalent to
$$
t^{(4)}_2(x\pm \frac{1}{2L+4}i) =t^{(5)}_2(x) 
   \bigl(  t^{(2)}_1(x\pm \frac{1}{2L+4}i)+  t^{(4)}_1(x) T_{B_3}(x)  \bigr ).
$$
To prove this, we prepare a few facts.

\begin{lemma} 
\label{tate2tate2}
\begin{eqnarray*}
\frac{1}{\phi(x-(m+\frac{3}{2})i) }  \Lambda_{(m,m-1)}(x)  &=& \frac{1}{\phi(x+(m+\frac{5}{2})i) }  
 \Lambda_{(m,m)/(m-1)}(x+i )   %\label{tate2tate21} 
\\
\frac{1}{\phi(x-(m+\frac{3}{2})i) }  \Lambda_{(m,1)}(x)  &=& \frac{1}{\phi(x+(m+\frac{5}{2})i) }  
 \Lambda_{(m,m)/(1)}(x+i )  % \label{tate2tate22} 
\end{eqnarray*}
\end{lemma}
Both of them are easily proved by the $a^{(2)}_2$ property.

\begin{lemma}
\label{L1t42t42L4kp1}
\begin{align}
&T_{D_{6k+1}}(x) T_{D_{2k}}(x+(4k+5)i) -T_{D_{6k+3}}(x+2i) T_{D_{2k-2}}(x+(4k+3)i)    \nonumber \\
& = \frac{\Lambda_{(2,1)}(x+(6k+4)i)}{\phi(x+(2k+\frac{11}{2})i)  \phi(x+(6k+\frac{1}{2})i) }  \times D 
\label{L1eqt42t42L4kp1}  \\
& {\rm where }     \nonumber   \\
& \quad D:=\quad  \Bigl( T_{6k+1}(x+2i) \Lambda_{(4k+3,4k+3)/(4k+2)}(x-(2k-1)i)     \nonumber   \\
  &\qquad  \qquad - T_{6k+3}(x+2i) \Lambda_{(4k+1,4k+1)/(4k)}(x-(2k-3)i)    \Bigr).
\nonumber
% \\
%&  & := \frac{\Lambda_{(2,1)}(x+(6k+4)i)}{\phi(x+(2k+\frac{11}{2}i)  \phi(x+(6k+\frac{1}{2}i) }  D
\end{align}
\end{lemma}

It is easily checked that the lhs is equal to 
$$
T_{6k+1}(x+2i) {\cal T}_{( 6k+3, 6k+2,6k+1)/( 4k+3, 4k+2)}(x+2i) -
T_{6k+3}(x) {\cal T}_{( 6k+1, 6k,6k-1)/( 4k+1, 4k)}(x+4i).
$$
The  diagram decomposition of this  then leads to  rhs of (\ref{L1eqt42t42L4kp1}).

\begin{lemma}
\label{L2t42t42L4kp1}
\begin{equation*}
D =  \frac{\phi(x+(2k+\frac{11}{2})i)}{\phi(x-(6k+\frac{1}{2})i)}    \Lambda_{(10k, 6k+2, 6k+1)/(2k,2k)}(x)    
%
%\label{L2eqt42t42L4kp1}
\end{equation*}
\end{lemma}

We use the two decompositions of    ${\cal T}_{(10k, 6k+2, 6k+1)/(2k,2k)}(x) $.
The comparison of these two leads to

\begin{eqnarray*}
 \Lambda_{(10k, 6k+2, 6k+1)/(2k,2k)}(x)  &=&
 \frac{\phi(x-(6k+\frac{1}{2})i)} {\phi(x+(2k+\frac{3}{2})i)}  T_{2k}(x-(8k+2)i) \Lambda_{(4k-1,1)}(x+(6k+2)i)  \\
&  & -T_{6k+3}  \Lambda_{(4k+1,4k)}(x+(-2k+2)i) .
\end{eqnarray*}

By the duality, the quantum Jacobi-Trudi formula and   Lemma \ref{tate2tate2}, we find
\begin{eqnarray*}
\Lambda_{(4k-1,1)}(x+(6k+2)i)  & =&   
\frac{\phi(x+(2k+\frac{3}{2})i)} {\phi(x+(2k+\frac{11}{2})i)} \Lambda_{(4k+3,4k+3)/(4k+2)}(x-(2k-1)i) \\
\Lambda_{(4k+1,4k)}(x+(-2k+2)i) &=&  
\frac{\phi(x-(6k+\frac{1}{2})i)} {\phi(x+(2k+\frac{11}{2})i)} \Lambda_{(4k+1,4k+1)/(4k+2)}(x-(2k-3)i).
\end{eqnarray*}
Finally by noticing $ T_{2k}(x-(8k+2)i) =T_{6k+1}(x+2i)$, we prove Lemma \ref{L2t42t42L4kp1}

The final lemma asserts a relation ,
\begin{lemma} \label{flL4kp1}
\begin{equation*}
 \Lambda_{(10k, 6k+2, 6k+1)/(2k,2k)}(x-(6k+5)i) = \Lambda_{(4k,4k,2k)/(4k-1,1)}(x) =
 \phi(x+(4k+\frac{5}{2})i) (t^{(4)}_1)^{\vee}(x-\frac{i}{2}).
%\label{flL4kp1}
\end{equation*}
\end{lemma}
The first equality is due to duality and the second comes from the definition.

Now we prove the relation.   
We conveniently consider, 
$$
\bigl(   t^{(4)}_2(x\pm \frac{i}{2L+4}) -t^{(2)}_1 (x \pm  \frac{i}{2L+4}) t^{(5)}_2(x) \bigr )|_{x \rightarrow x+\frac{3i}{2L+4}}
$$
and show that it is equal to $t^{(4)}_1 (x+\frac{3i}{2L+4})  t^{(4)}_3 (x+\frac{3i}{2L+4}) $.
First we rewrite the difference as 
\begin{equation}
t^{(5)}_2(x+\frac{3i}{2L+4})  \Bigl(
T_{D_{6k+1}}(x) T_{D_{2k}}(x+(4k+5)i) -T_{D_{6k+3}}(x+2i) T_{D_{2k-2}}(x+(4k+3)i)  \Bigr ).
\label{dif1L4kp1}
\end{equation}
then apply Lemma \ref{L1t42t42L4kp1}. 
 With the expression of $T_{B_3}(x)$ in (\ref{anotherdeftb3}),
 and the relation 
$t^{(4)}_3(x)=t^{(5)}_2(x) T_{B_3}(x)$, 
one concludes  that (\ref{dif1L4kp1}) coincides with 
$$
\frac{t^{(4)}_3(x+\frac{3i}{2L+4})}{\phi(x+(2k+\frac{11}{2})i)}  D.
$$
Therefore,  the relation is equivalent to 
\begin{equation}
\frac{D}{\phi(x+(2k+\frac{11}{2})i)} = t^{(4)}_1(x+\frac{3i}{2L+4}).
\label{toshow2}
\end{equation}

We will show this by rewriting  the lhs of  (\ref{toshow2}) , thanks to Lemma  \ref{L2t42t42L4kp1}, 
 $$
 \frac{1}{\phi(x-(6k+\frac{1}{2})i)}    \Lambda_{(10k, 6k+2, 6k+1)/(2k,2k)}(x).
$$ 
This can be further rewrite as, with aid of  Lemma \ref{flL4kp1},
$$
\frac{\phi(x+(10k+\frac{15}{2})i)}{\phi(x-(6k+\frac{1}{2})i)}  ( t^{(4)}_1)^{\vee} (x+(6k+5)i-i\frac{1}{2})
$$
which is readily shown to be equal to the rhs of   (\ref{toshow2}) for $k$ odd.
The case $k$ even can be treated similarly.
 %
%
%---------------------------------------------
%
\subsection{  the proof for    $ t^{(5)}_1(x\pm \frac{1}{2L+4}i) = t^{(3)}_1  t^{(4)}_1    t^{(6)}_1     +  t^{(5)}_2(x)$ } 
We assume  $k$ even. The  proof for the case $k$ odd  is similar.
One substitutes $t^{(5)}_1(x)=t^{(6)}_1(x \pm \frac{1}{2L+4}i)-t^{(6)}_2(x)$ and uses
 the relation $t^{(5)}_2(x)=t^{(6)}_2(x \pm \frac{1}{2L+4})$.

Then the  assertion is transformed into an equivalent form,
\begin{eqnarray*}
& &t^{(3)}_1(x)  t^{(4)}_1(x) = t^{(6)}_1(x) t^{(6)}_1(x \pm \frac{2}{2L+4}i)   \\
&  &\quad 
 -t^{(6)}_2(x+\frac{1}{2L+4}i )t^{(6)}_1(x-\frac{2}{2L+4}i )  -t^{(6)}_2(x-\frac{1}{2L+4}i )t^{(6)}_1(x+\frac{2}{2L+4}i ).   
\end{eqnarray*}

We conveniently make  a shift in $x$,
\begin{eqnarray}
& &t^{(3)}_1(x+\frac{1}{2L+4}i )  t^{(4)}_1(x+\frac{1}{2L+4}i ) = t^{(6)}_1(x+\frac{3}{2L+4}i ) t^{(6)}_1(x \pm \frac{1}{2L+4}i  )  \nonumber  \\
&  &\quad  -t^{(6)}_2(x )t^{(6)}_1(x+\frac{3}{2L+4}i )  
                 -t^{(6)}_2(x+\frac{2}{2L+4}i )t^{(6)}_1(x-\frac{1}{2L+4}i ).   
\label{shifted_to_prove_L4kp1}
\end{eqnarray}

We begin with the following lemma,

\begin{lemma}\label{lemt31t41L4kp1}
\begin{eqnarray}
& &t^{(3)}_1(x+\frac{1}{2L+4}i )  t^{(4)}_1(x+\frac{1}{2L+4}i ) =  
   t^{(6)}_1(x+\frac{3}{2L+4}i ) t^{(6)}_1(x \pm \frac{1}{2L+4}i  )    \nonumber  \\
& &\quad -t^{(6)}_2(x+\frac{2}{2L+4}i )t^{(6)}_1(x-\frac{1}{2L+4}i )-R
\label{t31t41L4kp1}
\end{eqnarray}

where

\begin{eqnarray}
R &= & \frac{T_{2k} (x) }{\phi(x+(6k+\frac{9}{2})i)}
  \Bigl(   \Lambda_{(6k+2,2k+2)/(1)}(x) t^{(6)}_1( x-\frac{1}{2L+4}i)   {\phi(x+(6k+\frac{9}{2})i)}  
                    \nonumber   \\
&   &    -
      \Lambda_{(4k-1,4k-1)/(4k-2)}(x+(2k+4)i) t^{(3)}_1( x+\frac{1}{2L+4}i)       \Bigl ).
\label{RdefL4kp1}
\end{eqnarray}
\end{lemma}

\begin{proof} Lemma \ref{lemt31t41L4kp1} \\
We first note a decomposition of  $t^{(4)}_1(x)$,
\begin{eqnarray*}
t^{(4)}_1(x+\frac{1}{2L+4}i ) &=& \frac{1}{\phi(x+(6k+\frac{9}{2}) i)}   T_{2k}(x) \Lambda_{(4k-1, 4k-1)/(4k-2) } (x+(2k+4)i)   \\ 
& &-   T_{2k-2}(x+2i)  t^{(6)}_1( x-\frac{1}{2L+4}i)    
\end{eqnarray*}

where we have used  the diagrammatic symmetry,

\begin{eqnarray*}
 t^{(6)}_1(x-\frac{1}{2L+4}i )  & = &  ( t^{(6)}_1(x+\frac{1}{2L+4}i)    )^*   \\
&=& \frac{1}{\phi(x+(6k+\frac{9}{2}) i)}  \Lambda_{(4k+1, 4k+1)/(4k) }(x+(2k+2)i)  .
\end{eqnarray*}

Then the product  in the lhs of (\ref{t31t41L4kp1}) is rewritten as,

\begin{eqnarray}
&  & t^{(3)}_1(x+\frac{1}{2L+4}i )  t^{(4)}_1(x+\frac{1}{2L+4}i )   \nonumber \\
&  &\quad =\frac{1}{\phi(x+(6k+\frac{9}{2}) i)}  
   T_{2k}(x) \Lambda_{(4k-1, 4k-1)/(4k-2) } (x+(2k+4)i) t^{(3)}_1(x+\frac{1}{2L+4}i )     \nonumber \\
&  & \quad  -  T_{6k+3}(x-(8k+2)i)    t^{(3)}_1(x+\frac{1}{2L+4}i ) t^{(6)}_1( x-\frac{1}{2L+4}i)
\label{t31t41_r}
\end{eqnarray}
where the duality relation  $ T_{2k-2}(x+2i)=T_{6k+3}(x-(8k+2)i) $ is applied.
Note that    $t^{(3)}_1(x+\frac{1}{2L+4}i )\equiv \Lambda_{(6k+2,2k)/(1)}(x)$.
Then we apply the decomposition rule to the second term in  (\ref{t31t41_r}),

\begin{eqnarray*}
&  &- T_{6k+3}(x-(8k+2)i)  t^{(3)}_1(x+\frac{1}{2L+4}i )    \\
&  &= t^{(5)}_1(x+\frac{P}{2}i-(4k+1)i) -T_{6k+1}(x-(8k+4)i) \Lambda_{(6k+2,2k+2)/(1)}(x)  \\
&  &= t^{(5)}_1(x+\frac{2i}{2L+4}i)-T_{2k}(x) \Lambda_{(6k+2,2k+2)/(1)}(x)  \\
& &= t^{(6)}_1(x+ \frac{3i}{2L+4}) t^{(6)}_1(x+\frac{i}{2L+4}) - t^{(6)}_2 (x+\frac{2i}{2L+4})   \\
& & \quad         -T_{2k}(x) \Lambda_{(6k+2,2k+2)/(1)}(x).
\end{eqnarray*}
The first equality follows from the quantum Jacobi-Trudi formula. Then the duality and
the established relation in \ref{subt6t6} lead to the last line.
By substituting this into  (\ref{t31t41_r}), one verifies  Lemma \ref{lemt31t41L4kp1}.
\end{proof}

From Lemma \ref{lemt31t41L4kp1}  and (\ref{shifted_to_prove_L4kp1}),  
 one only has to show 
  
  \begin{equation}
  R=t^{(6)}_1(x+i\frac{3}{2L+4})  t^{(6)}_2(x) .
  \label{toshow_R_L4kp1}
  \end{equation}

Our argument consists of two further steps.
  First we will show a lemma
  
  \begin{lemma} \label{altRL4kp1}
  \begin{eqnarray}
  R&=& \frac{T_{2k}(x)}{\phi(x-(2k+\frac{3}{2})i) \phi(x-(6k+\frac{1}{2})i)  \phi(x+(6k+\frac{9}{2})i) }  \nonumber \\
  & &   \times  \Lambda_{(L,1)} (x+(2k+1)i)   \Bigl (   
   T_1(x-(10k+5))i) \Lambda_{(2k,2k)/(2k-1)}(x-(8k+2)i)  \nonumber \\
&  &	 \quad -    T_0(x-(10k+6)i) \Lambda_{(2k-1,2k-1)/(2k-2)}(x-(8k+1)i)   \Bigr )  .
	\label{RalterLakp1}
   \end{eqnarray}
  \end{lemma}
Then the rhs of the above will be shown to be equal to  that of  (\ref{toshow_R_L4kp1}).

\begin{proof} Lemma \ref{altRL4kp1}\\

The definition of $R$ in (\ref{RdefL4kp1}) is not convenient in applying 
decomposition rules.

We utilize the duality and  rewrite
\begin{eqnarray*}
 t^{(6)}_1( x-\frac{1}{2L+4}i)   {\phi(x+(6k+\frac{9}{2})i)} &=&\Lambda_{(8k,4k)}(x-(2k+2)i)  \\
  \Lambda_{(4k-1,4k-1)/(4k-2)}(x+(2k+4)i)  &=&  \Lambda_{(8k,4k+2)}(x-(2k+2)i)  .
\end{eqnarray*}

Then they are located at right positions to apply the quantum Jacobi Trudi formula.
The resultant expression  reads
\begin{eqnarray*}
R&=& \frac{T_{2k}(x)   }
                  {\phi(x+(6k+\frac{9}{2})i)}      
     \\
&  &   \times    \Bigl (  T_{8k}(x-(2k+1)i) {\cal T}_{(8k+1,4k+1,4k)/(2k,2k-1)}(x-2 k i)     \\
&  & \qquad -  T_{8k+1}(x-(2k+2)i) {\cal T}_{(8k,4k,4k-1)/(2k-1,2k-2)}(x+(-2k+1)i)    \Bigr )  \\
&=&  \frac{T_{2k}(x)   }
                  {\phi(x+(6k+\frac{9}{2})i)}      \\
&  &     \times    \Bigl (   T_{1}(x-(10k+5)i) {\cal T}_{(8k+1,4k+1,4k)/(2k,2k-1)}(x-2 k i)     \\
&  & \qquad - T_{0}(x-(10k+6)i) {\cal T}_{(8k,4k,4k-1)/(2k-1,2k-2)}(x+(-2k+1)i)  \Bigr )  \\
\end{eqnarray*}
where the duality is applied in the last equality.

The "height 3" diagrams are decomposed into smaller pieces due to the tableaux rule, see fig. \ref{decomp3L4kp1 }.
\begin{figure}[hbtp]
\centering
\includegraphics[width=8cm]{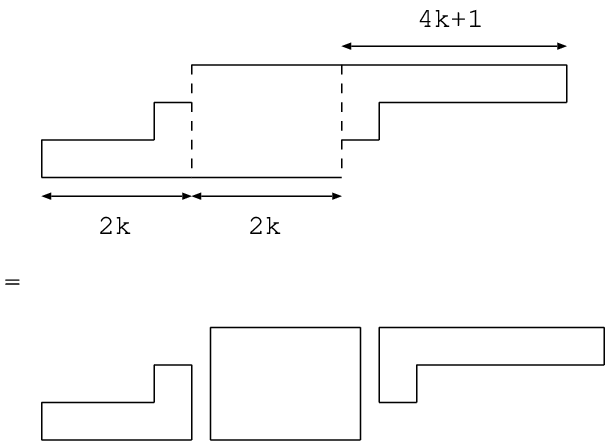}
\caption{ 
The decomposition of the    "height 3" diagram  corresponding to $   {\cal T}_{(8k+1,4k,4k-1)/(2k,2k-1)}  $ }
\label {decomp3L4kp1 }
\end{figure}

Accordingly,  $R$ is represented by

\begin{eqnarray*}
R&=& \frac{T_{2k}(x) \Lambda_{(4k+1,1) }(x+(2k+1)i) }
                  {\phi(x+(6k+\frac{9}{2})i)  \phi(x-(2k+\frac{3}{2})i)  \phi(x-(6k+\frac{1}{2})i)}      
     \\
&  &   \times    \Bigl (  T_{1}(x-(10k+5)i)    \Lambda_{(2k,2k)/(2k-1)}(x-(8k+2)i)     \\
&  & \qquad - T_0(x-(10k+6)i)  \Lambda_{(2k-1,2k-1)/(2k-2)}(x-(8k+1)i)  \Bigr ) .   %\label{Raltercfm1}
\end{eqnarray*}
Thus the lemma is established.
\end{proof}

The final step starts from rewriting
the first term in the bracket of   (\ref{RalterLakp1})  in Lemma \ref{altRL4kp1} as,

\begin{eqnarray*}
& &\frac{ \Lambda_{(1,1)}(x-(10k+5)i) }{\phi_2(x-(10k+5)i)}   \Lambda_{(2k,2k)/(2k-1)}(x-(8k+2)i)  \\
& & =\frac{\phi(x-(10k+\frac{7}{2})i)}{\phi(x-(10k+\frac{15}{2})i)}   \Lambda_{(2k+1,2k+1,1)/(2k)}(x-(8k+4)i)  \\
 & &\quad + T_0(x-(10k+6)i)  \Lambda_{(2k-1,2k-1)/(2k-2)}(x-(8k+1)i) .
\end{eqnarray*}

The second term in the rhs coincides with that in the bracket of  (\ref{RalterLakp1}).
Thus
\begin{eqnarray*}
R&=& \frac{T_{2k}(x) t^{(6)}_1(x+\frac{3i}{2L+4}  )  }
                  { \phi(x-(6k+\frac{1}{2})i)  \phi(x-(10k+\frac{15}{2})i) }     \\
&  &\times		  \Lambda_{(2k+1,2k+1,1)/(2k)}(x-(8k+4)i)  
\end{eqnarray*}		  
where  $\phi(x-(10k+\frac{7}{2})i)= \phi(x+(6k+\frac{9}{2})i)$ and the definition of $t^{(6)}_1$ is used.
Finally we apply  Lemma \ref{tate3tate2} to the above and reach the expression  (\ref{toshow_R_L4kp1}).
Thereby we prove the functional relation.

%###########################################
%-----------dA5 --------------------------
%
\section{A result for an exceptional case}
%
%
%#############################################

We sketch the result for  exceptional cases, $L=3 \sim 6$.

For  $L=3 , 4, 6$, the $Y$ system is neatly
described by 

$$
Y_{j}(x \pm  \frac{L-2}{2L+4} ) = \prod_{j \sim \ell }   \Xi_{\ell }(x) 
 ~~~~(j=1,\dots,r) 
$$
where the corresponding Dynkin diagrams are those for $E_r$,
and  $r=8,7,6$, respectively.

The solutions of $Y$ in terms of transfer matrices have been
reported in \cite{SuzE8} for $L=3$ and \cite{SuzE7}  for $L=4, 6$.

The case $L=5$, corresponding to 
 $M_{5,6}+\phi_{1,2}$  takes a little bit complicated as announced in \cite{annecyJS}.
 Below, we shall summarize the result then supplement outline of
 the proof as promised there.

In this case, there are more breathers in the system than the cases we discussed above.
We, therefore need to introduce more $sl_2$ origin objects.

\begin{eqnarray}
T_{B_1} (x) &:=& T_1 (x)         \label{defbr1} 
\\
T_{B_3} (x) &:=& \Lambda_{(8,1)}(x+\frac{13}{14}i)  /\phi(x-\frac{12}{7} i) \label{defbr2} 
\\
T_{B_5}(x) &:=& \Lambda_{(15,8,8)/(7,7)}(x)/\phi(x\pm \frac{3}{2} i)   \label{defbr3}
 \\
T_{B_7}(x) &:=& \Lambda_{(15,15,8,8)/(14,7,7)}(x+\frac{11}{14}i) /
 (\phi(x-\frac{12}{7} i)  \phi(x\pm \frac{9}{7}i))    \label{defbr4} 
 \\
T_{B_2}(x) &:=& T_7 (x)   \label{defbr5} 
\\
T^{(6)}(x) &:=& \Lambda_{(8,7)/(6)}(x+\frac{25}{14}i)  \label{defbr6} 
\end{eqnarray}
then the following relations hold.
\begin{eqnarray*}
T_{B_1} (x\pm \frac{3}{14} i)&=& T_0(x\pm \frac{11}{14}i) + \phi(x-\frac{12}{7} i) T_{B_3} (x)    \\
T_{B_3} (x\pm \frac{3}{14} i)&=& T_0(x) T_0(x \pm \frac{8}{14}i) + T_{B_1}(x) T_{B_5} (x)\\
T_{B_5} (x\pm \frac{3}{14} i)&=&  T_0(x\pm \frac{3}{14}i)  T_0(x\pm \frac{5}{14}i) +
   T_{B_3} (x) T_{B_7} (x)\\
T_{B_2} (x\pm \frac{3}{14} i)&=& T_0(x\pm \frac{1}{14}i)+ T^{(6)}(x) 
\end{eqnarray*}

For a magnon-like $t-$ system we assume,
\begin{eqnarray}
t^{(1)}_1 (x+\frac{i}{14})  t^{(1)}_1 (x-\frac{i}{14})  &=& 
   t^{(1)}_2 (x) +  t^{(3)}_1 (x) g^{(1)}_1(x)  \label{kinkt1A5} 
   \\
t^{(1)}_2 (x+\frac{i}{14})  t^{(1)}_2 (x-\frac{i}{14})  &=& 
   t^{(1)}_1 (x)  t^{(1)}_3 (x) +  t^{(3)}_2 (x) g^{(1)}_2(x)  \label{kinkt2A5}
   \\
t^{(1)}_3 (x+\frac{i}{14})  t^{(1)}_3 (x-\frac{i}{14})  &=& 
      t^{(1)}_2 (x)  t^{(1)}_4 (x)   \label{kinkt3A5}
	  \\
t^{(2)}_1 (x+\frac{i}{14})  t^{(2)}_1 (x-\frac{i}{14})  &=&
    t^{(2)}_2 (x) +  t^{(3)}_1 (x)  g^{(2)}_1(x)\label{kinkt4A5}
	\\
t^{(2)}_2 (x+\frac{i}{14})  t^{(2)}_2 (x-\frac{i}{14})  &=& 
    t^{(2)}_1 (x)  t^{(2)}_3 (x) +  t^{(3)}_2 (x)  g^{(2)}_2(x)  \label{kinkt5A5}
	\\
t^{(2)}_3 (x+\frac{i}{14})  t^{(2)}_3 (x-\frac{i}{14})  &=&  
      t^{(2)}_2 (x)  t^{(2)}_4 (x)  \label{kinkt6A5}
	  \\
t^{(3)}_1 (x+\frac{i}{14})  t^{(3)}_1 (x-\frac{i}{14})  &=& 
    t^{(3)}_2 (x) +  t^{(1)}_1 (x)  t^{(2)}_1 (x)  t^{(4)}_1 (x)  \label{kinkt7A5}
	\\
t^{(3)}_2 (x+\frac{i}{14})  t^{(3)}_2 (x-\frac{i}{14})  &=& 
   t^{(1)}_2 (x)  t^{(2)}_2 (x)  t^{(4)}_2 (x) \label{kinkt8A5}  
   \\
t^{(4)}_1 (x+\frac{i}{14})  t^{(4)}_1 (x-\frac{i}{14})  &=&
   t^{(4)}_2 (x) +  t^{(3)}_1 (x)   \label{kinkt9A5} 
   \\
t^{(4)}_2 (x+\frac{i}{14})  t^{(4)}_2 (x-\frac{i}{14})  &=&   t^{(3)}_2 (x)    \label{kinkt10A5}  
\\
g^{(a)}_1(x\pm \frac{i}{14}) &=& g^{(a)}_2(x), \qquad
a=1,2    \nonumber 
\end{eqnarray}
Remark: They are obtained from the T-system for $D_4$,
by requiring $t^{(3)}_3= t^{(4)}_3=0$ .  \par\noindent

The identification between $Y$ and $T$ is given by

\begin{eqnarray*}
Y_{B_1}&=&  \phi(x-\frac{12}{7} i) T_{B_3} (x) /  T_0(x\pm \frac{11}{14}i)   %\label{defUB1} 
\\
Y_{B_3}&=& T_{B_1} (x) T_{B_5} (x) /( T_0(x) T_0(x \pm \frac{4}{7}i))    %\label{defUB3}
\\
Y_{B_5}&=& T_{B_3} (x) T_{B_7}  /  (  T_0(x\pm \frac{3}{14}i)  T_0(x\pm \frac{5}{14}i)) 
%\label{defUB5}
\\
Y_{B_2}&=& T^{(6)} (x) / T_0(x\pm \frac{1}{14}i)  %\label{defUB2} 
 \\
Y_{K_1} (x)&=& \frac{t^{(1)}_2(x)}{t^{(3)}_1(x) g^{(1)}_1(x)}   %\label{defUK1}  
\\
Y_{K_2} (x)&= &\frac{t^{(2)}_2(x)}{t^{(3)}_1(x) g^{(2)}_1(x)}   %\label{defUK2}  
\\
 Y^{(1)} (x)&=& \frac{t^{(3)}_2(x)}{t^{(1)}_1(x) t^{(2)}_1(x) t^{(4)}_1(x)}  %\label{defU1}  
 \\
Y^{(2)} (x)&=& \frac{t^{(4)}_2(x)}{t^{(3)}_1(x) }.    %\label{defU2} 
\end{eqnarray*}

And the explicit solution is asserted as follows.
%\begin{proposition}
The following choice of $t^{(a)}_{1,2}(x)$ solves the kink T-system 
\begin{eqnarray}
t^{(1)}_1(x) &=& T^{(6)}(x)    \label{deft11} 
\\
t^{(2)}_1(x) &=& T_{B_7}(x)    \label{deft21}
\\
t^{(3)}_1(x) &=& \Lambda_{(12,8,7)/(5,4)}(x)  \label{deft31}
\\
t^{(4)}_1(x) &=& \Lambda_{(5,1)}(x+\frac{15}{14}i)  /\phi(x-\frac{13}{2}i+\frac{15}{14}i) \label{deft41}  
\\
t^{(1)}_2(x) &=&T_{B_5}(x) T_{B_2}(x\pm \frac{i}{7}) \label{deft12} 
\\
t^{(2)}_2(x) &=& T_{B_5}(x \pm \frac{i}{7}) T_{B_2}(x)     \label{deft22}
\\
t^{(3)}_2(x) &=&T_{B_5}(x\pm \frac{i}{14}) T_{B_2}(x\pm \frac{i}{14})  
\label{deft32}  
\\
t^{(4)}_2(x) &=& T_{B_5}(x) T_{B_2}(x)  \label{deft42}
\end{eqnarray}
with additional constraints,

\begin{eqnarray*}
t^{(1)}_3(x)&=& t^{(3)}_2(x),  %\label{cond1}  
\\
T_{B_3}  &=& \frac{t^{(2)}_3(x)}{t^{(3)}_2(x)}  %\label{cond2}  
\\
g^{(1)}_1(x) &=& T_0(x) , \qquad g^{(2)}_1(x)=T_0(x\pm \frac{2 i}{7}).  \nonumber
\end{eqnarray*}

%\end{proposition}

By these choice the above $Y$ satisfy the $Y-$ system proposed in \cite {DPT2},

\begin{eqnarray*}
Y_{B_1}(x\pm \frac{3}{14}i) &=& \Xi_{B_3}(x)    \\
Y_{B_3}(x\pm \frac{3}{14}i) &=& \Xi_{B_1}(x) \Xi_{B_5}(x)  \\
Y_{B_5}(x\pm \frac{3}{14}i) &=& \Xi_{B_3}(x) \Xi_{K_2}(x \pm  \frac{2}{14}   i) \Xi_{K_1}( x) 
\Xi^{(1)}(x \pm \frac{1}{14}  i) \Xi^{(2)}(x) \\
Y_{B_2}(x\pm \frac{3}{14}  i) &=&  \Xi_{K_1}(x \pm 2) \Xi^{(1)}(x \pm \frac{1}{14}  i) \Xi_{K_2}(x) 
\Xi^{(2)}(x)  \\ 
Y_{K_2}(x\pm \frac{1}{14}  i) &=& \Xi_{B_5}(x)  {\cal L}^{(1)}(x)  \\
Y_{K_1}(x\pm \frac{1}{14}  i) &=& \Xi_{B_2}(x) {\cal L}^{(1)}(x )  \\
Y^{(1)}(x\pm \frac{1}{14}  i) &=& {\cal L}^{(2)}(x) {\cal L}_{K_2}(x){\cal L} _{K_1}(x)  \\
Y^{(2)}(x\pm \frac{1}{14}  i) &=&  {\cal L}^{(1)}(x) .
\end {eqnarray*}

We first remark a few lemmas concerning about  equivalent expressions for transfer matrices
%-----------------
%
\begin{lemma} \label{manytb5}
$T_{B_5}(x)$  has  the following two equivalent  expressions in addition to (\ref{defbr3}).

\begin{equation*}
T_{B_5}(x)  = T_{D_2}(x) = 
    \frac{1}
	        {\phi(x\pm \frac{3}{2}i)}  \Lambda_{(8,8,1)/(7)} (x+\frac{12}{7}i) .
%
%\label{manytb5}
\end{equation*}  

\end{lemma}

The first equality has  already appeared in (\ref{anotherdeftb5} ).
The second is shown by the duality.
%-------------
\begin{lemma} \label{Lmanytb7}
$T_{B_7}(x) (=t^{(2)}_1(x))$  is also written as follows.

\begin{equation*}
T_{B_7}(x)  =
   \frac{1}
           {\phi(x+\frac{22}{14}i) } \Lambda_{(4,3,1)/(2)}(x-\frac{31}{14}i)  
%
%\label{manytb7}
\end{equation*}  
\end{lemma}

To prove this, we utilize  $ {\cal T}_{(15,15,8,8)/(14,7,7)}(x)$. 
The quantum Jacobi-Trudi formula and the duality relation conclude,
\begin{align}
{\cal T}_{(15,15,8,8)/(14,7,7)} (x)&={\Lambda}_{(15,15,8,8)/(14,7,7)} (x) \nonumber \\
&+T_0(x)T_0(x-2 i)\bigl ( T_1(x+17i) T_4(x-i) -T_0(x+16 i) T_3(x)  \bigr ).
\label{da5c1}
\end{align}

On the other hand, the same quantity is given by

\begin{equation}
{\cal T}_{(15,15,8,8)/(14,7,7)}(x)=
\phi(x+\frac{i}{2})  \phi(x+\frac{13i}{14})  T_7(x-11i)  \Lambda_{(8,8)/(7)}(x+10i) 
\label{da5c2}
\end{equation}
thanks to the semi-standard condition.
Again, by the  quantum Jacobi-Trudi formula and the duality relation,
$ \Lambda_{(8,8)/(7)}(x+10i) = T_1(x-7i) T_2(x+\frac{3}{7}i) -T_0(x-\frac{8}{7}i) T_1(x-2 i)$.
Using this expression, as well as with some rearrangement, we find
\begin{align*}
T_7(x-11i)  \Lambda_{(8,8)/(7)}(x+10i) 
&= T_1(x-7i) T_{D_2}(x-i) -T_0(x-\frac{8}{7}i) \Lambda_{(2,1)}(x)  \\
&+
T_0(x-i)(T_4(x-i)T_1(x-7i)  -T_0(x-\frac{8}{7}i)T_3(x)).
\end{align*}
 Substitute this into  (\ref{da5c2}) ,  compare the resultant expression with  (\ref{da5c1})
and we have,
$$
{\Lambda}_{(15,15,8,8)/(14,7,7)}(x)  = \phi(x+\frac{i}{2})  \phi(x+\frac{13i}{14}) 
  (T_1(x-7i) T_{D_2}(x-i)-T_0(x-\frac{8}{7}i) \Lambda_{(2,1)}(x) ).
$$
By the representation of  $T_1(x-7i)$ by  $2\times 1$ table, the diagrammatical rule  immediately leads to
$$
{\Lambda}_{(15,15,8,8)/(14,7,7)}(x)  =
\frac{ d_{B_7}(x-\frac{11}{14}i)}{\phi(x+\frac{11}{14}i)}     \Lambda_{(4,3,1)/(2)} (x-3i) , 
$$
where $d_{B_7}(x)$ specifies the denominator in  (\ref{defbr4}).
This proves the lemma after a shift in $x$ and the simplification in the coefficient.

\begin{lemma}
\label{manyt31}
$t^{(3)}_1(x)$  can be also represented by

\begin{equation*}
t^{(3)}_1(x)=
 \frac{1}{\phi(x \pm \frac{3}{2}i)} 
  \Lambda_{(7,7,5,1)/(6,2)}(x+\frac{12}{7}i).
%
%\label{manyt31}
\end{equation*}  
\end{lemma}
Just as in the proof of  Lemma \ref{Lmanytb7}, 
two expressions for ${\cal T}_{(12,8,7)/(5,4)}(x)$ leads to

\begin{align}
t^{(3)}_1(x) = \Lambda_{(12,8,7)/(5,4)}(x)=&
\frac{1}{\phi(x \pm \frac{3}{2}i)}  
 (
   \Lambda_{(5,1)}(x+8i) \Lambda_{(5,5)/(4)}(x+16 i)     \nonumber  \\
 & \quad   - \phi(x \pm \frac{3}{2}i) T_7(x) T_{D_2}(x) 
    ).
\label{t31c1}
\end{align}

The product   $\Lambda_{(5,1)}(x+8i) \Lambda_{(5,5)/(4)}(x+16 i)$ decomposes 
as follows.

\begin{align}
(\Lambda_{(5,1)}(x+8i) \Lambda_{(5,5)/(4)}(x+16 i) ) 
&=  (\Lambda_{(5,1)}(x-4i) \Lambda_{(5,5)/(4)}(x+4 i) )_{x->x+\frac{12}{7}i}   \nonumber \\
&= \Lambda_{(7,7,5,1)/(6,2)}(x+\frac{12}{7}i)+T_2(x+\frac{12}{7}i) \Lambda_{(8,8,1)/(7)} (x+\frac{12}{7}i)      \nonumber \\
&=  \Lambda_{(7,7,5,1)/(6,2)}(x+\frac{12}{7}i)+T_7(x)  T_{B_5}(x) \phi(x\pm \frac{3}{2}i )
\label{t31c2}
\end{align}
where we have used  Lemma \ref{manytb5}  in the last equality. 
Using (\ref{t31c2}) in (\ref{t31c1}) and remembering $T_{B_5}(x)=T_{D_2}(x)$, one easily establishes the lemma.

\begin{lemma}
\label{hugehuge}
The following equality can be easily verified,
\begin{equation*}
\Lambda_{( 6, 6, 4,3,1)/(5,3,2 )}(x)= 
\phi(x\pm \frac{3}{14}i)  T_{B_5}(x\pm \frac{i}{7}+\frac{12}{7}i).
%
%\label{hugehuge}
\end{equation*}  
\end{lemma}

The diagrammatic decomposition yields,
$$
\Lambda_{( 6, 6, 4,3,1)/(5,3,2 )}(x)=\frac{1}{f_2(x)} \Lambda_{(3,3,1)/(2)}(x \pm 5i) .
$$
The $m=3$ case of  Lemma \ref{tate3tate2} and $T_{B_5}(x)=T_{D_2}(x)$ then validates  the lemma..

With these preparations, we prove the magnon-like $t$ system. \par
\noindent  Proof of  (\ref{kinkt1A5} ) \par\noindent
With (\ref{deft11}), (\ref{defbr6}) and using the duality and the quantum Jacobi-Trudi formula, we can express
the product $t^{(1)}_1(x\pm \frac{1}{14}i) $ using only $T_0(x)$ and $T_2(x)$.
Similarly $t^{(3)}_1$ can be  rewritten by 
$T_0(x), T_2(x)$ and $T_4(x)$, according to (\ref{deft31}).
Then it is immediately seen,
$$
T_0(x) t^{(3)}_1(x) -t^{(1)}_1(x\pm \frac{1}{14}i)  =T_2(x\pm \frac{11}{7}i) T_{D_2}(x).
$$
Then the equality $T_{B_5}(x) = T_{D_2}(x), T_2(x+\frac{12}{7}i) =T_7(x)$ and the definition of $t^{(1)}_2(x)$ in
(\ref{deft12} ) concludes  (\ref{kinkt1A5} ).
\par
\noindent  Proof of  (\ref{kinkt2A5} ) \par\noindent

Consider the product $\Lambda_{(4,3,1)/(2)}(x-4i)  \Lambda_{(4,4,2)/(3,1)}(x+4i) $.
which is equal to 
\begin{equation}
\phi(x\pm \frac{3}{14}i) t^{(2)}_1(x\pm \frac{1}{14}i+\frac{12}{7}i)
\label{kt2c1}
\end{equation}
due to Lemma \ref{Lmanytb7}.

As is usual, the diagrammatical argument leads to
$$
\Lambda_{(4,3,1)/(2)}(x-4i)  \Lambda_{(4,4,2)/(3,1)}(x+4i) 
=\phi(x\pm \frac{3}{2}i)  \phi(x\pm \frac{5}{2}i)      \Lambda_{(7,7,5,1)/(6,2)}(x)
+
T_2(x) \Lambda_{( 6, 6, 4,3,1)/(5,3,2 )}(x).
$$
Then we apply Lemma \ref{manyt31} and  Lemma \ref{hugehuge} to the first and the second term in the rhs, respectively,
to reach
\begin{equation}
\phi(x\pm \frac{3}{14}i)( t^{(3)}_1(x+\frac{12}{7}i) T_0(x\pm \frac{2}{7}i+ \frac{12}{7}i) +
   T_{B_2}(x+ \frac{12}{7}i) T_{B_5}(x\pm \frac{1}{7}i+\frac{12}{7}i).
 \label{kt2c2}  
\end{equation}
By equating (\ref{kt2c1}) and   (\ref{kt2c2}), we show the validity of  (\ref{kinkt2A5} ) .

\noindent  Proof of  (\ref{kinkt4A5} ) \par\noindent

This is actually equivalent to  (\ref{t31c1}). 
Note that   $\Lambda_{(5,1)}(x+8i)= \phi(x+\frac{3}{2}i) t^{(4)}_1(x+\frac{1}{14}i  )$
and that the diagrammatic symmetry leads to 
$\Lambda_{(5,5)/(4)}(x-8i)= \phi(x-\frac{3}{2}i) t^{(4)}_1(x-\frac{1}{14}i)$.
Note also $T_7(x)=T_{B_2}(x)$.
Then one easily verifies  (\ref{kinkt4A5} ) using $t^{(4)}_2(x)$in  (\ref{deft42}).

\noindent  Proof of  (\ref{kinkt7A5} ) \par\noindent

This is the final nontrivial relation.
This relation can be simply shown by using the result for the $A_{4k+1}$ case.
We simply put $k=1$ and
identify $t^{(1)}_1,  t^{(2)}_1, t^{(3)}_1, t^{(4)}_1, t^{(3)}_2, g^{(1)}_1, g^{(2)}_1$ here with
$t^{(3)}_1,  t^{(4)}_1, t^{(5)}_1, t^{(6)}_1, t^{(5)}_2, t^{(1)}_1, t^{(2)}_1$ there
and use the duality $T_2=T^{\vee}_7$.

All remaining relations can be checked easily. Thus the above argument completes the proof of the
magnon-like $t$ system of the dilute $A_5$ model .

%###################################################
%
\section{Numerical Results for the dilute $A_8$ model in regime 2}
%
%####################################################
\label {numericsA8} 
We supplement some numerical data supporting the conjecture
of the analyticity of $Y$ functions in section \ref{Yanaly}
Throughout this appendix, we choose $\beta=0.4$ and $J=1$.

The solution to the Bethe ansatz equation, corresponding to 
the largest eigenvalue of QTM, assumes the form of two strings.
We list the explicit locations of roots for $N=10$, $q=0$.

\begin{tabular}{|l|}
\hline
BAE Roots $q=0$ \\
\hline
0.13548013966460354 $\pm$ 0.4464809553245548 i \\
0.03263577725129405 $\pm$ 0.4935687693509952 i    \\
0.002550182914375612  $\pm$  0.4961489632802383 i  \\
-0.02495383965703369 $\pm$   0.4946627475810687 i  \\
-0.09400161546158932 $\pm$   0.47389905045124336i \\
\hline
\end{tabular}

\begin{figure}[hbtp]
\centering
\includegraphics[width=8cm]{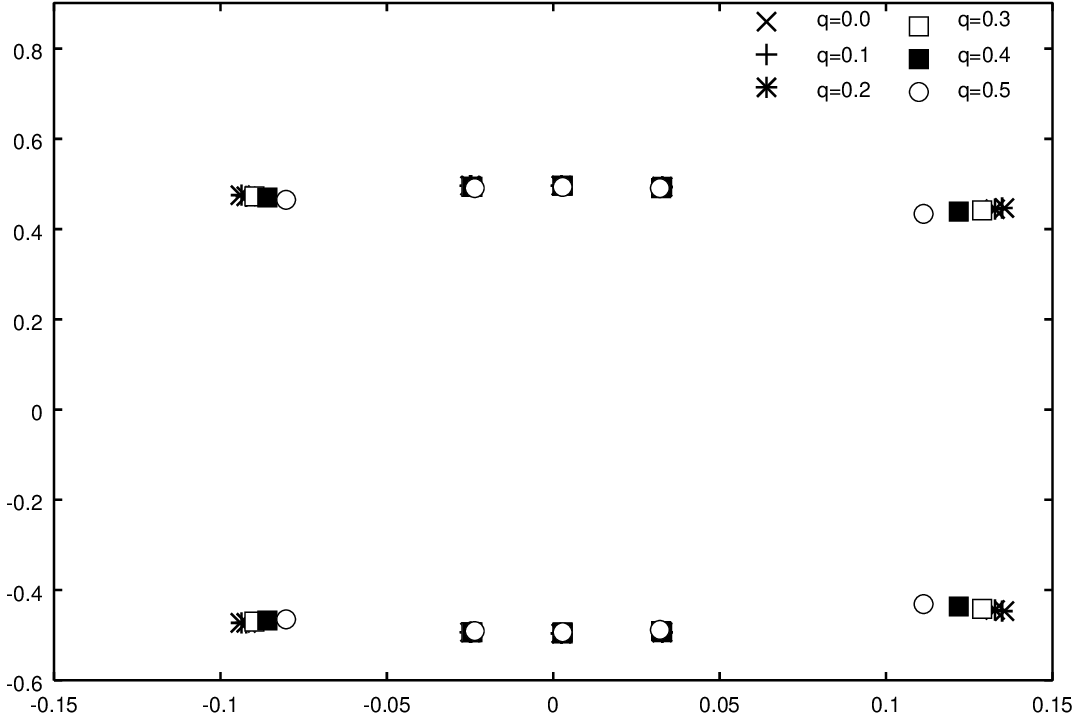}
\caption{The Bethe ansatz roots for $N=10$ for $q=0,0.1,0.2,0.3,0.4,0.5$.
Their location changes only slightly.}
\label{figroots}
\end{figure}

Their imaginary parts are approximately equal to $\pm\frac{1}{2}$,
 irrespective of $L, N, q, \beta$.
See figure \ref{figroots} .

\begin{figure}[hbtp]
\centering
{\includegraphics[width=6cm]{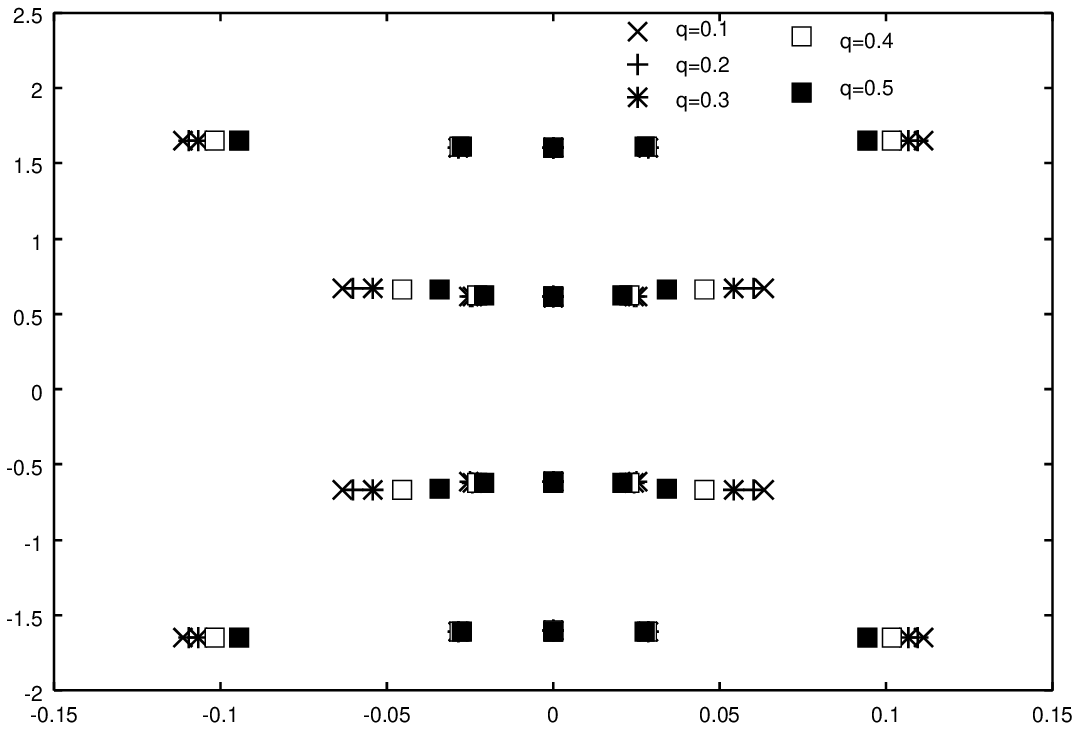} \hspace{1cm}
\includegraphics[width=6cm]{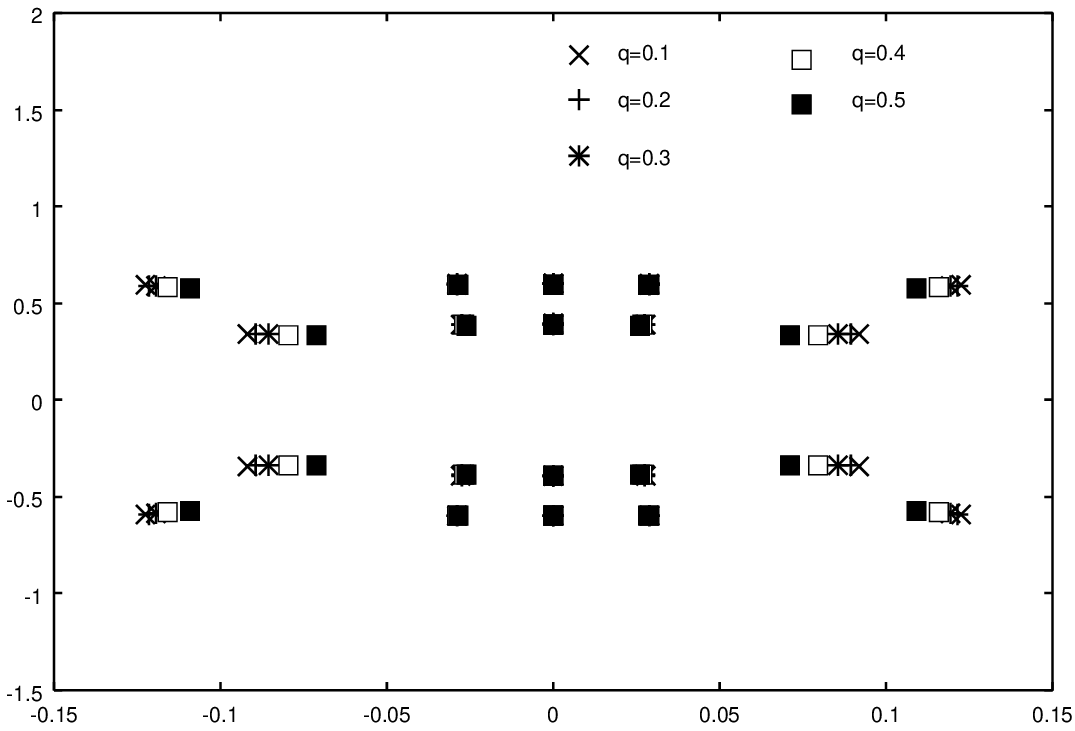}}
\caption{Zeros of $T_1(x)$(left)  and $T_2(x)$ (right)
for $N=10$ and $q=0.1,0.2,0.3,0.4,0.5$.
}
\label{t1N10}
\end{figure}

\begin{figure}[hbtp]
\centering
\includegraphics[width=6cm]{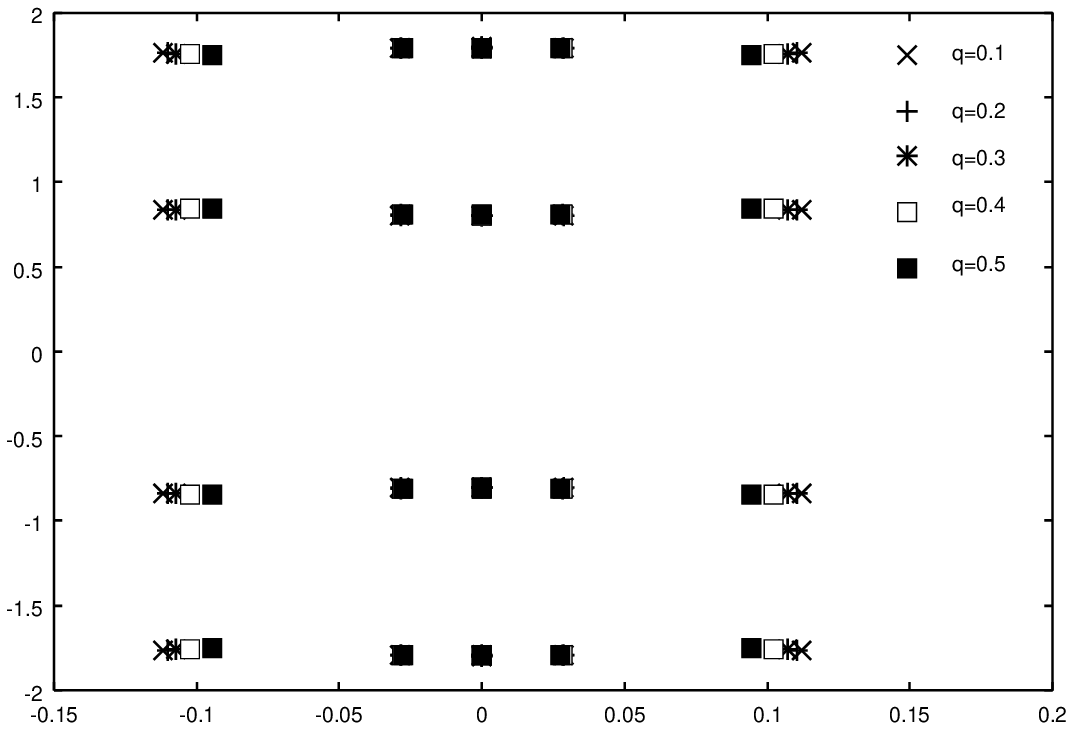} 
\caption{Zeros of $T_7(x)$
for $N=10$ and $q=0.1,0.2,0.3,0.4,0.5$.
}
\label{t7N10}
\end{figure}

The zeros for  symmetric fusion QTM, $T_m,\ (m=1,\cdots, 7)$, 
are plotted in fig. (\ref{t1N10})-fig. (\ref{t7N10}).
These figures clearly show that  the simple-minded choice of $Y$ in \ref{trialY1} is not appropriate:
 the choice requires $T_m$ should be analytic in the strip
$\Im x \in [-1,1]$. ( Or the arguments can be simultaneously shifted by half
period so that strip is $\Im x \in [0.8,1.8] \cup [-1.8,-0.8]$
Obviously, neither choice of strip results   set of $T_m$ free from zeros.
(Especially, $T_3$ AND $T_5$ break the analyticity in both strips.)
Therefore, the transformation of $Y$ system to TBA is not legitimate.

\begin{figure}[hbtp]
\centering
{\includegraphics[width=6cm]{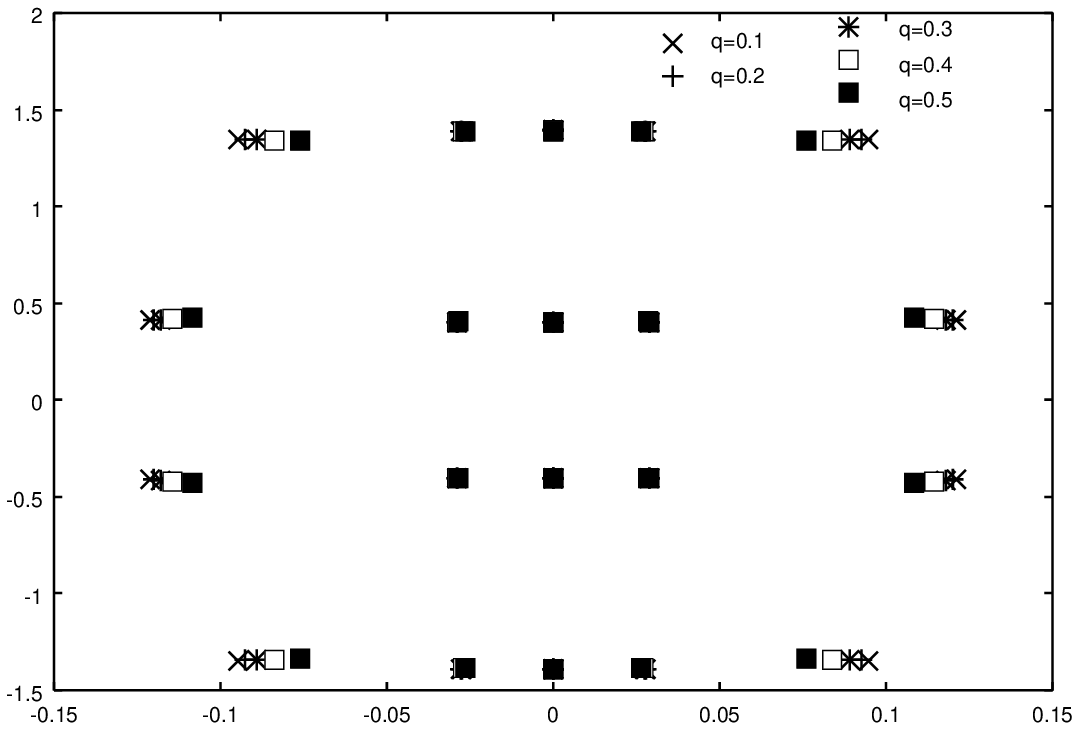} \hspace{1cm}
\includegraphics[width=6cm]{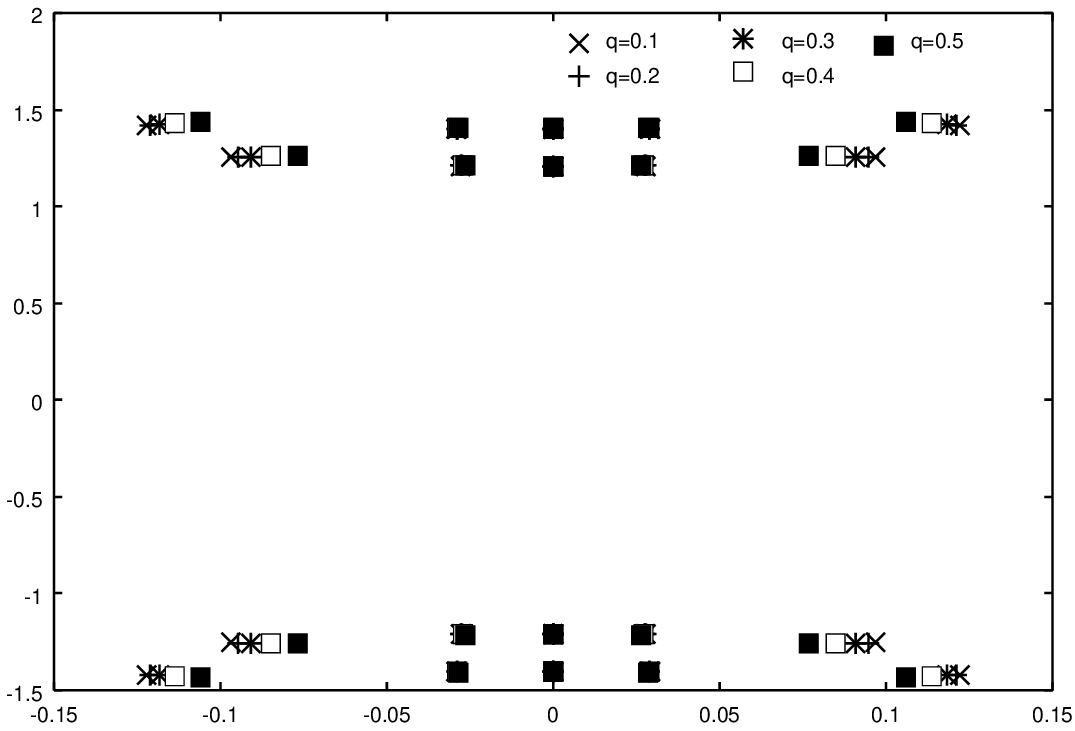}}
\caption{Zeros of $T_3(x)$(left)  and $T_4(x)$ (right) 
for $N=10$ and $q=0.1,0.2,0.3,0.4,0.5$.
}
\label{t3N10}
\end{figure}

\begin{figure}[hbtp]
\centering
{\includegraphics[width=6cm]{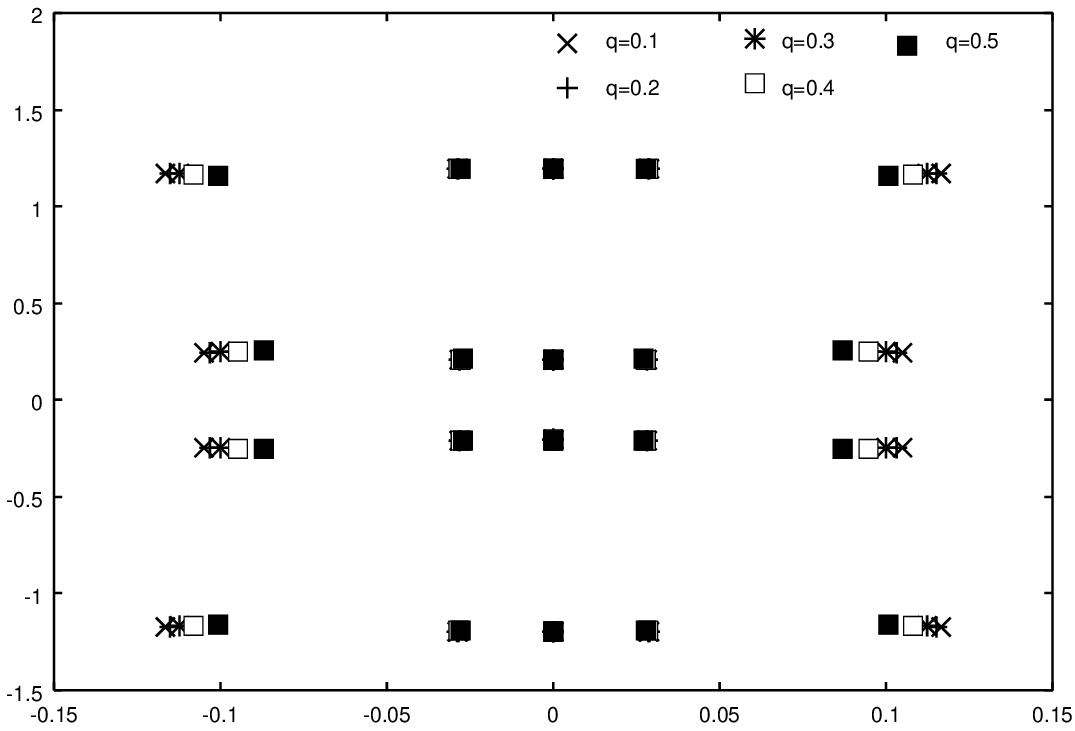} \hspace{1cm}
\includegraphics[width=6cm]{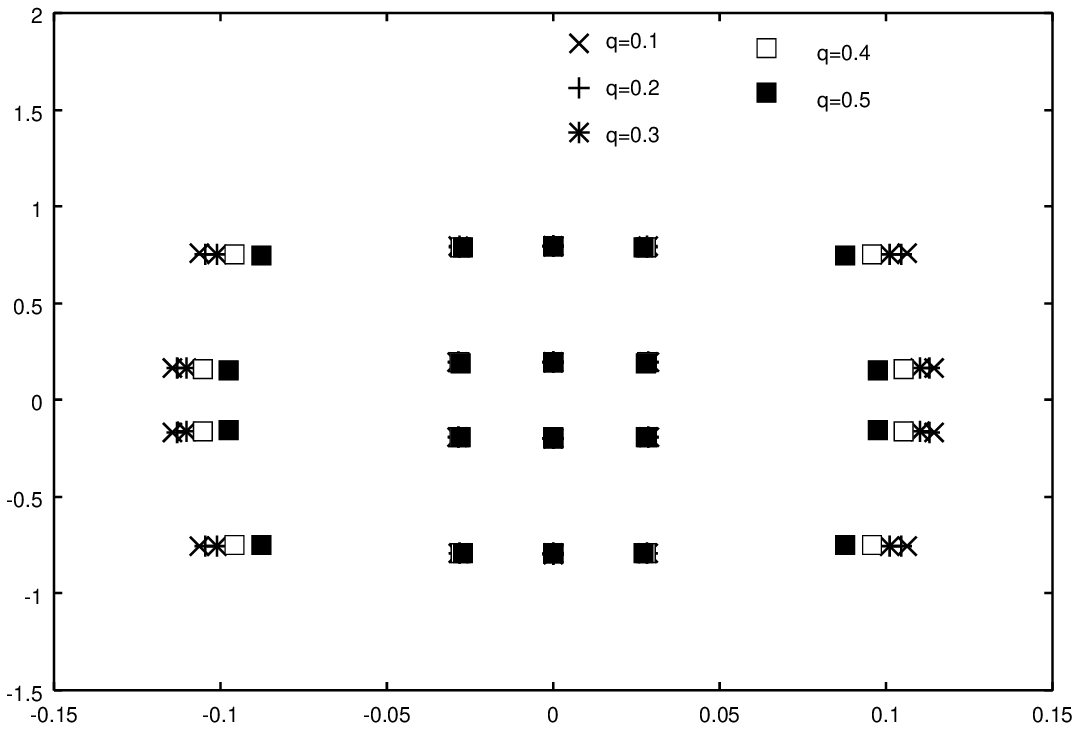}}
\caption{Zeros of $T_5(x)$(left)  and $T_6(x)$ (right) 
for $N=10$ and $q=0.1,0.2,0.3,0.4,0.5$.
}
\label{t5N10}
\end{figure}

The difference in the argument $x$ appearing in the breather $T$,
$sl_3$ $T$ and magnon-like $t$ is much smaller than unity, which is
crucial.

$T_{B_1}(x)$ is identified with $T_1(x)$. The functional relation
then implies that it should be analytic within $\Im x \in[-\frac{3}{10},
\frac{3}{10}]$. This can be easily checked from the figure \ref{t1N10}.
Among other $T_m$, $T_4(x)$ and $T_2(x+\frac{9}{5}i)$ 
are  identified with $t^{(1)}_1(x)$ and $t^{(3)}_2(x)$.
Their ANZC property in wider strip than $\Im x \in
[-\frac{1}{5},\frac{1}{5}]$
is clearly seen.

The breather related QTM, $T_{B_3}(x)$ is equal to $t^{(7)}_1(x)$,
and $T_{B_5}(x) (=T_{D_2}(x))$ is $t^{(4)}_2(x)$, thus the information
on their zeros is significant.
 Especially the latter appears frequently
in the solution of  other $t^{(a)}_m, (m=2,3)$.

\begin{figure}[hbtp]
\centering
{\includegraphics[width=6cm]{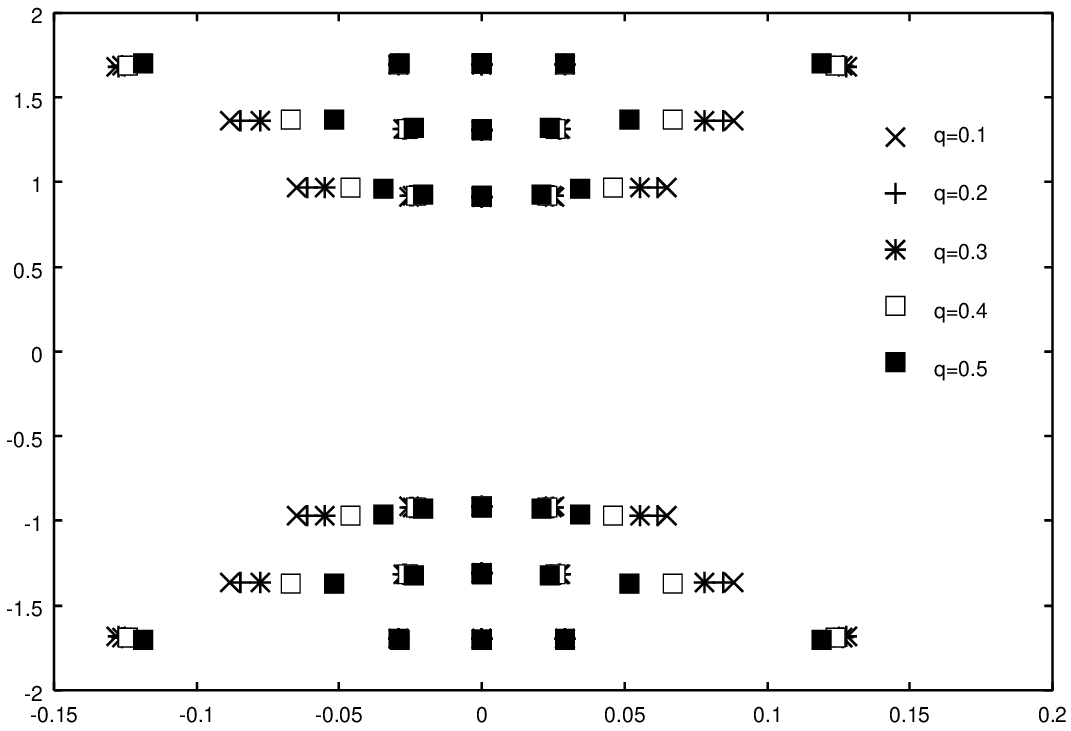} \hspace{1cm}
\includegraphics[width=6cm]{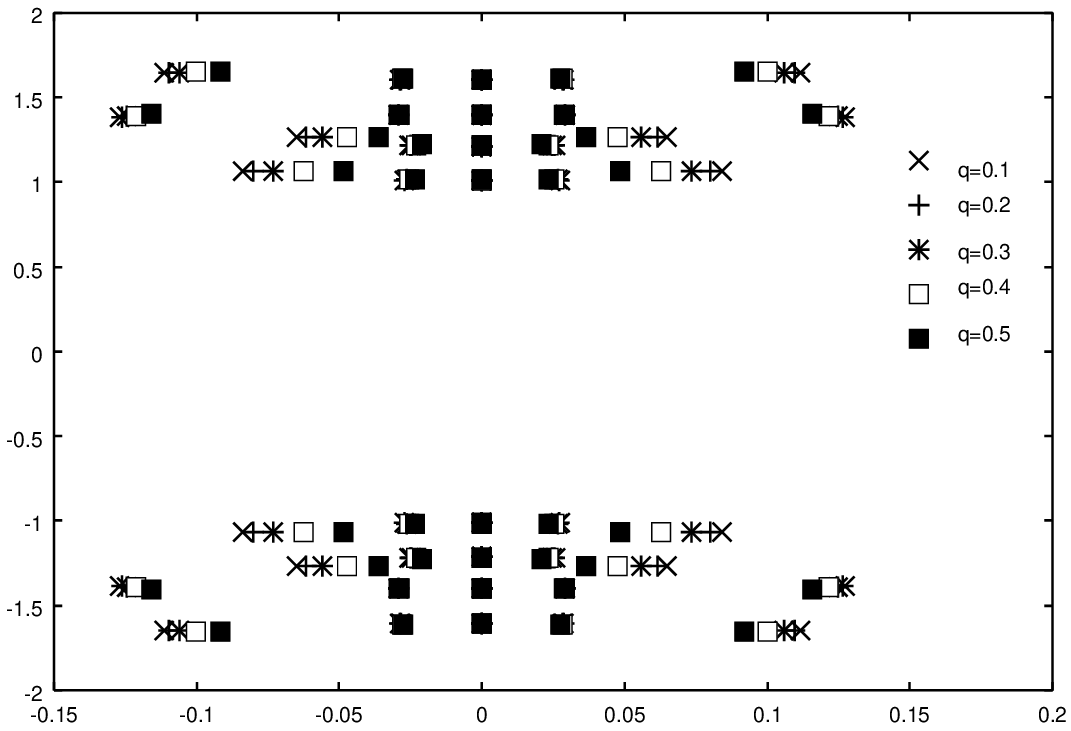}}
\caption{Zeros of $T_{B_3}(x)$(left)  and $T_{B_5}(x)(=T_{D_2})$ (right) 
for $N=10$ and $q=0.1,0.2,0.3,0.4,0.5$.
}
\label{breatherzeros}
\end{figure}

Their zeros are shown in fig. (\ref{breatherzeros})

\begin{figure}[hbtp]
\centering
{\includegraphics[width=6cm]{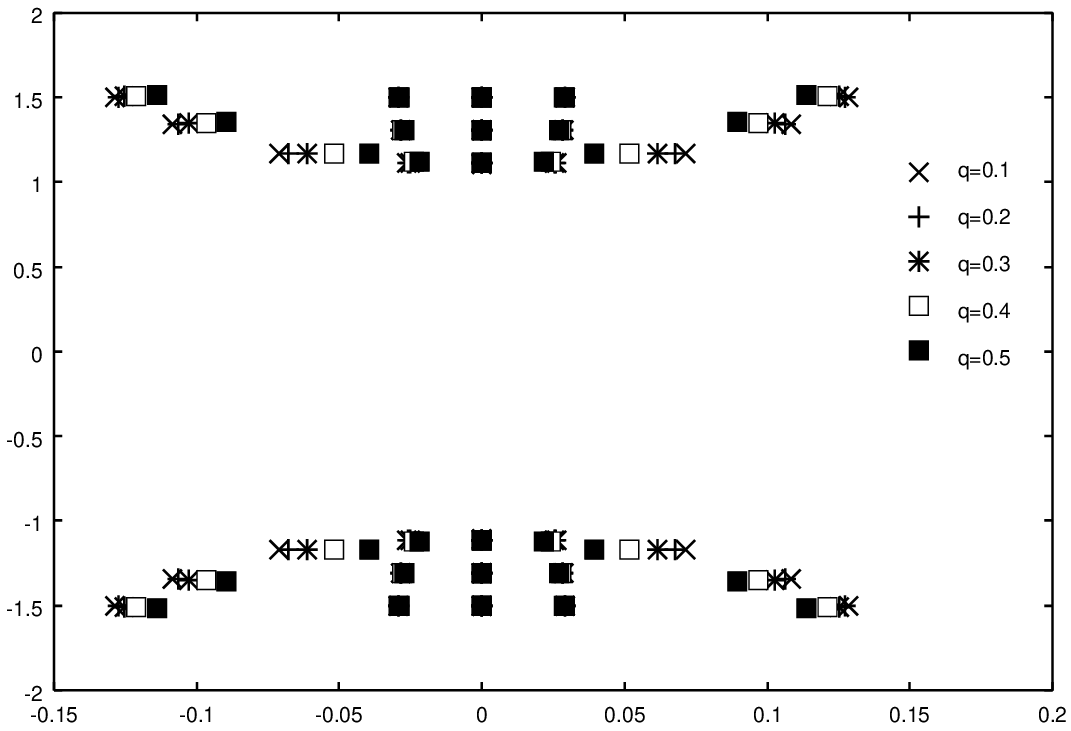} \hspace{1cm}
\includegraphics[width=6cm]{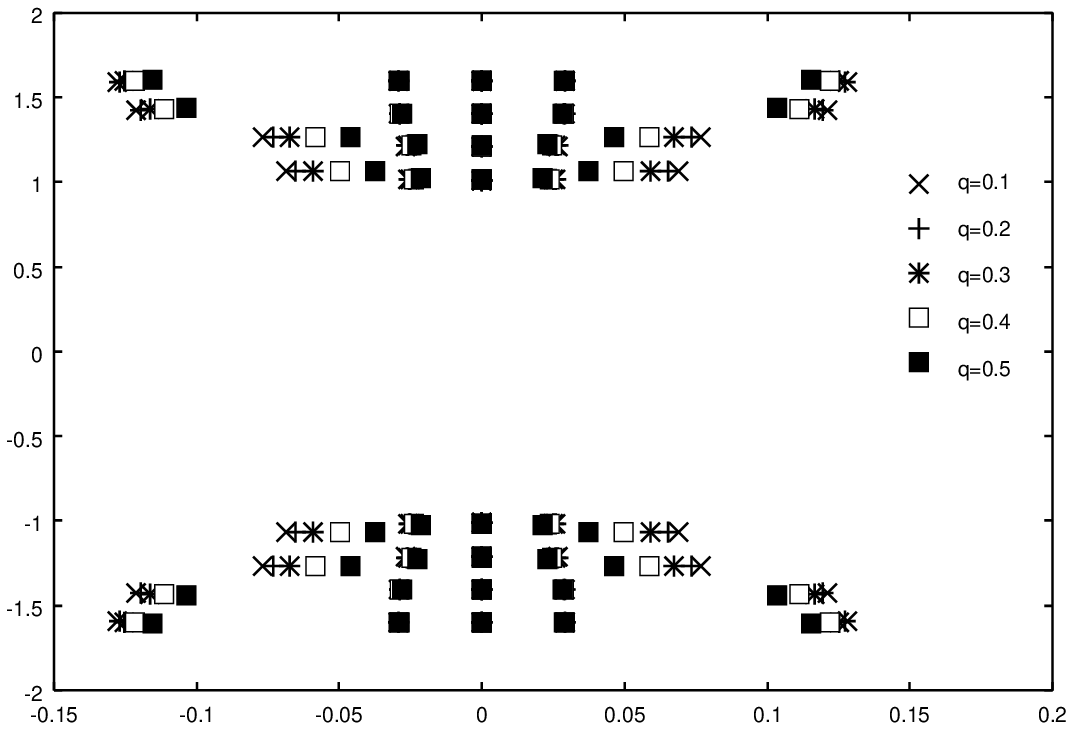}}
\caption{Zeros of $t^{(0)}_1(x)$(left)  and $t^{(2)}_1(x))$ (right) 
for $N=10$ and $q=0.1,0.2,0.3,0.4,0.5$.
}
\label{t01N10}
\end{figure}

\begin{figure}[hbtp]
\centering
{\includegraphics[width=6cm]{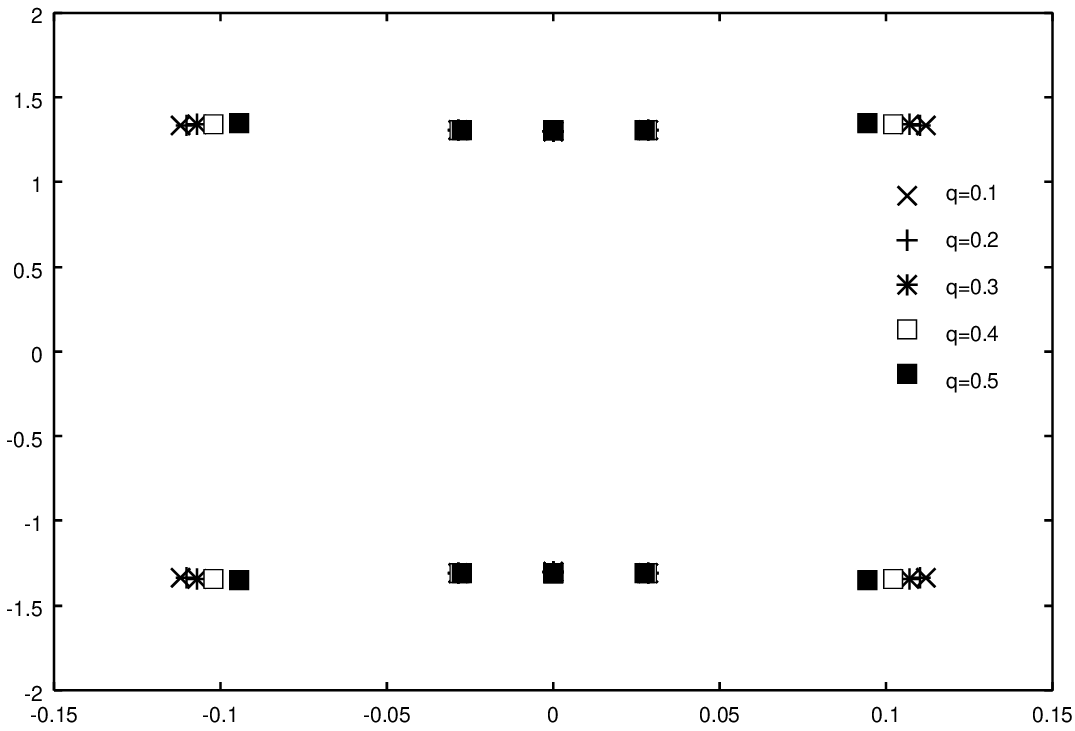} \hspace{1cm}
\includegraphics[width=6cm]{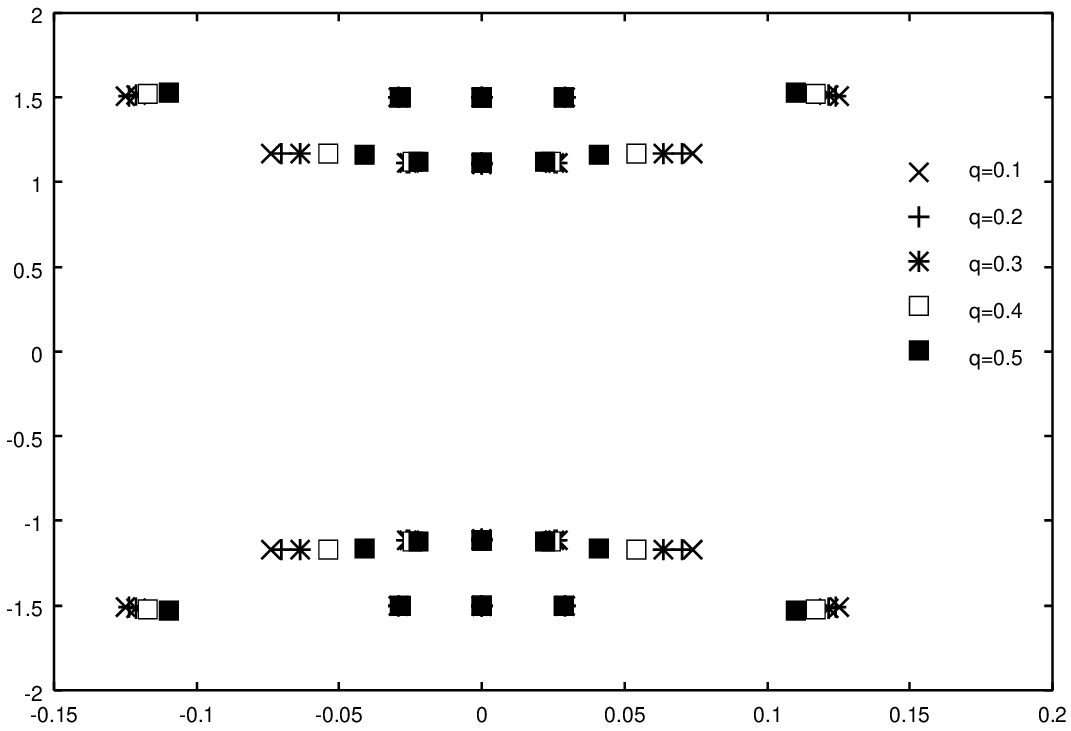}}
\caption{Zeros of $t^{(3)}_1(x)$(left)  and $t^{(4)}_1(x))$ (right) 
for $N=10$ and $q=0.1,0.2,0.3,0.4,0.5$.
}
\label{t31N10}
\end{figure}

The remaining zeros of QTM to be specified are those for
$t^{(0)}_1(x), t^{(2)}_1(x), t^{(3)}_1(x)$ and $t^{(4)}_1(x)$.
They are depicted in fig. (\ref{t01N10}) and fig.(\ref{t31N10}),
respectively.

By combining the above data on zeros, one can 
verify that all $Y$ and $1+Y$ possess no zeros or
singularities from QTM in their own strips.
Only trivial zeros (or singularity) 
of order $N$ exists for $Y_{B_1}$ simply coming
from normalization factor $\phi^{\vee}(x))$ (or $T_0(x)$).

%#############################################
%
\section{The kernel matrices for $L=5, 7$}
%
%
%#############################################
\label{kernelexample}

For $L=5$, we define the order as 
\begin{equation*}
^{t}  (   \widehat{\log} Y_{B_1},
                           \widehat{\log} Y_{B_3},
						   \widehat{\log} Y_{B_5} ,
							\widehat{\log} Y_{B_2},
						  \widehat{\log} Y_{K_1} ,
						    \widehat{\log} Y^{(1)} ,
							  \widehat{\log} Y^{(2)})
\end{equation*} 
then 
\begin{eqnarray*}
\widehat{M} &=&
\begin{pmatrix}
2\ch\frac{3}{14}k &        -1 &        0&     0&     0&     0&    0&    0  \\
-1      &   2\ch\frac{3}{14}k &      -1&      0&     0&    0&     0&   0   \\
0        &    -1    & 2\ch\frac{3}{14}k &     0&  -1&-2\ch\frac{2}{14}k&  -2\ch\frac{1}{14}k& , -1 \\
0&     0&   0&    2\ch\frac{3}{14}k & -2\ch\frac{2}{14}k& -1&  -2\ch\frac{1}{14}k&, -1     \\
0&     0&   0&    -1&  2\ch\frac{1}{14}k&  0&   0&   0   \\
0&   0&  -1& 0&  0&   2\ch\frac{1}{14}k&  0&  0  \\
0&  0&   0&   0&  0&   0&     2\ch\frac{1}{14}k& 0 \\
0&  0&   0&   0&  0&   0&    0&   2\ch\frac{1}{14}k\\ 
\end{pmatrix}  \\
\widehat{K}_0 &=&
\begin{pmatrix}
0 &        1 &        0&     0&     0&     0&    0&    0  \\
1      &  0 &      1&      0&     0&    0&     0&   0   \\
0        &    1    & 0 &     0&  1&2\ch\frac{2}{14}k&  2\ch\frac{1}{14}k& , 1 \\
0&     0&   0&    0 & 2\ch\frac{2}{14}k& 1&  2\ch\frac{1}{14}k&, 1     \\
0&     0&   0&    1&  0&  0&   -1&   0   \\
0&   0&   1&      0&  0&   0& -10&  0  \\
0&  0&   0&   0& -1&  -1&     0&-1 \\
0&  0&   0&   0&  0&   0&   -1& 0\\ 
\end{pmatrix}.
\end{eqnarray*}

For $L=7$, we define the order as 
\begin{equation*}
^{t}  (   \widehat{\log} Y_{B_1},
                           \widehat{\log} Y_{B_3},
						   \widehat{\log} Y^{(1)}_0 ,
							\widehat{\log} Y^{(2)}_0,
						  \widehat{\log} Y^{(1)}_1 ,
						    \widehat{\log} Y^{(2)}_1 ,
							  \widehat{\log} Y^{(3)}_1,
							   \widehat{\log} Y^{(4)}_1,
							    \widehat{\log} Y^{(5)}_1,
								 \widehat{\log} Y^{(6)}_1,
							  )
\end{equation*} 
then 

\begin{eqnarray*}
\widehat{M} &=&
\begin{pmatrix}
2\ch\frac{5}{18}k &        -1 &        0&     0&     0&     0&    0&    0&  0&  0 \\
-1      &   2\ch\frac{5}{18}k &      0&    -2\ch\frac{1}{18}k&     0&   -1&     0&   0&  0&  0   \\
0        &    0   & 2\ch\frac{4}{18}k &    -1&  0&0& 0& 0&0 &0 \\
0&     -1&   -1&    2\ch\frac{4}{18}k &  0&     0&    0&    0&  0&  0      \\
0&     0&   0&    0&  2\ch\frac{1}{18}k&  0&  -1&   0& 0&  0   \\
0&   0& 0& 0&  0&   2\ch\frac{1}{18}k&  0& -1&   0&  0  \\
0&  0&   0&   0&  -1&   0&     2\ch\frac{1}{18}k& 0 & -1& 0\\
0&  0&   0&   0&  0&   -1&    0&   2\ch\frac{1}{18}k& -1& 0\\
0& 0&  0&   0&   0&  0&   -1& -1&  2\ch\frac{1}{18}k& -1 \\
0& 0&  0&   0&   0&  0&   0& 0& -1&   2\ch\frac{1}{18}k\\
\end{pmatrix}  \\
\widehat{K}_0 &=&
\begin{pmatrix}
0&        1 &        0&     0&     0&     0&    0&    0&  0&  0 \\
1      & 0 &      0&     2\ch\frac{1}{18}k&     0&   1&     0&   0&  0&  0   \\
0        &    0   & 0 &    1&  -2\ch\frac{3}{18}k &0& -2\ch\frac{2}{18}k&-1& -2\ch\frac{1}{18}k&-1 \\
0&     1&   1&   0 &  0&     -2\ch\frac{3}{18}k &   -1&    -2\ch\frac{2}{18}k&  -2\ch\frac{1}{18}k&  -1     \\
0&     0& -1&    0&  0&  0&  1&   0& 0&  0   \\
0&   0& 0& -1&  0&   0&  0& 1&   0&  0  \\
0&  0&   0&   0&  1&   0&   0& 0 & 1& 0\\
0&  0&   0&   0&  0&   1&    0&   0& 1& 0\\
0& 0&  0&   0&   0&  0&   1& 1&  0& 1 \\
0& 0&  0&   0&   0&  0&   0& 0& 1&   0\\
\end{pmatrix} . \\
\end{eqnarray*}

%
%
%================================================================================
%

\end{document}